\documentclass[preprint,english,12pt,3p]{elsarticle}

\usepackage[T1]{fontenc}
\usepackage{color}
\usepackage{verbatim}
\usepackage{float}
\usepackage{mathrsfs}
\usepackage{amsbsy}
\usepackage{amstext}
\usepackage{amssymb}
\usepackage{graphicx}
\usepackage{url}
\usepackage{import}
\usepackage{hyperref}
\usepackage{babel}
\usepackage{amsmath}
\usepackage{array}
\usepackage{multirow}
\usepackage[table]{xcolor}
\usepackage{tabularx}
\usepackage{theorem}
\usepackage{tikz}
\usepackage{tikz-qtree}
\usepackage{pgfplots}
\usepackage{units}
\usepackage{xspace}
\usepackage{nomencl}
\usepackage{pdfpages}
\usepackage{graphics}
\usepackage{psfrag}
\usepackage{epstopdf}
\usepackage{latexsym}
\usepackage{color,soul}
\usepackage{etoolbox,ragged2e,siunitx}
\usepackage{subfig}

\makeatletter

\providecommand{\tabularnewline}{\\}

\usepackage{tikz} 
\usetikzlibrary{decorations}
\usetikzlibrary{shapes,arrows,chains}
\usetikzlibrary{trees}
\usepackage{pgfplots}
\pgfplotsset{compat=newest}
\usepackage{diagbox}
\usepackage{calc}
\usepackage{verbatim}
\usepackage{geometry}
\@ifundefined{showcaptionsetup}{}{%
 \PassOptionsToPackage{caption=false}{subfig}}
\usepackage{subfig}
\makeatother

\usepackage{babel}
\usepackage{listings}
\lstdefinelanguage{MyFortran}[08]{Fortran}{morecomment=[l]{\#},morestring=[d]',morekeywords={procedure,pass,deferred,non_overridable,generic,class,is}}
\lstdefinestyle{code}{%
  basicstyle=\footnotesize,%
  backgroundcolor=\color{gray},%
  language=MyFortran,%
  captionpos=b,%
  columns=fixed,%
  keepspaces=true,%
  xleftmargin=10pt,%
  numbers=none,%
  numberstyle={\tiny},%
  keywordstyle=\color{RoyalBlue},%
  texcl=true,%
  upquote=true,%
  commentstyle=\color{Maroon}%
}

\lstdefinestyle{codesimple}{%
  basicstyle=\footnotesize,%
  backgroundcolor=\color{gray},%
  captionpos=b,%
  columns=fixed,%
  keepspaces=true,%
  xleftmargin=10pt,%
  numbers=none,%
}

\begin{document}

\begin{frontmatter}
\title{\textit{UnDiFi-2D}: an Unstructured Discontinuity Fitting code for\\2D grids.}
\author[LC]{L. Campoli}
\ead{l.kampoli@spbu.ru}
\address[LC]{Saint-Petersburg State University, 7/9 Universitetskaya nab., St. Petersburg, %199034 
Russia}
\author[AS]{A. Assonitis}
\ead{alessia.assonitis@uniroma1.it}
\address[AS]{Dipartimento di Ingegneria Meccanica e Aerospaziale, Sapienza University, Rome, %Via Eudossiana 18, 00184,
Italy}
\author[MC]{M. Ciallella}
\ead{mirco.ciallella@inria.fr}
\address[MC]{Team CARDAMOM, INRIA, Univ. Bordeaux, CNRS, Bordeaux INP, IMB, UMR 5251, \\200 Avenue de la Vieille Tour, 33405 Talence cedex, France}
\author[AS]{R. Paciorri}
\ead{renato.paciorri@uniroma1.it}
%\address[RP]{Dipartimento di Ingegneria Meccanica e Aerospaziale, Sapienza University, Rome, %Via Eudossiana 18, 00184 
%Italy}
%
\author[AB]{A. Bonfiglioli}
\ead{aldo.bonfiglioli@unibas.it}
\address[AB]{Scuola di Ingegneria, Universit{\`a} degli Studi della Basilicata, Potenza, Italy}
\author[MC]{M. Ricchiuto}
\ead{mario.ricchiuto@inria.fr}
%\address[MR]{Team CARDAMOM, Inria Bordeaux Sud-Ouest, Talence, France}
%
\cortext[cor1]{Please address correspondence to L. Campoli}
\begin{keyword} CFD \sep Fortran \sep C \sep shock-fitting \sep shock-capturing, \sep unstructured grids
\end{keyword}
\begin{abstract}
\textit{UnDiFi-2D}, an open source (free software) {\textit{Un}structured-grid, \textit{Di}scontinuity \textit{Fi}tting} code, is presented. The aim of \textit{UnDiFi-2D} is to {model gas-dynamic discontinuities in {two-dimensional (\textit{2D})} flows as if they were \emph{true} discontinuities of null thickness that bound regions of the flow-field where a smooth solution to the governing PDEs exists}. {\textit{UnDiFi-2D} therefore needs to be coupled with an unstructured CFD solver that is used to discretize the governing PDEs within the smooth regions of the flow-field. Two different, in-house developed, CFD solvers are also included in the current distribution}. 

The main features of the \textit{UnDiFi-2D} software can be summarized as follows: \\
\vspace{-0.1cm}\\
\noindent \textbf{Programming Language} \textit{UnDiFi-2D} is written in standard Fortran 77/95; its design is highly modular in order to enhance simplicity of use, maintenance and {allow coupling with virtually any} existing CFD solver;\\
\vspace{-0.1cm}\\
\noindent \textbf{Usability, Maintenance and Enhancement} In order to improve the usability, maintenance and enhancement of the code also the documentation has been carefully taken into account. The \texttt{git} distributed versioning system has been adopted to facilitate collaborative maintenance and code {development};\\
\vspace{-0.1cm}\\
\noindent \textbf{Copyrights} \textit{UnDiFi-2D} is a free software that anyone can use, copy, distribute, change and improve under the GNU Public License version 3.\\

\noindent The present paper is a manifesto of {the first public release of the} \textit{UnDiFi-2D} code. {It} describes the currently implemented features, which are the result of more than a decade of still ongoing {CFD} developments. This work is focused on the computational techniques adopted and a detailed description of the main characteristics is reported. \textit{UnDiFi-2D} capabilities are demonstrated by means of examples test cases. The design of the {code allows to easily include} existing CFD codes and is aimed at ease code reuse and readability.
\end{abstract}
\end{frontmatter}{}

\tableofcontents{}

\subparagraph*{Program summary}

\begin{flushleft}
\emph{Program title}: \textit{UnDiFi-2D}
\par\end{flushleft}

\begin{flushleft}
\emph{Catalogue identifier}: 
\par\end{flushleft}

\begin{flushleft}
\emph{Program summary URL}: 
\par\end{flushleft}

\begin{flushleft}
\emph{Program obtainable from}: CPC Program Library, Queen's University, Belfast, N. Ireland
\par\end{flushleft}

\begin{flushleft}
\emph{Licensing provisions}: {GNU General Public Licence, version 3}
\par\end{flushleft}

\begin{flushleft}
\emph{No. of lines in distributed program, including test data, etc.}: 380222
\par\end{flushleft}

\begin{flushleft}
\emph{No. of bytes in distributed program, including test data, etc.}:  3734808
\par\end{flushleft}

\begin{flushleft}
\emph{Distribution format}: github repository
\par\end{flushleft}

\begin{flushleft}
\emph{Programming language}: Fortran; developed and tested with Intel
Fortran Compiler v. 18.0.3 and GNU gfortran.
\par\end{flushleft}

\begin{flushleft}
\emph{Computer}: Any computer system with a Fortran compiler is suited. \par\end{flushleft}

\begin{flushleft}
\emph{Operating system}:designed for POSIX architecture and tested
on GNU/Linux one. 
\par\end{flushleft}

\begin{flushleft}
\emph{Has the code been vectorized or parallelized?}: no.
\par\end{flushleft}

\begin{flushleft}
\emph{Classification}: 
\par\end{flushleft}

\begin{flushleft}
\emph{External routines}: the code depends on several libraries and third-party packages which are detailed in the corpus of the text.
%The proprietary libraries {[}1{]} must be already present on the machine to visualize solutions. Other required libraries are shipped with the code.
\par\end{flushleft}

\begin{flushleft}
\emph{Nature of problem}: numerical computation of flows with discontinuities.
\par\end{flushleft}

\begin{flushleft}
\emph{Solution method}: shock-fitting technique.
\par\end{flushleft}

\begin{flushleft}
\emph{Restrictions}: At present, \textit{UnDiFi-2D} is validated for inviscid  steady and unsteady two-dimensional flows without changes in the number of discontinuity lines and interaction points.
\par\end{flushleft}

\begin{flushleft}
\emph{Unusual features}: \textit{UnDiFi-2D} implements a shock-fitting algorithm and can be coupled with unstructured cell-vertex solvers, with an Arbitrary Langrangian-Eulerian (ALE) formulation.
\par\end{flushleft}

\begin{flushleft}
\emph{Additional comments}: \textit{UnDiFi-2D} project adopts \texttt{git} {[}1{]}, a free and open source distributed version control system. A public repository dedicated to \textit{UnDiFi-2D} project {[}2{]} has been created on \texttt{github} {[}3{]}, a web-based hosting service for software development projects using \texttt{git} versioning system. Finally, a comprehensive documentation is provided in the form of user manual developed in Pandoc {[}4{]}. %parsing source code comments by means of \texttt{doxygen} software {[}5{]}.
\par\end{flushleft}

\begin{flushleft}
%\begin{comment}

%\end{comment}
\par\end{flushleft}
\section{Introduction\label{sec:Introduction}}

\subsection{Background}
{Open-source Computational Fluid Dynamics (CFD) tools are nowadays gaining increasing popularity among both CFD users and developers. Software packages like \texttt{OpenFoam}~\cite{doi:10.1063/1.168744} and \texttt{SU2}~\cite{palacios_stanford_2013}, to name the two most popular ones, are not just CFD codes, but general-purpose multi-physics packages capable of simulating several fluid-related phenomena typically encountered in a wide range of engineering applications.  
Access to the source code makes them appealing also to CFD developers, because the cost of learning how to code new algorithms inside an existing infrastructure can be significantly less (and a {\em una-tantum} effort) than that required for developing a new in-house code from scratch}.
%Other public-domain CFD codes that are currently being actively developed and maintained include CoolFluid~\cite{coolfluid-web-page}, and PyFr~\cite{WITHERDEN20143028}, to name just a few}.   

{Regardless of whether commercial or open-source codes are used, when it comes to simulating high-speed compressible flows, all CFD codes rely upon the so-called shock-capturing approach to model gas-dynamic discontinuities, such as shock-waves and slip-lines}.
Shock-capturing, which can be traced back to the work by von Neumann and Ritchmyer~\cite{doi:10.1063/1.1699639} lays its foundations in the mathematical theory of weak solutions, which allows to compute all kind of flows, including those affected by shock-waves, using the same discretization of the conservation-law-form of the governing equations at all grid cells. This yields obvious consequences in terms of coding simplicity, since the same set of operations is repeated within all control volumes of the mesh, no matter how complicated the flow might be. {Coding simplicity comes not for free, however, and shock-capturing calculations of shocked-flows are plagued by several troubles, all of which
are rooted to the fact that ``the thickness of a von Neumann shock is unacceptable''~\cite{Moretti1998}.}

{Whether it was ``von Neumann's reputation [that] helped the shock-capturing myth to become a religion''~\cite{Moretti1998} or coding simplicity, shock-capturing modeling of gas-dynamic discontinuities is the {\em de-facto} standard in modern CFD codes and has overshadowed an even older shock-modeling technique, 
which dates back to a World War II NACA technical report by Emmons~\cite{emmons-TN-1944}, see also~\cite{emmons1948flow}.}
{Starting in the late 1960s~\cite{moretti:66}, Emmons' technique has been re-discovered and further developed by Gino Moretti and co-workers under the name of ``shock-fitting'' and, since the 1980s~\cite{glimm1981front,GLIMM1985259,chern1986front}, by James Glimm and co-workers under the name of ``front-tracking''}.

Fitting/tracking methods consist in first locating and then tracking the motion of the discontinuities, which are treated as boundaries between regions where a smooth solution to the governing partial differential equations (PDEs) exists. The flow variables on the two sides of the discontinuities can be analytically computed by using the Rankine-Hugoniot (R-H) jump relations, algebraic equations connecting the states on both sides of the discontinuity and its local speed. Then, this solution is used to compute the space-time evolution of the discontinuity, that is, to track its motion. 

{Shock-fitting and front-tracking methods share many similarities and differ primarily in the range of applications addressed, with shock-fitting leaning toward gas-dynamics and front-tracking toward other kind of interfaces.}
{We shall hereafter primarily refer to shock-fitting methods, simply because Gino Moretti has been very influential in the Italian academic community, despite the fact that most of his professional career took place in the Americas.}

In the early days of the CFD era, {when
computers were very slow compared to present-day standards,} shock-fitting methods enjoyed a remarkable popularity, {because~\cite{moretti2002thirty-six} ``the more discontinuities are fitted, the fewer grid points are needed''. In addition to this, shock-fitting methods are immune from most of the numerical troubles (such as carbuncles~\cite{pandolfi2001numerical} and accuracy degradation downstream of a captured shock~\cite{doi:10.2514/2.835}) incurred by shock-capturing}.
{Despite these clearly recognized advantages~\cite{Johnsen20101213,pirozzoli2011numerical}, the tracking of the shocks and their
interactions raise a number of topological and logical problems that make shock-fitting algorithms harder to code than shock-capturing ones. Over the years, algorithmic complexity seems
to have alienated the majority of CFD practitioners from shock-fitting or, put in Moretti's words~\cite{moretti2002thirty-six}: ``Analysts who are more familiar with calculus than logic shy
away from shock-fitting and prefer to pay the price of inaccuracy''}.

{As a consequence, only a handful of research teams are nowadays still developing and using shock-fitting algorithms. To the best of the authors' knowledge, and apart from isolated examples, the only groups that in recent years have published a significant amount of literature on the subject are: {\it i}) Xiaolin Zhong and co-workers from UCLA, who perform high-order shock-fitting DNS studies of high-speed flows~\cite{zhong2012direct}; {\it ii}) the LLNL group headed by Tariq Aslam, who uses high-order shock-fitting methods to simulate explosions~\cite{ROMICK2017210} and {\it iii}) Marcello Onofri and Francesco Nasuti~\cite{nasuti2017steady} from the University of Rome, ``La Sapienza'', who developed a second-order-accurate, multi-block, shock-fitting code to simulate the flow through propulsive nozzles}.
{The aforementioned references are clearly not exhaustive, but should only be considered as entry points to a much larger body of literature.

All three aforementioned research teams developed shock-fitting algorithms using structured meshes and the use of structured meshes is deemed responsible for at least some of the algorithmic difficulties encountered when coding shock-fitting algorithms; see for instance reference~\cite{Rawat20106744} which details the implementation of high-order differencing schemes within the shock-fitting framework.}

{Taking advantage of the gradual shift that has taken place in the CFD community from structured towards unstructured grids, two of the authors of this paper started developing a shock-fitting algorithm that inherits features of Moretti's technique while taking advantage of the geometrical flexibility offered by the use of unstructured triangular and tetrahedral meshes.}
The unstructured shock-fitting technique described in this paper made its first journal appearance in 2009~\cite{paciorri2009shock-fitting}; at that time the algorithm was capable of simulating steady, two-dimensional flows featuring {only one ``fitted''} shock-wave. Multiple shocks could be handled using a hybrid approach, whereby only one shock was fitted and all other discontinuities and their mutual interactions were captured. Later developments~\cite{ivanov2010computation, paciorri2011shock} made the algorithm capable of fitting contact discontinuities (beside shocks), shock--shock and shock--wall interactions. An order-of-accuracy analysis of the steady, two-dimensional scheme has been conducted in~\cite{bonfiglioli2014convergence} showing that shock-fitting allows to preserve the design order of the spatial discretization scheme within the entire shock-downstream region. The  unstructured, shock-fitting algorithm {was} also used to simulate real-gas effects in hypersonic, two-dimensional steady flows~\cite{pepe2014shock-fitting, pepe2014towards}, whereas time-accurate simulations of two-dimensional, unsteady flows {are} reported in~\cite{bonfiglioli2016unsteady,campoli2017and}. Finally, the technique has been recently used to simulate two-dimensional, laminar and turbulent viscous flows featuring shock-wave/boundary-layer interactions~\cite{Assonitis2020124}.

{A significant number of international collaborations have been established during the almost fifteen years of algorithmic development.}

The first collaboration was established with Mikhail Ivanov of Itam, Russian Academy of Sciences. It dealt with the numerical study of weak Mach reflections by means of the {shock-fitting} technique and led to the development of the numerical model for the treatment of the triple points~\cite{ivanov2010computation}. 

{Joint work~\cite{pepe2014towards} with Andrea Lani, at that time at the Von Karman Institute for Fluid Dynamics (VKI) in Belgium, led to a C++ version~\cite{lani2017sf} of the shock-fitting algorithm described in this paper and its coupling with the public-domain CFD code CoolFluid}~\cite{coolfluid-web-page}. 
This activity showed that the modular approach adopted in the shock-fitting code allows to {plug in} different gas-dynamic solvers with {a limited} coding effort.

{The ongoing collaboration with Mario Ricchiuto at INRIA Bordeaux focused, at first, on the coupling between the shock-fitting algorithm described in this paper and yet another CFD code, \texttt{NEO}, with the aim of improving the capability of the shock-fitting algorithm to deal with unsteady flows~\cite{bonfiglioli2016unsteady,campoli2017and}. More recent joint work borrows ideas from Shifted Boundary Methods (SBM)~\cite{Scovazzi3} and has led to what we call ``Extrapolated Discontinuity Tracking'' (EDiT). 
EDiT has a number of distinctive features compared
to the shock-fitting technique described in this paper;
for this reason it is not currently included in the
distribution and the interested reader is referred to~\cite{ciallella2020extrapolated} for details}.

{Even though all the developments described so far apply to two-dimensional flow configurations, a three-dimensional version of the algorithm appeared in 2013 in a journal publication~\cite{bonfiglioli2013three-dimensional}; it is capable of dealing with steady flows featuring one or more fitted shocks but, in contrast to the two-dimensional case, the interaction between different shocks cannot be ``fitted'', but only ``captured''. The aforementioned limitation is expected to be overcome thanks to an on-going collaboration with Carl Ollivier-Gooch from the University of British Columbia in
Vancouver. The algorithm that he has been developing~\cite{zaide2014inserting,zaide2016inserting,zaide2017inserting} for inserting a surface as an internal boundary into existing unstructured meshes, although originally motivated by different applications, 
naturally finds its way in the present shock-fitting algorithm for the reasons that will be clarified in Sect.~\ref{sec:algorithm}}. 

{Finally, Lorenzo Campoli, who contributed to the development of the time-accurate version of the algorithm~\cite{bonfiglioli2016unsteady,campoli2017and} while} {he was} {a PhD student at the University of Rome, ``La Sapienza''}, {now based} at St.\  Petersburg State University is the promoter of the project described in the present paper.

Apart from the aforementioned collaborations, {the increasing awareness within the CFD community of the benefits offered by shock-fitting is confirmed by the increasing number of research teams that, over the last few years, have been actively developing shock-fitting/tracking methods using unstructured grids.}

The group headed by Jun Liu at Dalian University of Technology, China, has developed a Mixed Capturing and Fitting Solver (MCFS) by combining a shock-fitting algorithm, {in many respects similar to the one described here}, with an existing shock-capturing, cell-centered Finite Volume (FV) solver~\cite{Liu2017,zou2017shock,Chang2019,ZOU2021104847}. 

{Shock-fitting/front-tracking ideas made their way also through the Finite Element community. We refer to the SUPG technique of~\cite{DAQUILA2021110096} and the Discontinuous Galerkin {(DG)} Finite Element methods (FEM) independently developed by two different research teams:~\cite{Zahr2018b,zahr2020implicit}
and~\cite{Corrigan2019moving,Corrigan2020}.
All three aforementioned techniques
simultaneously solve for the location of the grid-points,
in addition to the flow-variables, so as to constrain certain edges of the tessellation to be aligned with the discontinuities.
The use of shape-functions that are continuous across the element interfaces, which is the case with SUPG, or discontinuous, such as in DG, has implications on how discontinuities are fitted. Similarly to the algorithm described in this paper, in the SUPG-FEM of~\cite{DAQUILA2021110096} the discontinuities are internal boundaries of zero thickness: by doing so, a finite jump in the dependent variables can take place while crossing the discontinuity. On the other hand,
numerical methods that employ a data representation which is discontinuous across
the cell interfaces, which is the case with DG-FEM, but also with cell-centred FV methods,
allow to fit discontinuities as a collection of edges of the mesh, without introducing internal boundaries}.

The {rising interest towards shock-fitting/front-tracking techniques} for simulating compressible flows gave 
rise to the idea of making our shock-fitting code publicly accessible in order to further {promote its} collaborative development. 

{In its current stage, 
the project combines contributions from research teams at four different research institutions and is willing to further extend the list of contributors}. 

The present and future versions of the code can be downloaded  in the open-source repository made available at \url{https://github.com/UnDiFi/UnDiFi-2D}. {At present,} only the two-dimensional version of the algorithm is going to be released, because it is the one that has been {most actively} developed over the years and has reached a sufficient level of maturity and generality.

\subsection{Related {software packages}\label{subsec:other}}
%To the best of the author's knowledge, there are no similar initiatives publicly available, well documented, actively developed and maintained. Nevertheless, few groups have been working on similar but different approaches.

{As already mentioned},
a C++ implementation of the shock-fitting algorithm {described in this paper}, {has been} developed {in collaboration with} the Von Karman Institute for Fluid Dynamics (VKI). It is described in~\cite{lani2017sf} and it has been made publicly accessible at  \url{https://github.com/andrealani/ShockFitting}. {At the time of writing, however, it appears to be not yet fully operational}.

{The FronTier++ library package~\cite{Frontier++}, developed and maintained at Stony Brook University, demonstrates the capabilities achieved by the decades long activity in developing front-tracking algorithms by James Glimm and co-workers~\cite{glimm1981front, glimm1998front, glimm1998three, glimm1999simple, chern1986front, du2006simple}. According to the FronTier++ website~\cite{Frontier++}}, the library provides
''high resolution tracking for contact discontinuities and internal boundaries in continuum medium simulations''.

{An extensive set of sample problems\footnote{\url{http://www.ams.sunysb.edu/~linli/FTruns/index.html}}, both in two and three space dimensions, shows the wide range of applications that can be dealt with using FronTier++}. %Compared to the current project, however, FronTier++ appears to be not so much focused on tracking gas-dynamic discontinuities}.

{Moreover, according to~\cite{SHE2016383}, the front-tracking algorithm has been inserted in the plasma-physics code FLASH~\cite{flashcode}.}

\subsection{Motivations and aims}
{The \textit{UnDiFi-2D} code
has been developed} with the aim of having an effective method to definitely cure most if not all the issues and pathologies affecting traditional shock-capturing solutions and to satisfy the following requirements:
\begin{itemize}
\item it is publicly and freely available;
\item it is well documented; 
\item it allows easy maintenance and enhancement {within} a collaborative framework;
\item it is actively maintained and developed.
\end{itemize}
\subsection{Documentation\label{subsec:Documentation}}
Free, open source softwares have often poor documentation. This lack compromises the diffusion of this kind of codes and makes their usage very difficult. Since \textit{UnDiFi-2D} has also a didactic purpose,  its easiness and accessibility is taken into account by shipping it with an in-depth user-manual.
% \textit{UnDiFi} documentation is based on comprehensive comments placed directly into the source code files without the necessity of other external files. This choice keeps the tree of the project clean and lightweight. These documentation comments are parsed by means of \texttt{doxygen}\footnote{Doxygen is a documentation system for many programming languages freely available at \url{http://www.stack.nl/\textasciitilde{}dimitri/doxygen.}} free software producing high quality \texttt{html} and \texttt{latex} documentation pages. In particular, the in-code comments allow keeping together the theory (latex-style formula are allowed inside comments) and practice greatly simplifying the code understanding.

\subsection{Collaborative framework\label{subsec:Collaborative-framework} }
Nowadays, one of the keys for the success of a free software is its capability to be easily maintained and improved {within} a collaborative framework. For this reason, the \textit{UnDiFi-2D} code adopts \texttt{Git}\footnote{Git is a free and open source distributed version control system designed to handle everything from small to very large projects with speed and efficiency, see \url{https://git-scm.com.}}
as distributed versioning system and a public repository dedicated to {the} \textit{UnDiFi-2D} project has been created on \texttt{github}\footnote{Github is a web-based hosting service for software development projects using git versioning system and it is free for open source projects, see \url{https://github.com.}}. It facilitates the tracking of each modification while, by means of repositories, a worldwide collaboration is fostered.
\subsection{\textit{UnDiFi-2D}: general description of the directory tree\label{subsec:UnDiFi}}
The main directory \texttt{UnFiDi-2D} contains the following sub-directories:
\begin{enumerate}
    \item \texttt{bin}: where all the executables are {installed};
    \item \texttt{lib}: where {various} libraries and their source codes are stored;
    \item \texttt{doc}: which contains the documentation;
    \item \texttt{source}: where the source files of the \textit{UnDiFi-2D} code are {stored};
    \item \texttt{source\_utils}:  contains
    \begin{itemize} 
    \item the source files of various I/O {format} converters;
    \item the \texttt{Triangle}~\cite{Triangle,shewchuk1996engineering} mesh-generator;
    \end{itemize}
    \item \texttt{tests}: contains the various test-cases described in Sect.\ref{sec:applications};
    \item \texttt{tools}: contains the source code of the \texttt{f77split} and \texttt{f90split} programs~\cite{Burkardt++};
    \item \texttt{EulFS.3.7}: where the source files of the \texttt{EulFS}~\cite{bonfiglioli2000fluctuation,bonfiglioli2013mass-matrix} gas-dynamic solver {are stored}.
    \item \texttt{NEO}: where the source files of the \texttt{NEO} solver~\cite{arpaia2014ale, ricchiuto2010explicit, ricchiuto2015explicit} are stored;
\end{enumerate}
In addition to these sub-directories, the main  directory contains the script (\texttt{compile\_all.sh}) {which} compiles all the software packages.

Full description on how to download, compile and run the code can be found at the documentation page \url{https://github.com/UnDiFi/UnDiFi-2D/wiki}. %Here, only the main algorithmic features of \textit{UnDiFi} are presented.

%{\color{red}Qui venivano dette cose poi ripetute nella sezione \ref{sec:algorithm}; ho preferito snellire, anche perch\`e ci troviamo nell'introduzione, dopotutto. Si faceva anche riferimento alla Fig.~\ref{fig:workflow} che per\`o compare parecchie pagine dopo. Oppure bisognerebbe portare qui la  Fig.~\ref{fig:workflow} e la relativa descrizione. Il testo precedente l'ho comunque solo commentato e si pu\`o recuperare facilmente.}
%The unstructured discontinuity fitting algorithm is described in detail in Sect.~\ref{sec:algorithm} while Fig.~\ref{fig:workflow} shows a typical workflow. The flowchart structure resembles very closely the function call sequence performed by the code, I/O files and branches for unsteady test-cases are highlighted. Depending on which gasdynamic solver is used, not all files and function calls may be present. The name of the functions is quite self-explaining, nevertheless, an in-depth description is provided in the code and in the documentation. The main purpose here is to help the reader to follow the main steps of the \textit{UnDiFi-2D} code, to recognise the coupling points with the gasdynamic solver, the conversion calls and to familiarize with I/O files.
%\input{workflow}
%{\color{red}Nella versione precedente veniva inoltre riassunto l'algoritmo, parlando di ''computational mesh'' e ''shock-mesh'', cose che vengono descritte nel dettaglio solo pi\`u oltre.}

{The directory tree highlights} the fact that the software is made up of three key components:
\begin{enumerate}
    \item the shock-fitting module \textit{UnDiFi-2D} which handles the motion of the discontinuities and their interactions (if any), but also drives the other two components, i.e.
    \item the gas-dynamic solver, either \texttt{EulFS} or \texttt{NEO}, which is used to discretize the governing PDEs in smooth regions of the flow-field; 
    \item the meshing software \texttt{Triangle}, which is used to locally re-mesh while the discontinuities move throughout the computational domain.
\end{enumerate} 
Communication among the driver \textit{UnDiFi-2D}, the gas-dynamic solver and the meshing software is handled using format converters (to be found in the \texttt{source\_utils} folder) that rely on disk I/O.
This programming approach {is certainly} not the best from the standpoint of computational efficiency, {one of the reasons being that} one has to switch among the different data-structures used by the three different modules. However, this approach is very convenient, since it allows us to use {off-the-shelf} gas-dynamic solvers and mesh generation tools that are treated as black boxes and can be replaced by similar ones only by changing the {format converters}, with a modest coding effort.
\subsection{Copyrights\label{sec:copyrights}}
\textit{UnDiFi-2D} is a free software. The authors encourage anyone to use, copy, distribute, study, change and improve the code. \textit{UnDiFi-2D} is distributed under the GNU Public License version 3. Users are kindly requested to cite the present paper when publishing results obtained by means of \textit{UnDiFi-2D}.
\subsection{Paper layout\label{sec:layout}}
The present article is organized as follows. The mathematical model and the numerical methods adopted are presented in Sect.~\ref{sec:numerics}. The unstructured shock-fitting algorithm is described in Sect.~\ref{sec:algorithm}. A selected set of representative numerical results is given in Sect.~\ref{sec:applications}, whereas Sect.~\ref{sec:development} addresses some currently unsolved issues and presents ongoing work aimed at overcoming these limitations.

\section{Fluid flow model and CFD codes\label{sec:numerics}}
This section is devoted to the description of the compressible flow equations solved in \textit{UnDiFi-2D}, as well as of the numerical methods implemented in the CFD solvers included in the distribution. We will detail these aspects sufficiently for the reader to appreciate {the} numerical results discussed in {Sect~\ref{sec:applications}}. For more information, the reader is referred to the extensive bibliography provided {hereafter}, as well as to the documentation included with the distributed codes.

\subsection{Governing equations\label{sec:equations}}
The current version of \textit{UnDiFi-2D} allows to simulate compressible, non-viscous, non-heat-conducting, perfect gases modelled by the Euler equations which can be written in   conservative form as:
\begin{equation}
\mathbf{U}_{t}+\nabla\cdot \mathbf{\mathcal{F}} =0 
\label{eq:Euler0}
\end{equation}
where, $U$ and $\mathbf{\mathcal{F}}$ denote the arrays of conservative variables and fluxes defined as: 
\begin{equation}
\mathbf{U}=\left(\begin{array}{c}
\rho \\
\rho E \\
\rho\mathbf{u}
\end{array}\right),
\qquad \mathcal{F}=\left(\begin{array}{c}
\rho\mathbf{u}\\
\rho\mathbf{u}H\\
\rho\mathbf{uu}+p\mathbf{I}
\end{array}\right)
\label{eq:Euler1}
\end{equation}
with standard notation for all physical variables.
%%%%%%%%%%%%%%%%%%%%%%
%%%%%%%%%%%%%%%%%%%%%%
The numerical methods implemented in the CFD codes distributed in \textit{UnDiFi-2D} often require the use of the flux Jacobians. If $d$ denotes the number of space dimensions, given a direction $\boldsymbol{\xi}\in{\mathbb{R}}^{d}$, it is useful to define the matrix:
\begin{equation}
\mathbf{K}_{\boldsymbol{\xi}}^{\mathbf{U}}= \sum_{i=1}^{d}\mathbf{A}_{i}^{\mathbf{U}}\xi_i \qquad\mbox{where}\qquad \mathbf{A}_{i}^{\mathbf{U}}=\frac{\partial\mathcal{F}_{i}}{\partial\mathbf{U}}
\label{eq:Jacobian0}
\end{equation}
The hyperbolic character of {\eqref{eq:Euler0}} guarantees that $\mathbf{K}_{\boldsymbol{\xi}}$ admits
a set of real eigenvalues, with a set of eigenvectors which are linearly independent almost everywhere. In particular, one can show that:
\begin{equation}
\mathbf{K}_{\boldsymbol{\xi}}^{\mathbf{U}}=
\mathbf{R}_{\xi}^{\mathbf{U}}\mathbf{\mathbf{\boldsymbol{\Lambda}}}_{\xi}\mathbf{L}_{\xi}^{\mathbf{U}}
\label{eq:Jacobian1}
\end{equation}
%%%%%%%%%%%%%%%%%%%%%%
having denoted by  $\mathbf{\boldsymbol{\Lambda}}_{\xi}$ the diagonal
matrix of the eigenvalues, and with $\mathbf{R}_{\xi}^{\mathbf{U}}$ and $\mathbf{L}_{\xi}^{\mathbf{U}}=\left(\mathbf{R}_{\xi}^{\mathbf{U}}\right)^{-1}$ the matrices whose columns/rows represent the corresponding eigenvectors. For system~\eqref{eq:Euler0}-\eqref{eq:Euler1} the eigenvalues can be shown to be given by the wave speeds:
\begin{equation}
\mathbf{\Lambda}_{\xi}=\textrm{diag}\left(\mathbf{u\cdot \boldsymbol{\xi},\,u\cdot \boldsymbol{\xi},\,u\cdot \boldsymbol{\xi},\,u\cdot \boldsymbol{\xi}}+a,\,\mathbf{u\cdot \boldsymbol{\xi}}-a\right)\label{eq:-20}
\end{equation}
For brevity, we omit the explicit form of the eigenvectors which can be computed  by standard means,
see e.g.~\cite{IDOLIKE}. The above eigenvalues represent the components along $\boldsymbol{\xi}$ of the wave speeds
carrying the information in the flow. If $\boldsymbol{\xi}$  is the velocity direction, one recovers the well known 
key role played by the Mach number $\|\mathbf{u}\|/a$ in determining whether all information travels downstream,
or some of it goes back upstream via the acoustic wave travelling at speed $\mathbf{u\cdot \boldsymbol{\xi}}-a$.
This is  one of the acoustic Riemann invariants,   will be referred to
as the negative acoustic Riemann invariant~\cite{IDOLIKE,anderson}, and denoted by 
$R^-$ .
\subsection{Shocks,  Rankine-Hugoniot relations, and weak solutions\label{subsec:RH}}
An important property of system {\eqref{eq:Euler0}} is that, as all nonlinear conservation laws, discontinuous solutions may occur in finite times even with smooth initial/boundary conditions. In this case, the differential form of the model is not the relevant one.
Weak discontinuous solutions are defined as solutions to the differential form in the sub-domains
empty of discontinuities, connected across each point of the  discontinuity by the Rankine-Hugoniot (or jump) relations:
\cite{anderson,IDOLIKE,onofri2017shock}
{\begin{equation}
w_{n}\,[\![\mathbf{U}]\!]=[\![\mathcal{F}]\!] \cdot \boldsymbol{n}
\label{eq:RH}
\end{equation}
}
where by $[\![\cdot]\!]$ we have denoted the jump of a quantity, with {%$\mathcal{F}_{\boldsymbol{n}}
$\mathcal{F} \cdot \mathbf{n}
$}
the flux in the direction normal to the {discontinuity}, and with {$w_{n}$} the speed  {of the discontinuity along the direction} $\boldsymbol{n}$.

Compressible flows modelled by the Euler system develop discontinuities of two kinds:
\begin{itemize}
    \item shock waves for which the flow goes into the discontinuity.
The (relative) upstream  Mach number is higher than unity, so in~\eqref{eq:RH} all the upstream quantities are given data. The  (relative) downstream Mach number is lower than one, which means that  the downstream negative acoustic Riemann invariant is also transported toward the discontinuity. So {constantness of $R^-$, together with Eq.~\eqref{eq:RH},} allows to have a nonlinear algebraic system with enough relations to compute the downstream state, and the shock speed;
    \item {slip-lines (or slip-streams)} with respect to {which} the (relative) normal flow speed is zero. 
      In this case one finds trivially that that there is no jump in the pressure. The main unknowns
      are in this case the (unique) value of the pressure and of the {discontinuity} speed/normal flow component
      which are compatible with the given data, and in particular with the two acoustic Riemann invariants transported into the shock on either side of the discontinuity, thus again this is a closed algebraic system.
\end{itemize}
In simple {flow} configurations, {such as those addressed in Sect.~\ref{subsec:SS1}, \ref{subsec:SS2} and~\ref{subsec:RR}}, the above rules allow to compute exact solutions to discontinuous flows.

The resulting solutions can be also characterized in terms of an entropy which is transported along {the} streamlines and has a finite jump across discontinuities. In particular, for steady flows with homo-entropic boundary conditions, we know that the entropy is constant before the discontinuity,
and conserved along streamlines after the shock. Moreover, entropy increases along physically acceptable discontinuities (see e.g.~\cite{TADMOR1986211,anderson}). For the Euler equations with perfect gas equation of state a possible definitions of entropy  is:
\begin{equation}
    \label{eq:entropy}
    S = \dfrac{p}{\rho^{\gamma}} + S_0
\end{equation}
with $\gamma$ the ratio of the specific heats at constant pressure/volume, and $S_0$ a reference value depending on the boundary/initial conditions.

\subsection{Residual distribution methods\label{sec:RDmethods}}
As noted in Sect.~\ref{subsec:UnDiFi}, \textit{UnDiFi-2D} has been coupled to %several   different gas-dynamic solvers. This paper however focuses on the results obtained with 
{two different codes}, \texttt{EulFS} and \texttt{NEO}, bearing many similarities in terms of numerical methods used, and both  included in the \textit{UnDiFi-2D} repository. 

\texttt{EulFS} and \texttt{NEO} provide  different implementations of a class of numerical methods known
as  Residual Distribution (RD) or Fluctuation Splitting (FS) schemes~\cite{ar2017,dr2017}. 
In its most classical formulation, the  RD approach provides discrete approximations of the compressible Euler equations  on simplicial grids, starting from values of the dependent variables  stored at the vertices of the  mesh. The second order variant of the methods exploits a classical continuous piece-wise linear finite element interpolation of the unknowns, for  which we will  use the standard notation $\mathbf{U}^h$, with $h$ the mesh size.
  
The main idea behind these methods is summarized in Fig.~\ref{fig:RDconcept}: discrete equations for the steady state values of the unknowns are assembled over each element. The steady solution can be obtained as the limit of the (pseudo-)time iteration:
  \begin{equation}
      \label{eq:RD0}
    |C_i| \dfrac{\mathbf{U}^{n+1}-\mathbf{U}^{n}}{\Delta t} + \sum\limits_{T\ni i}\Phi_i^T = \text{B.C.s}
  \end{equation}
  with $\Delta t$ a (pseudo-)time step, and $|C_i|$ usually taken as the area of the median dual obtained joining the gravity centers of the cells surrounding $i$ to the mid-points of the edges running into the node (the polygonal-shaped boundaries in Fig.~\ref{fig:RDconcept}). The {\em nodal} fluctuations (or nodal residuals) verify
in each triangle, {$T$}   the consistency constraint:
\begin{equation}
      \label{eq:RD1}
      \sum\limits_{j \in T}\Phi_j^T = \Phi^T := \int\limits_T \nabla\cdot \mathcal{F}(\mathbf{U}^h) \, {\mathrm{d}V}
  \end{equation}
  In Eq.~\eqref{eq:RD0} the right hand side represents the boundary condition terms, which are left out of the discussion.
Several design criteria exist for \eqref{eq:RD0}-\eqref{eq:RD1}, impacting both the practical evaluation of the \emph{element} fluctuation (or residual) $\Phi^T$, as well as the definition of the nodal residuals \eqref{eq:RD0}. Among the most important we mention: \begin{figure}[H]
\centering
        \subfloat[The flux balance (or cell residual) of triangle $T$ is scattered among its vertices.]{\includegraphics[width=0.45\linewidth]{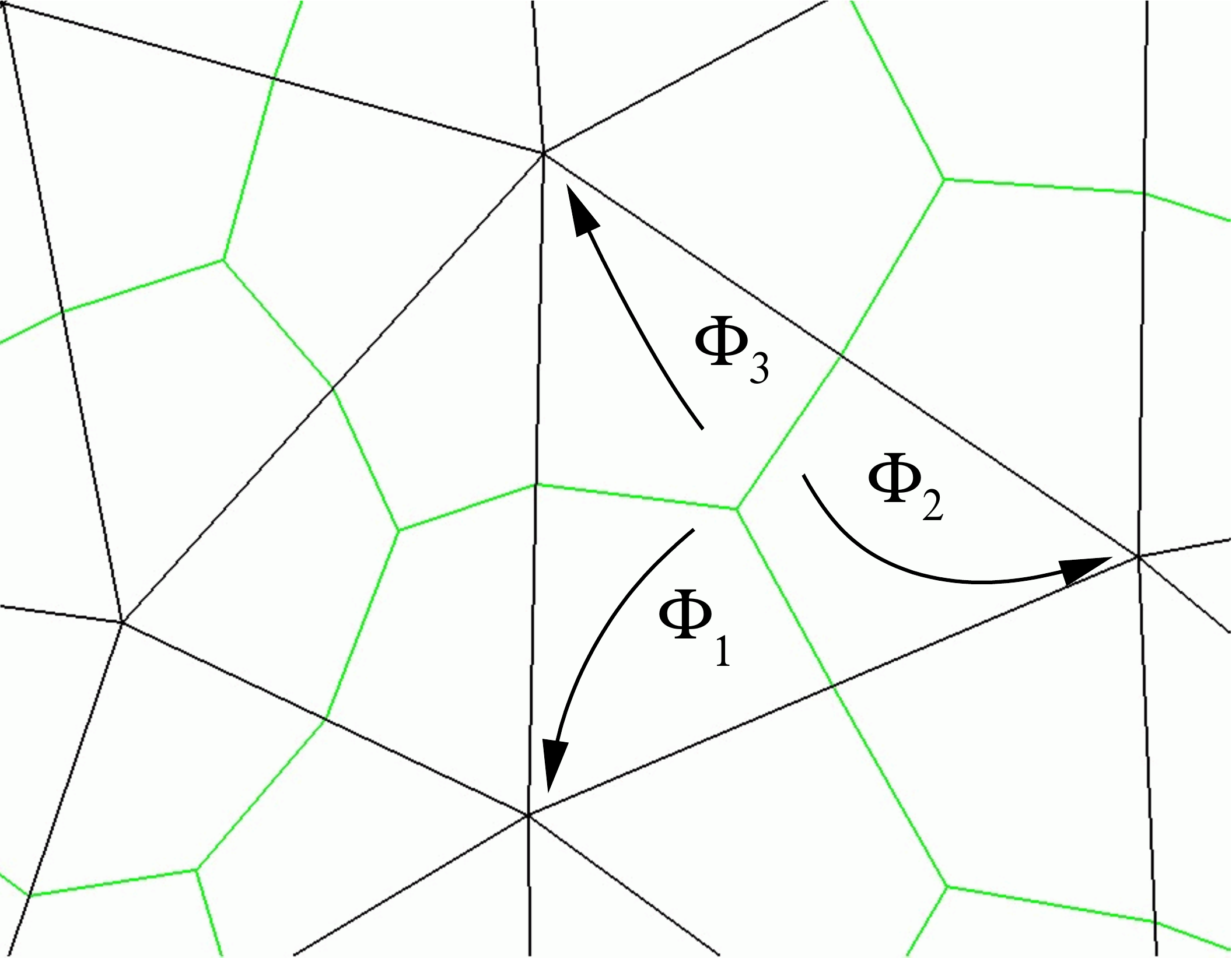}}\quad
        \subfloat[Grid-point $i$ gathers the fractions of cell residuals from the surrounding triangles.]{\includegraphics[width=0.45\linewidth]{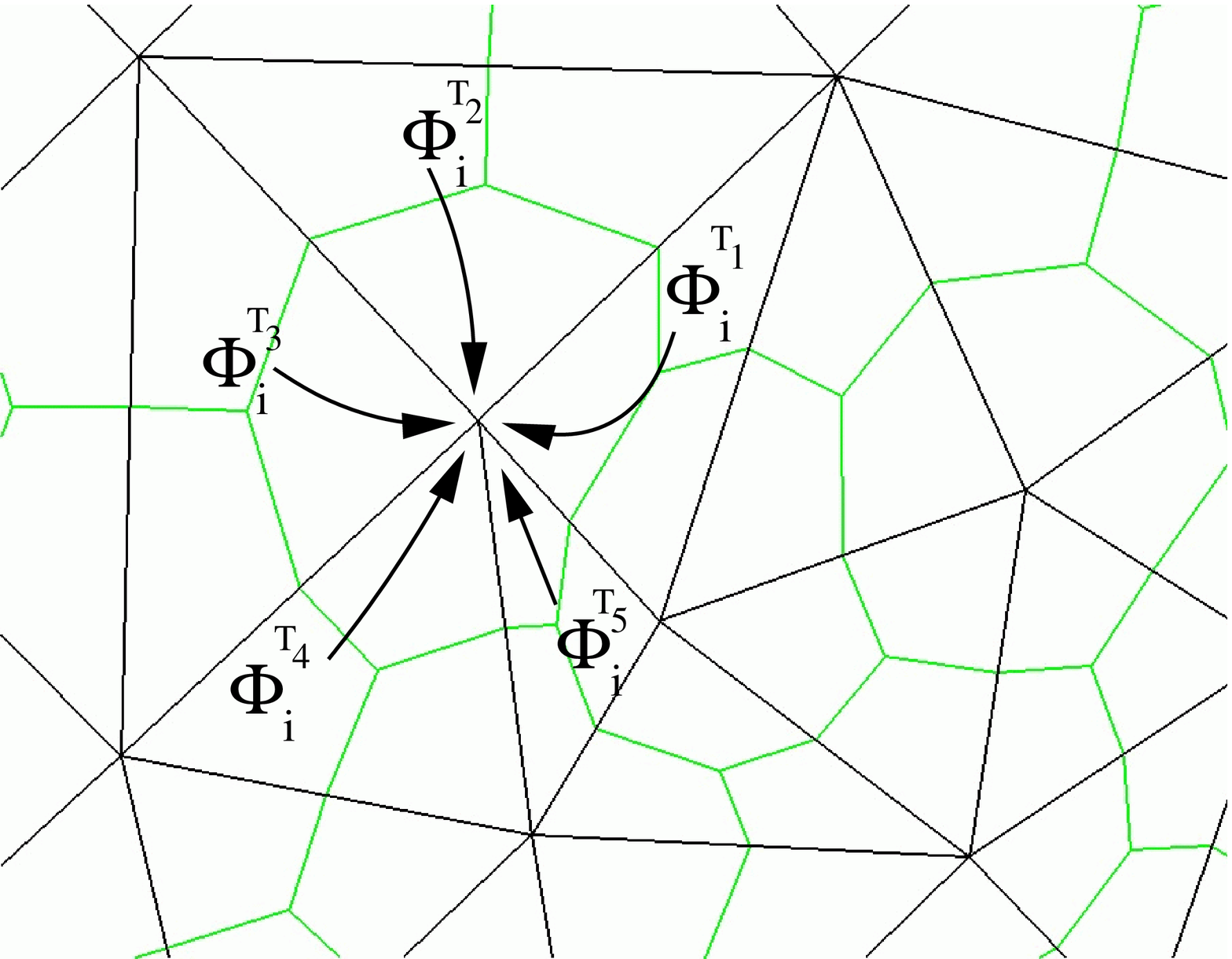}}
\caption{Residual distribution concept.}
\label{fig:RDconcept}
\end{figure}

\begin{description}
\item[Conservation] The consistency condition \eqref{eq:RD1} implies conservation if  the evaluation of the element residual is such that the equality:
\begin{equation}
      \label{eq:RD2}
\Phi^T =\oint\limits_{\partial T}%\mathcal{F}^h_n
{\mathcal{F}\left(\mathbf{U}^h\right)\cdot \mathbf{n}\,\mathrm{d}S}
\end{equation}
{holds} for some continuous approximation of the  normal  flux.  The condition above can be shown to be essential for solutions of 
scheme \eqref{eq:RD0}-\eqref{eq:RD1} to converge to  weak solutions  \cite{abg99b,abg2001d}.
In practice there exist two approaches to fulfill this condition. The first is to introduce a local conservative linearization of the equations. For the Euler equations with perfect gas equation of state, this can be achieved by means of a multi-dimensional extension of Roe's linearization~\cite{Roe:81,deconinck93}.  
This is the approach used in \texttt{EulFS}, which also
uses the conservative linearization to evaluate all quantities necessary for the splitting. 

In alternative, one can compute the element residual by approximating directly the contour integral on $\partial T$ by some quadrature formula, and with some assumption on the (continuous) polynomial interpolation used to perform the flux evaluation in quadrature points \cite{crd02,rcd05}. This is the formulation implemented in \texttt{NEO}, which makes use of a set of physical variables (pressure, density, velocity) for both   the interpolation, and the averaging of the flux Jacobians where necessary. 
\item[Accuracy] The consistency/truncation error condition originally introduced in \cite{abg99f} (see also \cite{dr2017,ar2017}) provides a necessary condition for second order of  accuracy. Schemes verifying this condition \emph{at steady state} can be generally cast in a {FEM} like form:
\begin{equation}
      \label{eq:RD3}
\Phi_i^T = \int\limits_\Omega\pmb{\omega}_i\nabla\cdot\mathcal{F}{\left(\mathbf{U}^h\right)\,\mathrm{d}V}\qquad\qquad\sum\limits_{j\in T}\pmb{\omega}_{{j}}\big|_T =\pmb{I}
\end{equation}
where $\pmb{\omega}_i$ is a bounded test function, and {the second relation equivalent 
to the consistency condition \eqref{eq:RD1}}.  
\item[Monotonicity]  The theory of positive coefficient schemes  has  been used systematically to study these schemes in the scalar case, proving a discrete maximum principle equivalent to the preservation in time of the initial bounds on the discrete solution (cf. \cite{dr2017} and references therein).  Formal matrix generalizations of this condition have been considered in several works \cite{vanderweide,abg2001c_steady}. Linear monotone schemes are only first order accurate, and several different non-linear approaches  exist to combine  second (or higher) order of accuracy with a monotonicity preserving property.
\item[Upwinding and multidimensional upwinding] A bias of the residual distribution in the direction of propagation of the information is often present. This notion is clear in one space dimension \cite{Roe:81}, and a geometrical  
generalization  to the multi-dimensional case  can be provided for scalar problems on linear finite elements~\cite{Roe:87,Roe:90,sidilkover}, see also~\cite{dr2017}.  
For  multidimensional systems, this notion becomes less clear,  unless one focuses on steady two-dimensional supersonic flows 
for which exact decompositions in scalar waves exist~\cite{bonfiglioli2000fluctuation,pd97b}. 
In practice, this notion is embedded either by means of such decompositions, as it is the case in \texttt{EulFS}, or in a  formal matrix generalization of the scalar discretizations \cite{vanderweide}, as done in \texttt{NEO}.
%deconinck1993multidimensional,dr2017}
\end{description}

Two families of distribution methods are available 
in \texttt{EulFS} and \texttt{NEO}. Both codes  implement the most classical multidimensional upwind methods, known as the high order Low Diffusion A (LDA) scheme, and the monotone Narrow (N) scheme.
In the scalar case, the first corresponds to a bounded area (volume in 3D) weighted splitting  \cite{bonfiglioli1997}, while the second is the optimal low diffusion upwind first order positive coefficient scheme on simplicial meshes  \cite{Roe:87,Roe:90}.  For flows with
discontinuities, the non-linear blending of the above two schemes, called LDAN \cite{abg99f,crd02}, is available in both codes. 

The codes also provide implementations of non-upwind methods, in the form of a centered distribution with an upwind biasing term bearing close similarities to the classical streamline upwind stabilization \cite{hughes2010stabilized}. More particularly, \texttt{EulFS} provides an implementation of the cell-vertex Lax-Wendroff method
\cite{hubbard}, while \texttt{NEO}
features implementations of both the streamline-upwind (SU) method, as well as of a  nonlinear variant of the latter in the form of a 
 blended central (Bc) distribution between the SU method and a non-linear  limited Lax-Friedrich's distribution. We refer the reader to \cite{ricchiuto2010explicit,ricchiuto2015explicit,ar2017} and references therein for more details. For more information, 
the interested reader may also refer to the repository documentation at \url{https://github.com/UnDiFi/UnDiFi-2D/wiki}.

\emph{Extension to time dependent flows.} For time dependent solutions, the prototype \eqref{eq:RD0} is in general inconsistent with the accuracy conditions provided e.g.  in \cite{dr2017}.
For example,  the application of definition~\eqref{eq:RD3} to the time dependent case leads in general to the introduction of a mass matrix associated to the term $\int_\Omega\boldsymbol{\omega}_i\,\partial_t\mathbf{U}\,\mathrm{d}V$.   This matrix has a non-diagonal structure,
and may depend on the solution, which leads to a discrete prototype substantially more expensive than \eqref{eq:RD0},  which is closer to what is usually obtained in finite volume methods. To  cope with the cost of  inverting the resulting (non-linear) system of algebraic equations several approaches have been proposed:
\begin{itemize}
    \item fully implicit methods, often based on dual time stepping or space-time formulations \cite{csikjournal,doru_CF,abg2001c,rcd05,HUBBARD2011263,bonfiglioli2013mass-matrix};
    \item predictor-corrector and defect-correction strategies which allow  to keep a resolution cost very close to that of \eqref{eq:RD0}, while accounting for the presence of the mass matrix in the correction iterations~\cite{ricchiuto2010explicit,ABGRALL-dec,ABGRALL2019274,ar2017}
    \item Lax-Wendroff formulations \cite{hubbard}  relying on the classical truncated Taylor series development in time to achieve high order accuracy, and on  a central  approximation in space coupled with mass lumping which is  compatible with \eqref{eq:RD0} and second order of accuracy,
\end{itemize}

The code \texttt{EulFS} features  both implicit time integration with mass matrix, and an implementation of the explicit Lax-Wendroff scheme in a modified Arbitrary Lagrangian-Eulerian (ALE) formulation. Conversely, in \texttt{NEO} a second order predictor-corrector formulation of linear and non-linear schemes is implemented, both in a fixed and in an ALE form. Please refer to~\cite{arpaia2014ale}, for additional details on the schemes.

\section{Unstructured Shock-Fitting: Algorithmic Features\label{sec:algorithm}}
{Figure~\ref{fig:workflow} shows the algorithmic workflow of the \textit{UnDiFi-2D} code, including the sequence of subroutines and external programs being called during a typical run. The CFD codes (either \texttt{NEO} or \texttt{EulFS}) and the \texttt{Triangle} mesh-generator are invoked as black boxes, communication being handled through disk I/O, see the 1 and 5 circles in  Fig.~\ref{fig:workflow}. Figure~\ref{fig:workflow} also includes those optional parts that ensure the time-accurate integration (circled points 2, 3, 4)}.
\tikzstyle{startstop} = [rectangle, rounded corners, minimum width=3cm, minimum height=1cm,text centered, draw=black, fill=red!30, font=\sffamily]

\tikzstyle{io} = [trapezium, trapezium left angle=70, trapezium right angle=110, minimum width=3cm, minimum height=1cm, rounded corners, text centered, draw=black, fill=blue!30]

\tikzstyle{process} = [rectangle, minimum width=6cm, rounded corners, minimum height=1cm, text centered, text width=3cm, draw=black, fill=orange!15, font=\ttfamily]

\tikzstyle{unsprocess} = [rectangle, minimum width=6cm, rounded corners, minimum height=1cm, text centered, text width=3cm, draw=black, fill=blue!30, font=\ttfamily]

\tikzstyle{decision} = [diamond, minimum width=2cm, minimum height=0.25cm, rounded corners, text centered, draw=black, fill=green!30, aspect=2.5, text width=5em]

\tikzstyle{arrow} = [thick,->,>=stealth, >=triangle 60]

% -------------------------------------------------
% Set up a new layer for the debugging marks, and make sure it is on top
\pgfdeclarelayer{marx}
\pgfsetlayers{main,marx}
% A macro for marking coordinates (specific to the coordinate naming
% scheme used here). Swap the following 2 definitions to deactivate
% marks.
\providecommand{\cmark}[2][]{%
  \begin{pgfonlayer}{marx}
    \node [nmark] at (c#2#1) {#2};
  \end{pgfonlayer}{marx}
  } 
\providecommand{\cmark}[2][]{\relax} 

\begin{figure}[H]
\centering
\scalebox{0.75}{
\begin{tikzpicture}[
node distance=2cm, 
>=triangle 60,                  % Nice arrows; your taste may be different
start chain=going below,        % General flow is top-to-bottom
node distance=21.5mm and 100mm, % Global setup of box spacing
every join/.style={norm},       % Default linetype for connecting boxes
]

\tikzset{
  %base/.style={draw, on chain, on grid, align=center, minimum height=4ex},
  %proc/.style={base, rectangle, text width=8em},
  %test/.style={base, diamond, aspect=2, text width=5em},
  %term/.style={proc, rounded corners},
  % coord node style is used for placing corners of connecting lines
  coord/.style={coordinate, on chain, on grid, node distance=6mm and 25mm},
  % nmark node style is used for coordinate debugging marks
  nmark/.style={draw, cyan, circle, font={\sffamily\bfseries}},
    % -------------------------------------------------
  % Connector line styles for different parts of the diagram
  %norm/.style={->, draw, lcnorm},
  %free/.style={->, draw, lcfree},
  %cong/.style={->, draw, lccong},
  %it/.style={font={\small\itshape}}
}

\node [startstop] (start) {./run.sh};

%\node [xshift=3.5cm, yshift=0.25cm] {(Input files)};
\node [xshift=8.5cm, yshift=1.8cm] {(Input files)};

\node (infiles) [process, right of=start, xshift=5cm, yshift=0cm, fill=red!30, minimum width=5cm] 
{na00.1.* \\
sh00.dat \\
input.dat \\
type.dat \\
.petscrc \\
inputfile-exp.txt};

\node (pro1) [process, below of=start, yshift=-0.3cm, text width=5cm] {
readmesh \\ readmap \\ mesh(0)$\rightarrow$mesh(2) \\ re\_inp\_data \\ re\_sdw\_info};
                                     
\node (dec1) [decision, below of=pro1, yshift=-0.5cm] {Unsteady?};

\node (pro2a) [process, below of=dec1, yshift=-0.75cm] {
fnd\_phps \\ co\_norm \\ interp\_sp \\ co\_pnt\_dspl \\ fx\_msh\_sps};

\node (dec2) [decision, below of=pro2a, yshift=-0.5cm] {Unsteady?};

\node (pro2b) [unsprocess, right of=dec1, xshift=5cm, fill=blue!30] {
re\_dt\_data \\ co\_norm \\ co\_state\_dps \\ fx\_state\_dps};
                                
\node (pro3a) [process, below of=dec2, yshift=-0.5cm, text width=5cm] {
wtri \\ triangle \\ na00xtovvvv \\ triangle2grd/triangle2dat};                                                    
\node (pro3b) [unsprocess, right of=dec2, xshift=5cm, fill=blue!30] {
sh(0)$\rightarrow$sh(1) \\ calc\_vel}; 

\node [process, below of=pro3a, yshift=-1.2cm, text width=5cm] (pro4a) {
NEO/EulFS \\ neo2triangle/dat2triangle \\ readmesh \\ zroe(1)$\rightarrow$zroe(0) \\ co\_state\_dps \\ fx\_state\_dps \\ zroesh(0)$\rightarrow$zroesh(1)};

\node [xshift=1.75cm, yshift=-14.65cm, text width=3.3cm] {(NEO 1st iter.)};

\node (dec3) [decision, below of=pro4a, yshift=-0.95cm] {Unsteady?};

\node (pro4b) [unsprocess, right of=dec3, xshift=5cm, yshift=4.25cm, fill=blue!30, text width=4.5cm] {
calc\_vel \\ triangle \\ na2vvvv \\ triangle2grd \\ NEO/EulFS \\ 
neo2triangle \\ mv\_grid \\ readmesh \\ roe(1)$\rightarrow$zroe(0) \\ co\_state\_dps \\ fx\_state\_dps \\ zroesh(0)$\rightarrow$zroesh(1) \\ wsh\_mean};                         
\node (pro5a) [process, below of=dec3, yshift=-0.7cm] {
mv\_dps \\ fx\_dps\_loc \\ interp \\ wtri0 \\ wrt\_sdw\_info};

\node [xshift=8.75cm, yshift=-20.75cm] {(Output files)};

\node (outfiles) [process, right of=pro5a, xshift=5cm, yshift=-1cm, fill=red!30, text width=5cm] 
{na0000?.1.* \\
file00[123].dat \\
file0010.dat \\
neogrid.grd \\
vvvv.dat \\
vvvv\_input.dat \\
na99.node \\
sh99.dat \\
shocknor.dat \\
vel.dat};

\draw [arrow] (infiles) -- (start);
\draw [arrow] (start) -- (pro1);
\draw [arrow] (pro1) -- (dec1);
\draw [arrow] (dec1) -- node[anchor=east] {no} (pro2a);
\draw [arrow] (dec1) -- node[anchor=south] {yes} (pro2b);
\draw [arrow] (pro2b) |- (0,-6.3);
\draw [arrow] (pro2a) -- (dec2);
\draw [arrow] (dec2) -- node[anchor=east] {no} (pro3a);
\draw [arrow] (dec2) -- node[anchor=south] {yes} (pro3b);
\draw [arrow] (pro3b) |- (0,-11.75);
%\draw [arrow] (dec3) -- node[anchor=south] {yes} (pro4b);
\node [xshift=1.85cm, yshift=-19.25cm] {yes};
\draw[arrow] (dec3.east) -- ++(0,0) -- ++(1.75,1) |- (pro4b.west);	
\draw [arrow] (dec3) -- node[anchor=east] {no} (pro5a);
\draw [arrow] (pro3a) -- (pro4a);
\draw [arrow] (pro4a) -- (dec3);
\draw [arrow] (pro4b) |- (0,-20.75);
%\draw [arrow] (pro5a) -- (outfiles);
\draw[arrow] (pro5a.east) -- ++(0,0) -- ++(1.1,0);
\draw[->] (pro5a.west) -- ++ (-1.0,0) |- (dec1);

\node[rectangle, blue, very thick, dashed, xshift=8.75cm, yshift=-3.25cm] (un1) {(Unsteady part)};

\draw[blue, very thick, dashed] (3.5,-19.) rectangle (10.5,-3.5);

\node [coord, right=of infiles, xshift=0.5cm, yshift=1.25cm] (c1) {}; \cmark{1}
\node [coord, right=of dec1] (c2)  {}; \cmark{2}  
\node [coord, right=of dec2] (c3)  {}; \cmark{3}
\node [coord, right=of dec3] (c4)  {}; \cmark{4}
\node [coord, right=of outfiles, xshift=1cm, yshift=2.25cm] (c5) {}; \cmark{5}

\end{tikzpicture}}
\caption{Typical \textit{UnDiFi-2D} workflow.} 
\label{fig:workflow}
\end{figure}
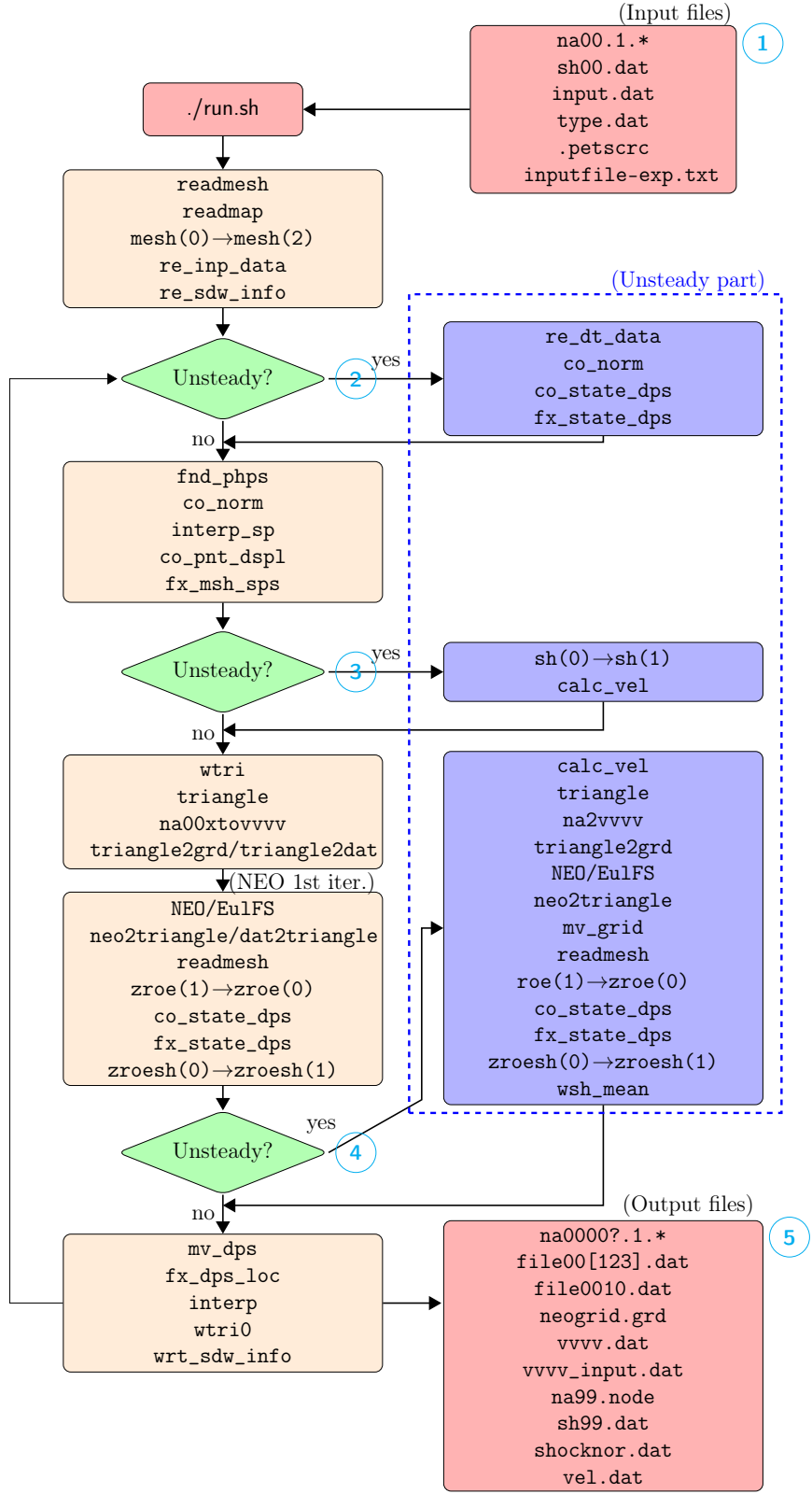
{Even though a thorough description of the various procedures listed in Fig.~\ref{fig:workflow} is available in the \textit{UnDiFi-2D} documentation, it is appropriate to give a brief description of the data storage and key algorithmic ingredients of the shock-fitting algorithm}.  %
%which have been summarized in the seven steps described in} 

{Regardless of whether steady or time-accurate simulations are performed}, the approach is inherently time-dependent, because both the solution and the grid change with time, due to the displacement of the fitted discontinuities. When a steady solution exists, the shock speed will asymptotically vanish and the tessellation of the flow domain will not any longer change. 

\begin{figure}[H]
\centering
        \subfloat[Shock front moving over the background triangular mesh at time $t$.]{\includegraphics[width=0.38\linewidth]{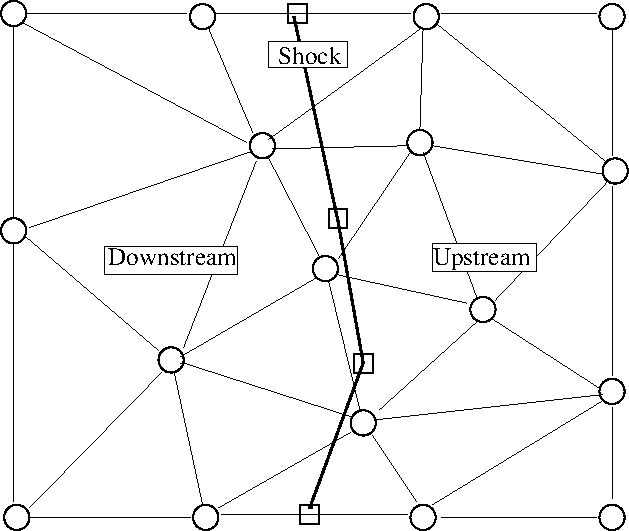}\label{fig:f1a}}\quad
        \subfloat[Dashed lines mark the cells to be removed; dashed circles denote the phantom nodes.]{\includegraphics[width=0.38\linewidth]{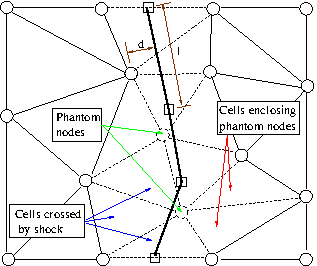}\label{fig:f1b}}\\
        \subfloat[The background mesh is split into disjoint sub-domains by a hole which encloses the shock.]{\includegraphics[width=0.38\linewidth]{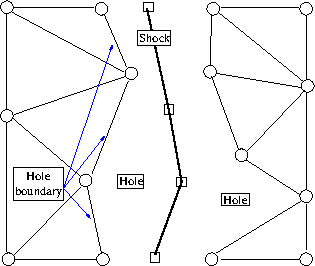}\label{fig:f2a}}\quad
        \subfloat[The triangulation around the shock is rebuilt.]{\includegraphics[width=0.38\linewidth]{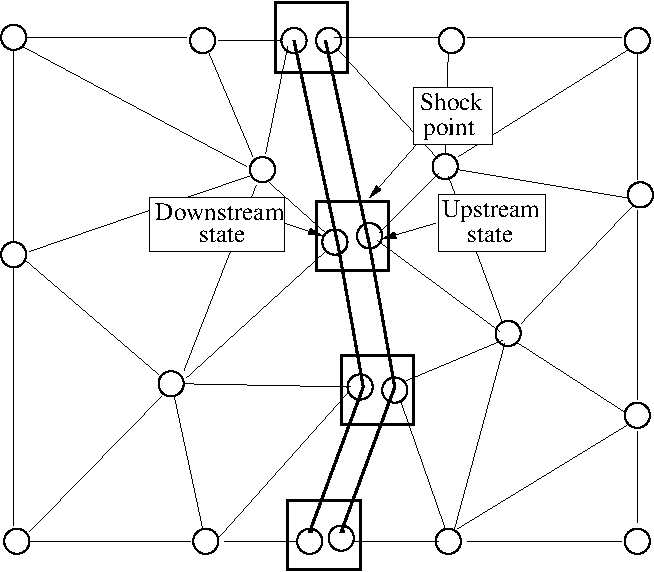}\label{fig:f3}}\\
        \subfloat[Calculation of the shock-tangent and shock-normal unit vectors.]{\includegraphics[width=0.38\linewidth]{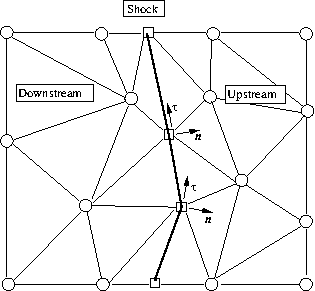}\label{fig:fig1b}}\quad
        \subfloat[The shock displacement induces mesh deformation.]{\includegraphics[width=0.38\linewidth]{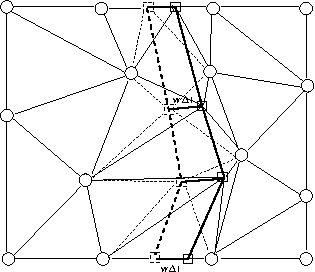}\label{fig:f70}}
\caption{Unstructured shock-fitting: summary of the key algorithmic steps.}
\label{fig:algorithm}
\end{figure}

{As far as data storage is concerned,}
the dependent variables and grid velocity vector are available within all grid-points of a two-dimensional triangulation that covers the entire computational domain; this is what we call the {\em background} mesh. {In addition to the background mesh, the fitted discontinuities (either shocks or slip-lines) are discretized using a collection of grid-points (the shock-points) which are mutually joined to form a connected series of line segments (the shock-edges); shock-points and shock-edges make up what we call the {\em shock}-mesh. In contrast to the grid-points of the background mesh, where a single set of dependent variables is stored, the shock-points are duplicated items that share the same geometrical location, but store two different sets of dependent variables, corresponding to the two sides of the discontinuity. This is schematically shown in Fig.~\ref{fig:algorithm}d. Shock-edges, which connect the shock-points on both sides of the discontinuity (see Fig.~\ref{fig:algorithm}d where the width of the discontinuity has been increased to improve readability) also overlap, so that each fitted discontinuity behaves like a double-sided internal boundary of zero thickness}. 
%Because the width of the discontinuity is negligible, its two sides are discretized using the same polygonal curve or triangulated surface; each pair of nodes that face each other on the two sides of the discontinuity share the same geometrical location, but store different values of the dependent variables, one corresponding to the upstream state and the other to the downstream one. Moreover, a velocity vector normal to the discontinuity is assigned to each pair of grid-points on the fitted discontinuity, representing its displacement velocity}. 
As shown in Fig.~\ref{fig:algorithm}a), the spatial location of the fitted discontinuities is independent of the location of the grid-points that make up the background grid. 

The {sequence of operations} that leads from the available mesh and solution at time $t$ to an updated mesh and solution at time $t+\Delta t$ can be split into the seven steps that will be described in Sects.~\ref{subsec:cell_removal} to~\ref{subsec:phantom_nodes}. 

\subsection{Cell Removal Around the Shock Front\label{subsec:cell_removal}}
In this first step, the fitted discontinuities are laid on top of the background mesh, as shown in Fig.~\ref{fig:algorithm}a. All those cells that are crossed by the fitted discontinuities and those mesh points that are located too close to it are temporarily removed from the background mesh, as shown in Fig.~\ref{fig:algorithm}b. This operation is performed in the subroutine \texttt{fnd\_phps}, whose interface is shown in Listing \ref{lis:cell_removal}. Here, the arguments referring to the background mesh are passed with index zero. We call phantom those grid-points of the background mesh (shown using dashed circles in Fig.~\ref{fig:algorithm}b) that have been temporarily removed. All cells having at least one phantom node among their vertices are also removed from the background triangulation; these are the cells shown using dashed edges in Fig.~\ref{fig:algorithm}b. Further details concerning the criteria used to identify and remove
the phantom nodes can be found in~\cite{paciorri2009shock-fitting}.

\begin{lstlisting}[caption={Cell removal around the shock front.},label={lis:cell_removal},language={Fortran}]

call fnd_phps(
.    nedge(0),           ! number of faces
.    istak(lbndfac(0)),  ! boundary faces index pointer
.    nbfac(0),           ! number of boundary faces
.    istak(lcelnod(0)),  ! cell to node index pointer
.    nvt,                ! number of vertices of a cell
.    nelem(0),           ! number of elements of the background grid
.    dstak(lcorg(0)),    ! point coordinates
.    xysh,               ! shock point coordinates
.    istak(lnodcod(0)),  ! node code flag
.    npoin(0),           ! number of points of the background grid
.    istak(lnodptr(0)),  ! node index pointer
.    nbpoin(0),          ! number of boundary points
.    nshocks,            ! number of shocks
.    nshockpoints,       ! number of shock points
.    nshocksegs,         ! number of shock segments
.    nphanpoints,        ! number of phantom points
.    istak(lpmap(0)))    ! pointer to integer array
\end{lstlisting}
{It is important to underline that all the quantities that refer to the background or the computational mesh (such as the coordinates of the grid-points)  are stored inside a large double precision array (\texttt{dstak}) where integers (\texttt{istak}) can also be stored by using an \texttt{EQUIVALENCE} statement. Therefore, when calling the various subroutines, these quantities are passed providing the corresponding pointers inside the \texttt{dstak} array; for example: \texttt{dstak(lcorg(0))} represents the pointer to the coordinates of the background mesh.
On the contrary, the quantities that refer to the discontinuities (for example the coordinates of the shock-points of the shock-mesh) are stored in specific arrays; for example: \texttt{xysh} contains the coordinates of the shock-points.
}
\subsection{Local Re-Meshing Around the Shock Front\label{subsec:local_remesh}}
Following the cell removal step, the background triangulation has been split into two or more disjoint sub-domains, as shown in Fig.~\ref{fig:algorithm}c. The hole dug by the fitted front is then re-meshed using a Constrained Delaunay Tessellation (CDT): the edges 
that make up the fitted discontinuity and the boundary of the hole are both constrained to be part of the final tessellation; this is illustrated in Fig.~\ref{fig:algorithm}d. This operation is performed in the subroutines \texttt{fx\_msh\_sps}, \texttt{wtri} and by calling \texttt{Triangle} whose interfaces are shown in Listings~\ref{lis:local_remesh}, \ref{lis:wtri}  and~\ref{lis:triangle}. Observe that re-meshing {should} be localized around the discontinuities. {The code currently included in the repository is not fully compliant with the algorithm described above, because the CDT is applied to the entire computational domain and not only within the hole carved around the shock.
Future releases will fix this open issue}.
Upon completion of this stage, the computational domain is discretized using what we call the {\textit{computational}} mesh, which differs from the background mesh only in the neighbourhood of the fitted discontinuities. Further details concerning the software used to construct the CDT have been given in Sect.~\ref{subsec:UnDiFi}.
\begin{lstlisting}[caption={Local fix of special points.},label={lis:local_remesh},language={Fortran}]
call fx_msh_sps(
.    istak(lbndfac(0)), ! boundary faces index pointer
.    istak(lnodcod(0)), ! node code flag
.    nbfac(0),          ! number of boundary faces
.    nbfac_sh,          ! number of shock faces
.    nvt,               ! number of vertices of a cell
.    nelem(0),          ! number of elements of the background grid
.    dstak(lcorg(0)),   ! point coordinates
.    xysh,              ! shock point coordinates
.    dstak(lcorg(0)+npoin(0)*ndim),                     ! upstream
.    dstak(lcorg(0)+npoin(0)*ndim+nshmax*npshmax*ndim), ! downstream
.    npoin(0),          ! number of points of the background grid
.    nshocks,           ! number of shocks
.    nshockpoints,      ! number of shock points
.    nshocksegs,        ! number of shock  segments
.    nspecpoints,       ! number of special points
.    typespecpoints,    ! type of special points
.    shinspps,          ! special points in shock
.    ispclr)            ! special points color
\end{lstlisting}

\begin{lstlisting}[caption={Local re-meshing around the shock front.},label={lis:wtri},language={Fortran}]
call wtri(
.    istak(lbndfac(0)), ! boundary faces index pointer
.    nbfac(0),          ! number of boundary faces
.    nbfac_sh,          ! number of shock faces
.    istak(lcelnod(0)), ! cell to node index pointer
.    nvt,               ! number of vertices of a cell
.    dstak(lcorg(0)),   ! point coordinates
.    xysh,              ! shock point coordinates
.    dstak(lcorg(0)+npoin(0)*ndim),                     ! upstream coor.
.    dstak(lcorg(0)+npoin(0)*ndim+nshmax*npshmax*ndim), ! downstream coor.               
.    dstak(lzroe(0)),                                   ! roe variables
.    dstak(lzroe(0)+npoin(0)*ndof),                     ! upstream state
.    dstak(lzroe(0)+npoin(0)*ndof+nshmax*npshmax*ndof), ! downstream state
.    istak(lnodcod(0)),          ! upstream node code flag
.    istak(lnodcod(0)+npoin(0)), ! downstream node code flag
.    npoin(0),          ! number of points of the background grid
.    fname(1:7),        ! mesh input file name to Triangle 
.    nshocks,           ! number of shocks
.    nshockpoints,      ! number of shock points
.    nshocksegs,        ! number of shock segments
.    nphanpoints)       ! number of phantom points
\end{lstlisting}

\begin{lstlisting}[caption={Mesh generation with Triangle.},label={lis:triangle},language={Fortran}]
write(*,1001,advance='no')'triangle --> '
execmd = bindir(1:10)//'triangle_'//hostype(1:6)//
.        ' -nep '//fname(1:7)//' > log/triangle.log'    
ifail = system(execmd)                                 
call flush(6)
if (ifail /= 0) then
  write(6,*)'Triangle has returned an error code ifail = ', ifail
  call exit(ifail)
endif
write(*,1002)' ok'
\end{lstlisting}

\subsection{Calculation of the Unit Vectors Normal to the Shock Front\label{subsec:normal_vectors}}
In order to apply the jump relations, normal ($\mathbf{n}$) and tangent ($\boldsymbol{\tau}$) unit vectors are needed within each pair of grid-points located along the discontinuities, see Fig.~\ref{fig:algorithm}e. These unit vectors are computed using finite-difference (FD) formulae which involve the coordinates of the shock-point itself and those of its neighboring
shock-points. This operation is performed in the subroutine \texttt{co\_norm} whose interface is shown in Listing \ref{lis:normals}. Depending on the local, shock-downstream flow regime, it may be necessary to use upwind-biased formulae to avoid the appearance of geometrical instabilities along the fitted discontinuity. Full details describing how to {select the stencil} can be found in~\cite{paciorri2009shock-fitting}. %, see also~\cite{salas20091st}. 
{Be aware that the FD formulae reported in~\cite{paciorri2009shock-fitting} contain a typo and should be replaced by those published in~\cite{ciallella2020extrapolated}}.
%for the 2D case. %and in~\cite{bonfiglioli2013three-dimensional} for the 3D case.

\begin{lstlisting}[caption={Calculation of the unit vectors normal to the shock front.},label={lis:normals},language={Fortran}]

 call co_norm(
 .    xysh,             ! shock point coordinates
 .    dstak(lzroe(0)+npoin(0)*ndof),                     ! upstream
 .    dstak(lzroe(0)+npoin(0)*ndof+nshmax*npshmax*ndof), ! downstream 
 .    norsh,            ! normal vectors
 .    nshocks,          ! number of shocks
 .    nshockpoints,     ! number of shock points
 .    typeshocks,       ! type of shock
 .    nspecpoints,      ! number of special points
 .    typespecpoints,   ! type of special points
 .    shinspps,         ! special points in shock
 .    ispclr,           ! special points color
 .    istak(lia(0)),    ! pointer for integer arrays in setbndrynodeptr
 .    istak(lja(0)),    ! pointer for integer arrays in setbndrynodeptr
 .    istak(liclr(0)),  ! pointer for integer arrays in setbndrynodeptr
 .    nclr(0),          ! colours of boundary patches
 .    dstak(lcorg(0)))  ! point coordinates
\end{lstlisting}

\subsection{Solution Update Using the Shock-Capturing Code\label{subsec:solutionupdate}}
Using the {computational mesh} as input, a single time step calculation is performed using one of the available shock-capturing solvers which returns updated nodal values at time $t+\Delta t$, as shown in Listing \ref{lis:solver} for the \texttt{EulFS} solver. Since the discontinuities are seen by the shock-capturing code as internal boundaries (of zero thickness) moving with the velocity of the discontinuity, there is no need to modify the spatial discretization scheme already implemented in the PDEs solver to account for the presence of the fitted discontinuities. In practice, the shock-capturing solver is used as a black-box: it receives in input the {computational} grid, the nodal values of the solution and grid velocity at time $t$ and returns the updated solution at time $t+\Delta t$. The solution returned by the shock-capturing solver at time $t+\Delta t$ is however {missing some boundary conditions}. 
%on one or both sides of each discontinuity, depending on whether it is a shock or a contact. 
These missing pieces of information will be determined as described in Sect.~\ref{subsec:jump_relations}.

\begin{lstlisting}[caption={Call to the gasdynamics solver.},label={lis:solver},language={Fortran}]
execmd = bindir(1:10)//"eulfs_"//hostype(1:6)//"-itmax 1 > log/eulfs.log"
ifail = system(execmd) 
call flush(6) 
if (ifail /= 0) then
  write(6,*) "Eulfs has returned an error code ifail = ", ifail
  call exit(1)                  
endif
\end{lstlisting}

\subsection{Enforcement of the Jump Relations\label{subsec:jump_relations}}
{As explained in Sect.~\ref{subsec:RH}}, the missing pieces of information that are needed to correctly update the solution within all pairs of grid-points located along the discontinuities are obtained by enforcing the R-H jump relations; this also provides the local velocity of the discontinuity along its normal.
The R-H jump relations are a set of non-linear algebraic equations that can be solved within each pair of grid-points located along the discontinuities by means of Newton-Raphson's algorithm. In order to match the number of unknowns with the available equations, one or more additional pieces of information are required within both or either of the two sides of the fitted discontinuity, depending on whether this is a shock or a contact discontinuity. These additional pieces of information are obtained from the characteristic formulation of the Euler equations and correspond to those characteristic quantities that are convected towards the discontinuity from the sub-domain that is attached to that side of the discontinuity. Using an upwind-biased discretization within the shock-capturing solver, one can reasonably assume that the spatial and temporal evolution of these characteristic
quantities has been correctly computed. 

The {enforcement of the R-H jump relations} is performed in the subroutine \texttt{co\_state\_dps} whose interface is shown in Listing \ref{lis:RH}. Full algorithmic details concerning the practical implementation of the jump relations for shocks and contact discontinuities are reported elsewhere~\cite{ivanov2010computation, paciorri2009shock-fitting, paciorri2011shock} and will not be repeated here. % for the 2D case. %and~\cite{bonfiglioli2013three-dimensional} for the 3D case. 
Furthermore, an ad-hoc treatment is required within those special points where different discontinuities interact {(triple or quadruple points) or whenever a discontinuity interacts with a solid or free boundary.} The algorithmic details are described in~\cite{ivanov2010computation, paciorri2011shock} and their implementation can be found in the subroutines \texttt{co\_utp}, \texttt{co\_uqp}, respectively. %{\color{blue}Dovremmo dire che non tutti i tipi di interazione vengono attualmente gestiti.}
\begin{lstlisting}[caption={Enforcement of the jump relations.},label={lis:RH},language={Fortran}]

call co_state_dps(
.    xysh,          ! shock point coordinates
.    dstak(lzroe(0)+npoin(0)*ndof),                     ! upstream
.    dstak(lzroe(0)+npoin(0)*ndof+nshmax*npshmax*ndof), ! downstream
.    zroeshuold,    ! upstream zroe states at previous iteration
.    zroeshdold,    ! downstream zroe states at previous iteration
.    norsh,         ! normal vectors
.    wsh,           ! shock velocity
.    nshocks,       ! number of shocks
.    nshockpoints,  ! number of shock points
.    nshocksegs,    ! number of shock segments
.    typeshocks,    ! type of shock
.    iter)          ! current iteration
\end{lstlisting}

\subsection{Shock Displacement\label{subsec:shock_displacement}}
The enforcement of the jump relations provides the speed, $w$, %{\textit{\textbf{w}}}, 
at which each pair of grid-points located on the discontinuity move along its local normal unit vector, {\textit{\textbf{n}}}. The position of the discontinuity at time $t+\Delta t$ is computed in a Lagrangian manner by displacing all its grid-points, as shown in Fig.~\ref{fig:algorithm}f where the dashed and solid lines represent the discontinuity at time $t$, resp. $t+\Delta t$. When simulating steady flows, this can be accomplished using the following first-order-accurate (in time) integration formula:
\begin{equation}
\boldsymbol{P}_{i}^{t+\triangle t}=\boldsymbol{P}_{i}^{t}+{w_{i}^{t}}\boldsymbol{n}_{i}^{t}\Delta t\label{eq:move:shock}
\end{equation}
which returns the spatial coordinates of the $i$-th {shock}-point at time $t+\Delta t$. The low temporal accuracy of Eq.~(\ref{eq:move:shock}) does not affect the spatial accuracy of the steady state solution which only depends on the spatial accuracy of the gas-dynamic solver and that of the tangent and normal unit vectors. 

On the contrary, when dealing with unsteady flows, the temporal accuracy of the shock motion has to be the same as that of the spatial discretization, i.e. second order accurate in our case. This can be accomplished using a predictor-corrector type temporal integration scheme, or a Runge-Kutta multi-step scheme. {In the former case}, the predictor step estimates the position of the {discontinuity} at time level $n+1/2$ using the explicit Euler scheme: 
\begin{equation}
\boldsymbol{P}_{i}^{n+\frac{1}{2}}=\boldsymbol{P}_{i}^{n}+w_{i}^{n}{\boldsymbol{n}_{i}^{n}}\frac{\Delta t}{2}\label{eq:predictor}
\end{equation}
The speed {of the discontinuity} $w_{i}^{n+\frac{1}{2}}$ and the normal unit vector
${\boldsymbol{n}_{i}^{n+\frac{1}{2}}}$ at time level $n+1/2$ are then computed using the intermediate  position of the {discontinuity} $P_{i}^{n+\frac{1}{2}}$ and, finally, the position of each shock-point is updated at time level $n+1$ in the corrector step: 
\begin{equation}
\boldsymbol{P}_{i}^{n+1}=\boldsymbol{P}_{i}^{n}+w_{i}^{n+\frac{1}{2}}{\boldsymbol{n}_{i}^{n+\frac{1}{2}}}\Delta t\label{eq:corrector}
\end{equation}
This operation is performed in the subroutine \texttt{mv\_dps} whose interface is shown in Listing~\ref{lis:move_shock}.

\begin{lstlisting}[caption={Shock displacement.},label={lis:move_shock},language={Fortran}]

call mv_dps(
.    xysh,          ! shock point coordinates
.    dstak(lzroe(0)+npoin(0)*ndof+nshmax*npshmax*ndof), ! downstream
.    wsh,           ! shock velocity
.    i,             ! current iteration
.    nshocks,       ! number of shocks
.    nshockpoints,  ! number of shock points
.    nshocksegs,    ! number of shock segments
.    typeshocks)    ! type of shock
\end{lstlisting}
{One further observations is in order concerning the discontinuity displacement step}.
Figure~\ref{fig:algorithm}f shows that even when the background mesh is fixed in space, the triangular cells that abut on the discontinuity have one of their edges that moves with the discontinuity, thus deforming the cell. This implies that the shock-capturing solver used in Step~\ref{subsec:solutionupdate} must be capable of handling moving meshes, i.e. it must be capable of solving the governing PDEs written using an Arbitrary Eulerian Lagrangian (ALE) formulation. 
\subsection{Interpolation of the Phantom Nodes\label{subsec:phantom_nodes}}
Upon completion of the previous steps, all {grid-points} of the {computational mesh} have been updated at time $t+\Delta t$. The {computational mesh} is made up of {shock-mesh} and all {grid-points} of the background mesh, except those that have been declared {\textit{phantom}}. Therefore, the nodal values within the phantom nodes have not been updated to time $t+\Delta t$. However, during the current time step, the discontinuities might have moved sufficiently far away from their previous position, that some of the phantom nodes may re-appear in the {computational mesh} at the next time step. It follows that also the nodal values within the phantom nodes need to be updated to time $t+\Delta t$. This is easily accomplished by transferring the available solution at time $t+\Delta t$ from the current {computational mesh} to the grid-points of the background one, using linear interpolation. This operation is performed in the subroutine \texttt{interp} whose interface is shown in Listing~\ref{lis:interpolate}. Once the phantom nodes have been updated, the {computational mesh} used in the current time interval has completed its task and can be {discarded}. At this stage the numerical solution has correctly been updated at time $t+\Delta t$ within all grid-points of the background {and shock meshes}. 

The next time interval can be computed re-starting from  step~\ref{subsec:cell_removal} of the algorithm.
\begin{lstlisting}[caption={Interpolation of the phantom nodes.},label={lis:interpolate},language={Fortran}]

call interp(
.    istak(lbndfac(1)), ! boundary faces index pointer
.    nbfac(1),          ! number of boundary faces
.    istak(lcelnod(1)), ! cell to node  index  pointer
.    nvt,               ! number of vertices of a cell
.    nelem(1),          ! number of elements of the shocked grid
.    dstak(lcorg(1)),   ! point coordinates
.    dstak(lzroe(1)),   ! point roe states
.    xysh,              ! shock point coordinates
.    dstak(lcorg(1)+npoin(0)*ndim),                     ! upstream
.    dstak(lcorg(1)+npoin(0)*ndim+nshmax*npshmax*ndim), ! downstream
.    nphampoints,       ! number of phantom points
.    dstak(lcorg(0)),   ! background point coordinates
.    dstak(lzroe(0)),   ! background points roe states
.    istak(lnodcod(0)), ! background grid node code flag
.    npoin(0),          ! number of points of the background grid
.    nshocks,           ! number of shocks
.    nshockpointsold,   ! old number of shock points
.    nshockpoints,      ! new number of shock points
.    istak(lia(1)),     !pointer for integer arrays in setbndrynodeptr
.    istak(lja(1)),     ! pointer for integer arrays in setbndrynodeptr
.    istak(liclr(0)),   ! pointer for integer arrays in setbndrynodeptr
.    nclr(0))           ! colours  of  boundary  patches
\end{lstlisting}

% TODO: flow chart of calling functions and files ...
%In order to deal with both steady and unsteady flows, some modification to the original discontinuity fitting algorithm must be introduced. Such modification are highlighted in Fig.~\ref{fig:calling}, where the black arrows represent the steps done in the steady case while the green ones are the additions for the unsteady case. In addition to the algorithmic steps explined above, conversion modules between the shock-fitting code and the CFD solver are necessary.
%\begin{figure}[H]
%\centering
%\includegraphics[scale=0.75]{figs/calling_sequence.png}
%\caption{Calling sequence in the steady and unsteady case.}
%\label{fig:calling} 
%\end{figure}

%\subsection{Algorithm summary %\label{subsec:Algorithm_summary}}

\section{Test cases\label{sec:applications}}
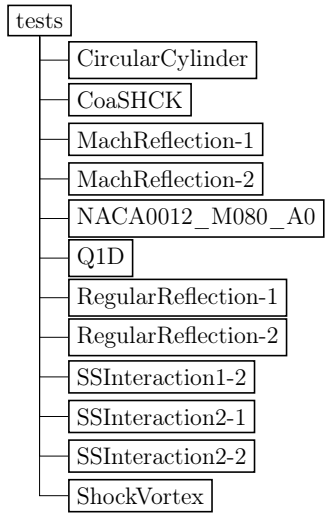
\begin{figure}[H]
\centering
\scalebox{0.75}{
\begin{tikzpicture}[%
  grow via three points={one child at (0.5,-0.7) and
  two children at (0.5,-0.7) and (0.5,-1.4)},
  edge from parent path={(\tikzparentnode.south) |- (\tikzchildnode.west)}]
  \node {tests}
    child { node {CircularCylinder}}		
    child { node {CoaSHCK}}
    child { node {MachReflection-1}}
    child { node {MachReflection-2}}
    child { node {NACA0012\_M080\_A0}}
    child { node {Q1D}}
    child { node {RegularReflection-1}}
    child { node {RegularReflection-2}}
    child { node {SSInteraction1-2}}
    child { node {SSInteraction2-1}}
    child { node {SSInteraction2-2}}
    child { node {ShockVortex}};
    %child { node [selected] {tex}
    %  child { node {generic}}
    %  child { node [optional] {latex}}
    %  child { node {plain}}
    %}
    %child [missing] {}				
    %child [missing] {}				
    %child [missing] {}				
    %child { node {texdoc}};
%    
\end{tikzpicture}}
\caption{Test-cases {available} in the \texttt{tests} directory.} 
\label{fig:TC}
\end{figure}

In order to illustrate the capabilities of \textit{UnDiFi-2D}
several test-cases can be run  {within each of the sub-directories (listed in Fig.~\ref{fig:TC}) of the folder \texttt{tests}}.
In addition to this, these test-cases are intended to test different parts of the code and, therefore, the successful execution of all test-cases (which can be performed with the script \texttt{run\_all\_x86.sh}) represents a verification of the code. These checks are particularly important before a new version of the code is released to verify that changes and modifications of the code have not produced unwanted side-effects elsewhere in the code.

{More specifically, in the shock-shock or shock-wall interaction test-cases addressed in Sect.\ref{subsec:SS1}, \ref{subsec:SS2} and~\ref{subsec:RR}, the various discontinuities bound regions of uniform flow, where an analytical solution can be computed using the jump relations.
Therefore, these test-cases provide a powerful code validation tool, because  the shock-fitting algorithm is expected to return the exact solution everywhere, even if a first-order accurate discretization of the governing PDEs is used.

An analytical solution is also available for the transonic test-case addressed in Sect.~\ref{subsec:Q1D} which features spatially variable fields both upstream and downstream of the shock, so that it can be used to measure the order-of-convergence of the spatial discretization}. 
\tikzstyle{every node}=[draw=black,thick,anchor=west]
\tikzstyle{selected}=[draw=red,fill=red!30]
\tikzstyle{optional}=[dashed,fill=gray!50]

\subsection{Hypersonic flow past a Circular cylinder (\texttt{CircularCylinder-1})\label{subsec:ccylinder} }
The first test-case deals with the hypersonic flow past a circular cylinder at free-stream Mach number, $M_{\infty} = 20$. It {was} one {of the first flow configurations to be set-up and tested during the} \textit{UnFiDi-2D} code development, because this {apparently simple test-case allows to check} the basic subroutines of the code, {such as those in charge of calculating the shock normal unit vectors and applying the R-H relations, under all possible post-shock conditions. This is because by} moving along the bow shock is like sweeping the {entire} shock polar, starting from a normal shock {at the symmetry axis}, which then turns into a strong oblique shock and finally becomes a weak wave. %since it transforms first from normal shock wave to strong oblique shock and, then, to weak oblique shock.

\begin{figure}[H]
\centering
\includegraphics[scale=0.30]{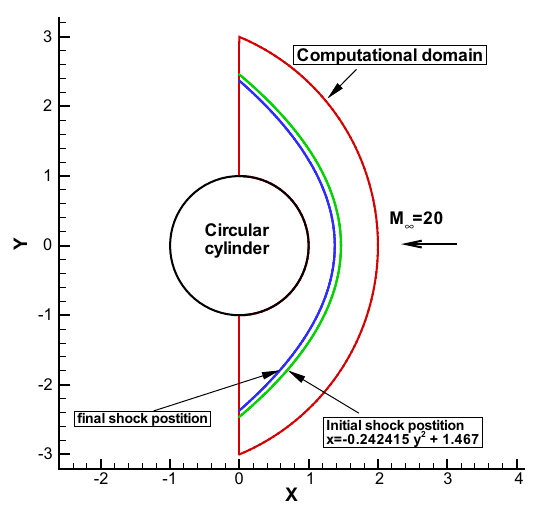}
\includegraphics[scale=0.30]{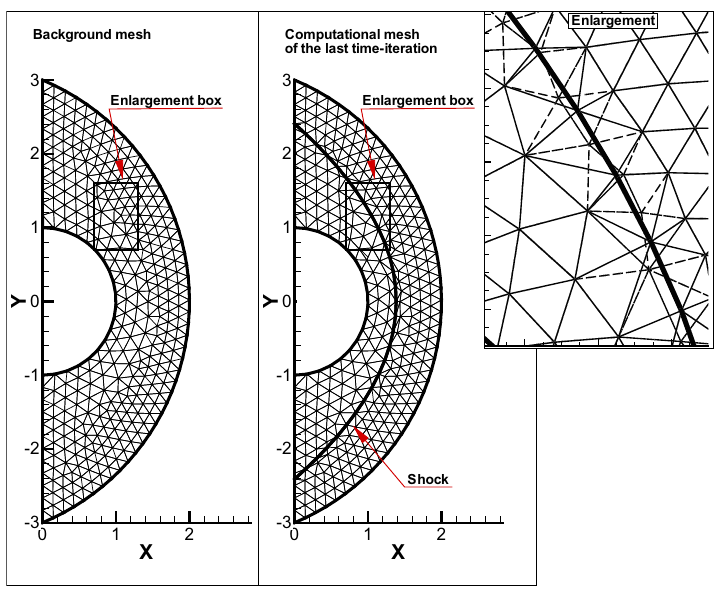}
\includegraphics[scale=0.35]{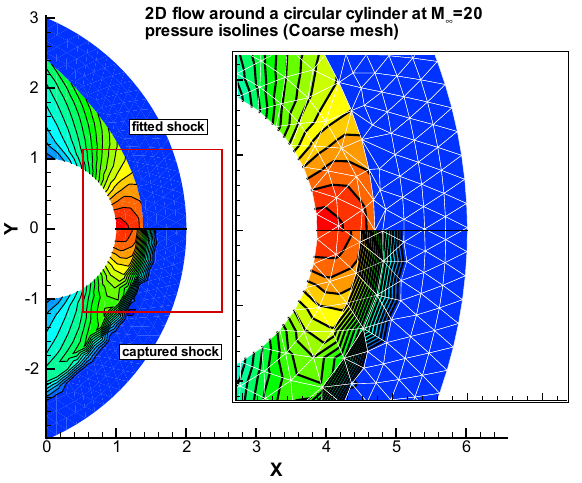}\\
 a) \hspace{5cm} b) \hspace{5cm} c)
\caption{Computational domain: initial and final shock position. Comparison between the coarse background and the computational mesh at  {steady-state}. Comparison between the shock-fitting and shock-capturing solutions.}
\label{fig:domain} 
\end{figure}

As shown in  Fig.~\ref{fig:domain}, the computational domain surrounds the {fore} half of a circular cylinder having radius $R = 1$. The background mesh has been created using the \texttt{delaundo}~\cite{mullerdelaundo} mesh generator by specifying a evenly spaced distribution of boundary nodes with spacing $h = 0.08\, R$.  The numbers of triangles and nodes of the background mesh are reported in Tab.~\ref{tab:IC}.
\begin{center}
\begin{table}[H]
\caption{\label{tab:CCgrids}Nodes and cells of the background and computational meshes}
\label{tab:IC}
\begin{centering}
\begin{tabular}{cccccc}
\hline 
%\multirow{2}{*}{} 
\multicolumn{2}{c}{Background} & & \multicolumn{3}{c}{Computational mesh}\tabularnewline
\multicolumn{2}{c}{Mesh} & &\multicolumn{3}{c}{at steady-state}  \tabularnewline
\cline{1-2} \cline{4-6} 
 Triangles & Nodes & & Triangles & Nodes & Shock nodes\tabularnewline
\hline 
610 & 351 & &624 & 358 & 29\tabularnewline
\hline 
\end{tabular}
\par\end{centering}
\end{table}
\par\end{center}

The solution computed on the background mesh using the unstructured code in shock-capturing mode has been used to initialize the flow-field and to determine the parabolic {shape} of the initial position of the shock front, see Fig.~\ref{fig:domain}. 

In the shock-fitting simulation, the initial upstream state in the shock-points has been set equal to the free-stream conditions, while the initial shock-downstream state has been computed from the upstream state and the local shock slope, assuming zero shock speed: $w = 0$. The flow-field and shock position are then integrated in {pseudo-}time until steady-state is reached. Fig.~\ref{fig:domain}a shows the shock displacement that occurs between the initial shock-position and the one reached at steady state. During the computation, the number of grid-points {and triangles} of the computational mesh varies with respect to that of the background mesh, as shown in Tab.~\ref{tab:IC}. Figure~\ref{fig:domain}b also displays the background mesh and the computational mesh {at steady-state}. The smaller frame of Fig.~\ref{fig:domain}b, which shows an enlargement of the near-shock region, allows comparing the two meshes. It can be seen that these are superimposed everywhere except in the region adjacent to the shock, where the differences between the background (dashed lines) and the computational mesh (solid lines) are due to the addition of the shock-points and shock-edges. Fig.~\ref{fig:domain}b, {as well as  Tab.~\ref{tab:IC}}, clearly show that the re-meshing technique does not significantly increase the number of grid-points and triangles with respect to those of the background mesh.
Figure~\ref{fig:domain}c allows to compare the solutions computed by the~\texttt{EulFS} code working in shock-capturing and shock-fitting mode; similar results can be obtained using \texttt{NEO}. {Inside this folder (as well as the other folders that will be described below) there is a script (run.sh) that allows to run the shock-capturing or shock-fitting solutions with either \texttt{EulFS} or \texttt{NEO}.} Figure~\ref{fig:domain}c also includes a detailed view of the stagnation point region showing the modifications introduced by the shock-fitting algorithm to the background mesh in order to take into account the presence of the shock front. 
A detailed analysis of the numerical results of this test-case can be found in~\cite{paciorri2009shock-fitting}.
\subsection{Interaction between two shocks of opposite families (\texttt{SSInteraction1-2})\label{subsec:SS1} }

\begin{figure}[H]
\centering
\includegraphics[scale=0.30]{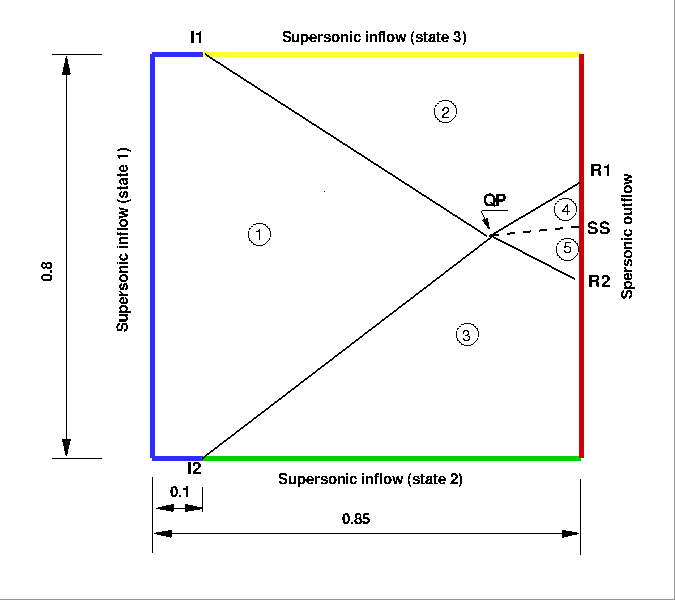}
\caption{Interaction between two shocks of opposite families: flow configuration.}
\label{fig:SS1domain} 
\end{figure}

A uniform supersonic ($M_{\infty} = 3$) stream is crossed by two incident shocks, I1 and I2, of opposite families and different strength which interact in the  {quadruple point, QP,} giving rise to two outgoing reflected shocks, R1 and R2, and a slip-line (or slip-stream), SS, located between the two reflected shocks. {These five discontinuities bound as many uniform flow regions (numbered as shown in Fig.~\ref{fig:SS1domain})} surrounding the quadruple point.
Given the {free-stream Mach number and 
flow deflections ($\delta_2 = -15^{\circ}$ and $\delta_3 = 20^{\circ}$) through the two incident shocks,
the constant states in regions 2 to 5 can be analytically computed and
are listed in Tab.~\ref{ss1-t1}, which includes: the Mach number ($M$), the
deflection angle ($\delta$) measured with respect to the horizontal axis and the pressure ($p/p_1$) and density ($\rho/\rho_1$) ratios}.
\begin{table}[!htb]
\caption{Interaction between two shocks of opposite families: analytical steady state solution.\label{ss1-t1}}
\centerline{
\begin{tabular}{l|rrrrr}
               & 1    & 2      & 3      & 4      & 5 \\ \hline
M              & 3.0  & 2.255  & 1.994  & 1.461  & 1.432  \\
$\delta$       & 0.0  & -15$^{\circ}$   & 20$^{\circ}$   & 4.796$^{\circ}$ & 4.796$^{\circ}$   \\
$p/p_1$        & 1.0  & 2.822  & 3.771  & 8.353  & 8.353 \\
$\rho/\rho_1$   & 1.0  & 2.034  & 2.420 & 4.263   &  4.211
\end{tabular}
}
\end{table}
{This test-case} checks the code capability of computing multiple shocks and slip-lines, but it {also} verifies the computation of the quadruple point. 

The numerical solutions have been computed inside a rectangular domain ($0.85\,L \times 0.8\,L$) with {boundary conditions prescribed as shown in Fig.~\ref{fig:SS1domain}}. The background mesh is made of 7988 grid-points and 15644, {mostly equilateral,} triangles. The characteristic mesh spacing is about $0.01\,L$.

The shock-fitting simulation has been initialised using the shock-capturing solution obtained on the {background} mesh. This solution has also been used to supply the approximate position of the four shocks, the slip-stream and the quadruple point which are needed to construct the internal boundaries corresponding to the various discontinuities.
Starting from this initial flow-field, the {shock-fitting calculation has been} integrated in pseudo-time until steady state is reached.
\begin{figure}[ht]
\centering
\subfloat[Shock-capturing]{
\includegraphics[scale=0.3]{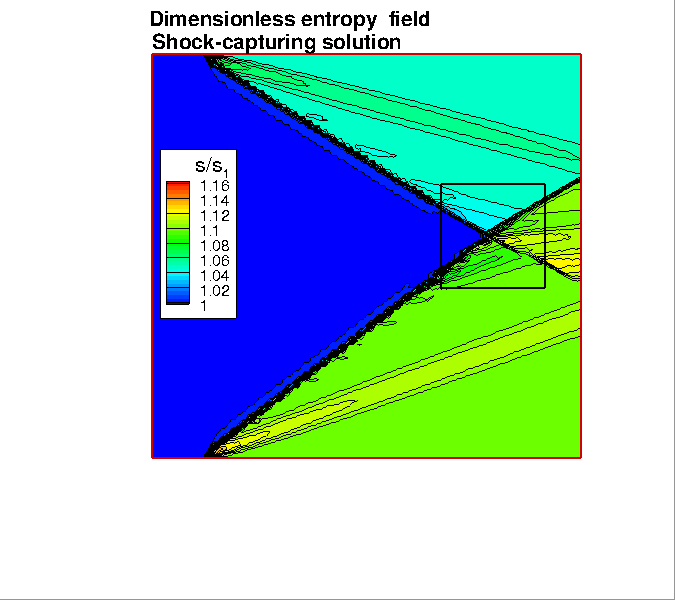}\label{fig:SSsols1}}\quad\quad
\subfloat[Shock-fitting]{
\includegraphics[scale=0.3]{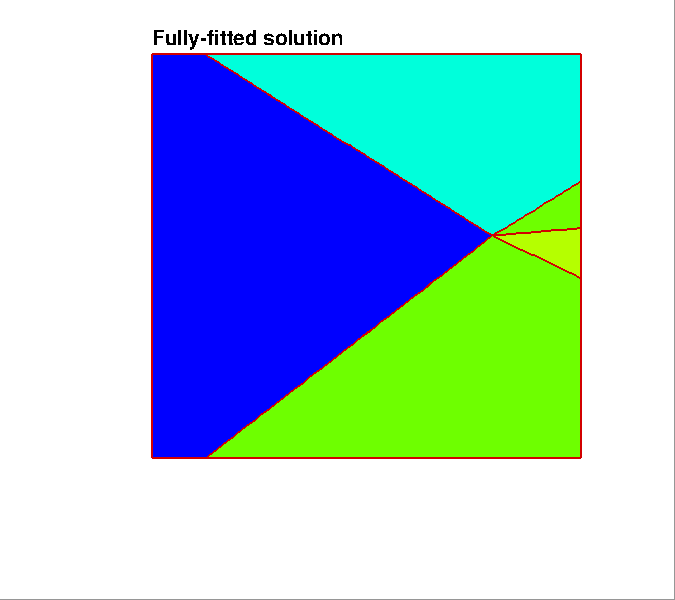}\label{fig:SSsols2}}
\caption{Interaction between two shocks of opposite families: computed entropy iso-contours.}
\label{fig:SSsols} 
\end{figure}
The numerical solution obtained using the \texttt{eulfs} shock-capturing solver is compared in Fig.~\ref{fig:SSsols} with that obtained in fully-fitted mode.
Since the exact solution is uniform within each of the five regions bounded by the discontinuities, the shock-fitting calculation {is expected to return the exact solution of Tab.~\ref{ss1-t1}. This is indeed the case, as revealed by Fig.~\ref{fig:SSsols2}}.
Further analyses dealing with this particular test-case can be found in~\cite{paciorri2011shock}, where the analytical and algorithmic treatment of the  slip-lines and the quadruple point are also described in detail.
\subsection{Interaction between two shocks of the same families (\texttt{SSInteraction2-1}, \texttt{SSInteraction2-2})\label{subsec:SS2} }
\begin{figure}[ht]
\centering
\subfloat[Computational domain and boundary conditions.]{\includegraphics[width=0.45\linewidth]{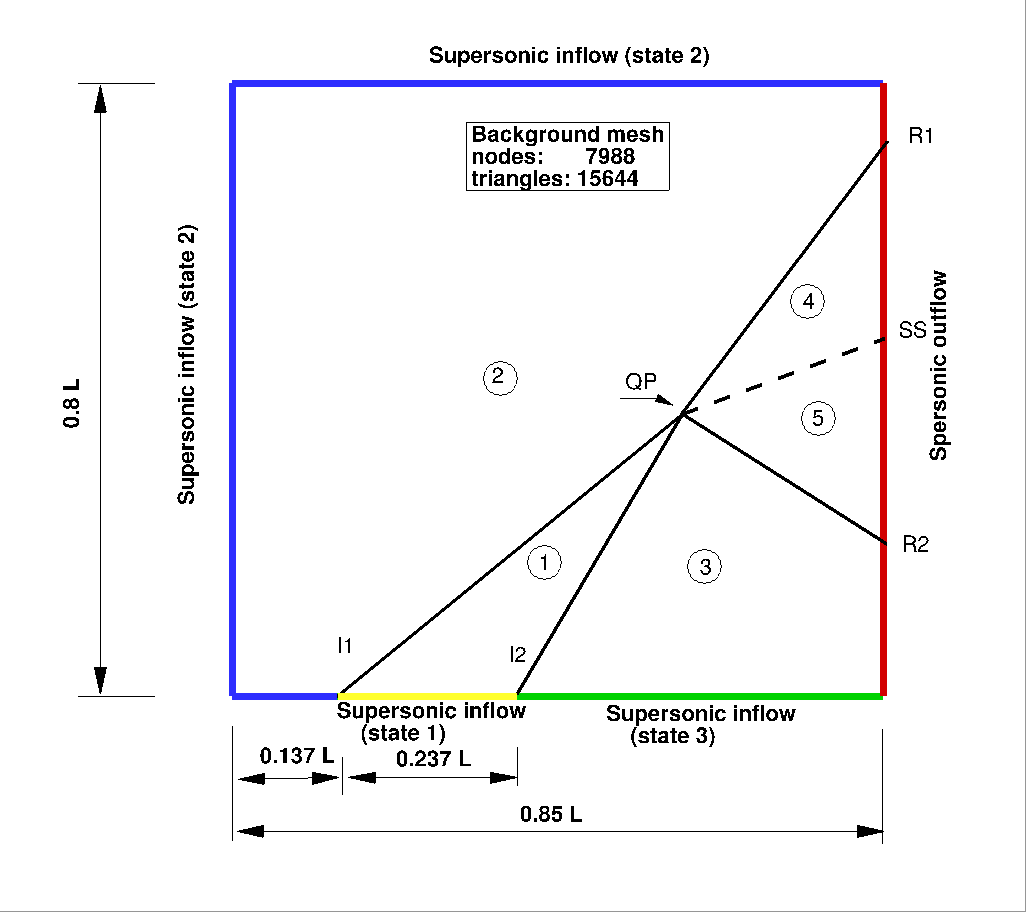}\label{fig:SS2-sketch1}}\quad
\subfloat[Motion of the discontinuities during pseudo-time integration.]{\includegraphics[scale=0.3]{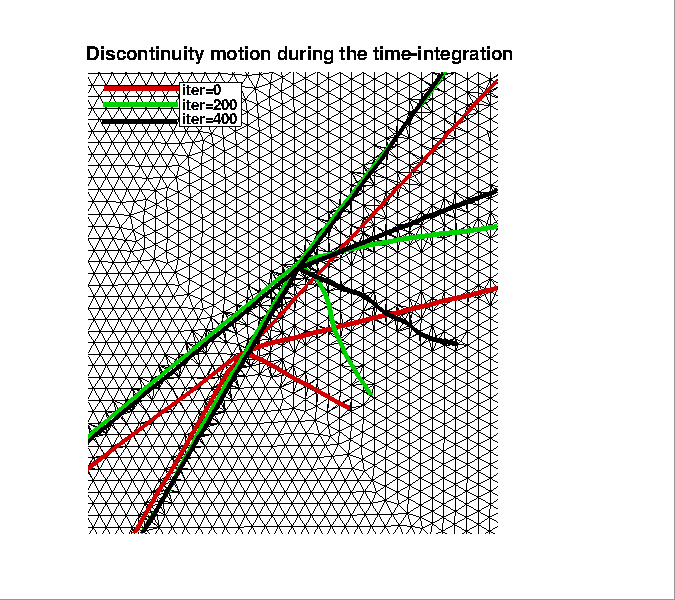}\label{fig:SS2-sketch2} }
\caption{Interaction between two shocks of the same family.}
\label{fig:SS2-sketch} 
\end{figure}

This test-case {is similar to the one addressed in Sect.~\ref{subsec:SS1}, except that} the two {incident shocks}, I1 and I2, belong to the {\em same} family. {Their interaction} in the quadruple point, QP, gives rise to a reflected shock, R1, which is stronger than the two incident ones, a slip-stream, SS, and a second reflected shock wave, R2, of nearly negligible strength. 
%From a topological standpoint, this kind of shock-shock interaction is very similar to the one described in Sect.~\ref{subsec:SS1}. 
If we name the various zones and discontinuities as shown in Fig.~\ref{fig:SS2-sketch1}, the only topological difference between the interaction of shocks {of the same or opposite families lies} in the orientation of the unit vector normal to I1, which points from zone 1 to zone 2 in the present case, {whereas the orientation is reversed, see Fig.~\ref{fig:SS1domain}, when the interacting shocks belong to opposite families. Because of this, the same subroutine \texttt{co\_uqp} is used to compute the QP, regardless of whether the interacting shocks belong to the same or opposite families}.
{The analytical steady state solution, which is listed in Table~\ref{ss2-t1}, has been} analytically computed from the known free-stream state ($M_2 = 2.0$) and the known flow deflections through the incident shocks: $\delta_1 = 10^{\circ}$ and $\delta_3 = 20^{\circ}$.
\begin{table}[ht]
\caption{Interaction between two shocks of the same  family: analytical steady state solution \label{ss2-t1}}
\centerline{
\begin{tabular}{l|rrrrr}
               & 1    & 2      & 3      & 4      & 5 \\ \hline
M              & 1.641         & 2.000  & 1.285  & 1.218  & 1.280 \\
$\delta$       & $10^{\circ} $ &  $0^{\circ}$ &  $20^{\circ}$ &  $19.87^{\circ}$ & $19.87^{\circ}$  \\
$p/p_1$        & 1.707         &  1.000 & 2.803 & 2.822 & 2.822 \\
$\rho/\rho_1$   &  1.460 & 1.000 & 2.074 & 2.035 & 2.084 
\end{tabular}
}
\end{table}
The present interaction has been numerically simulated both in hybrid and fully-fitted mode: {the two different set-ups are available in the \texttt{SSInteraction2-1} and \texttt{SSInteraction2-2}} folders. The computational domain and the background triangular grid are identical to those described in Sect.~\ref{subsec:SS1}, whereas the boundary conditions have been changed as described in Fig.~\ref{fig:SS2-sketch1}.

\begin{figure}[p]
\begin{center}
%\subfloat[]
{\includegraphics[scale=0.3]{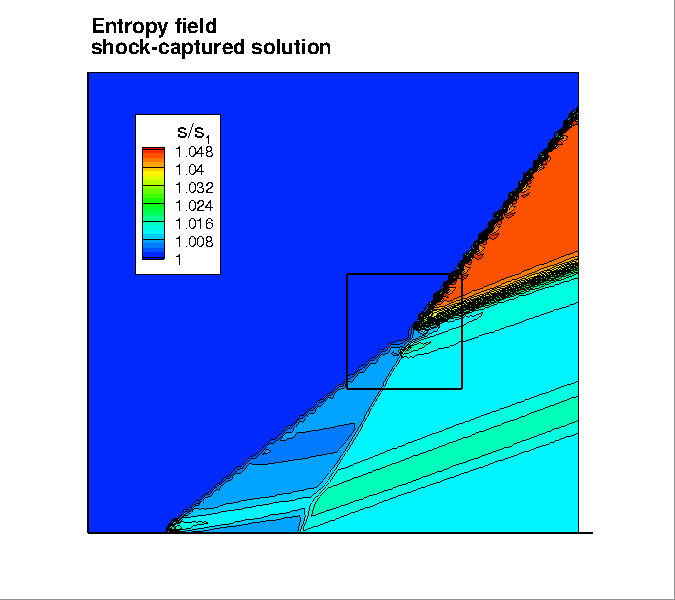}}
%\subfloat[]
{\includegraphics[scale=0.3]{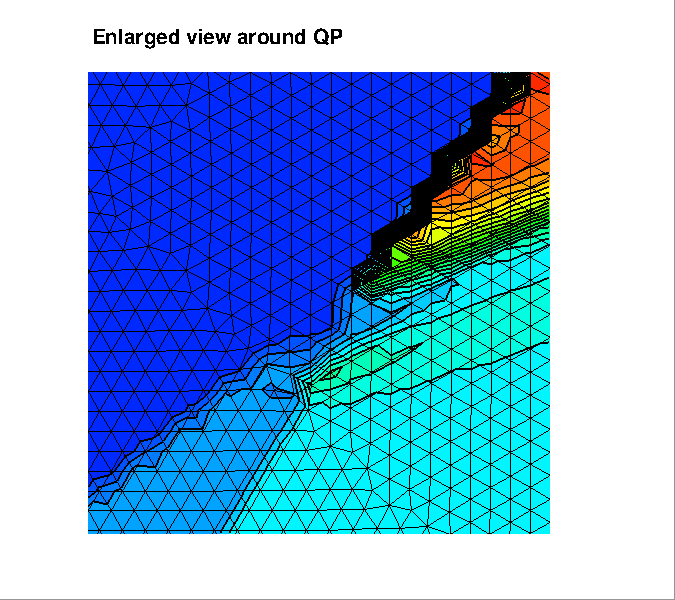}}\\ 
%\subfloat[]
{\includegraphics[scale=0.3]{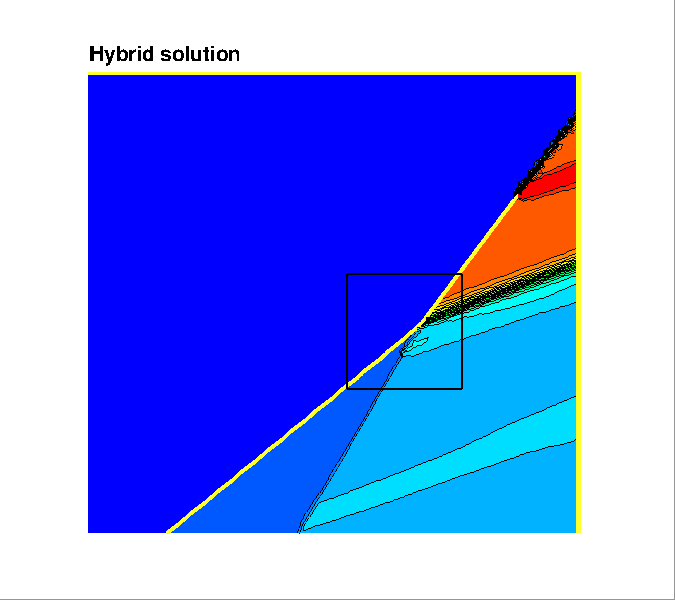}}
%\subfloat[]
{\includegraphics[scale=0.3]{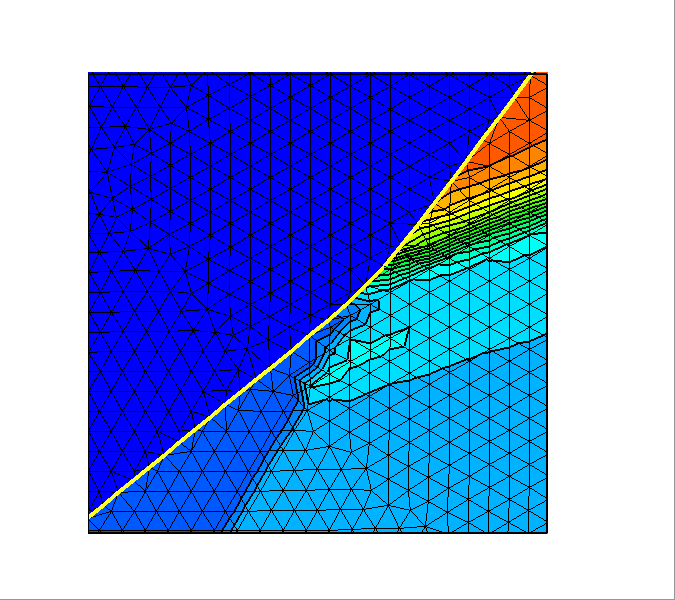}}\\
%\subfloat[]
{\includegraphics[scale=0.3]{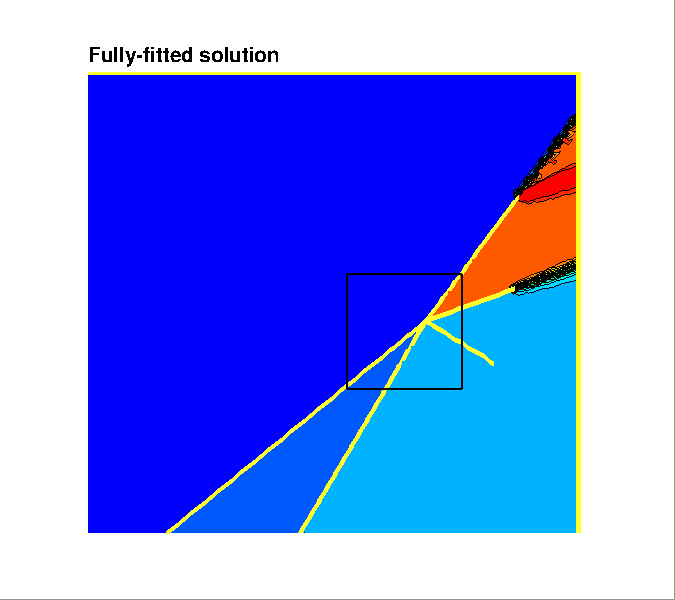}}
%\subfloat[]
{\includegraphics[scale=0.3]{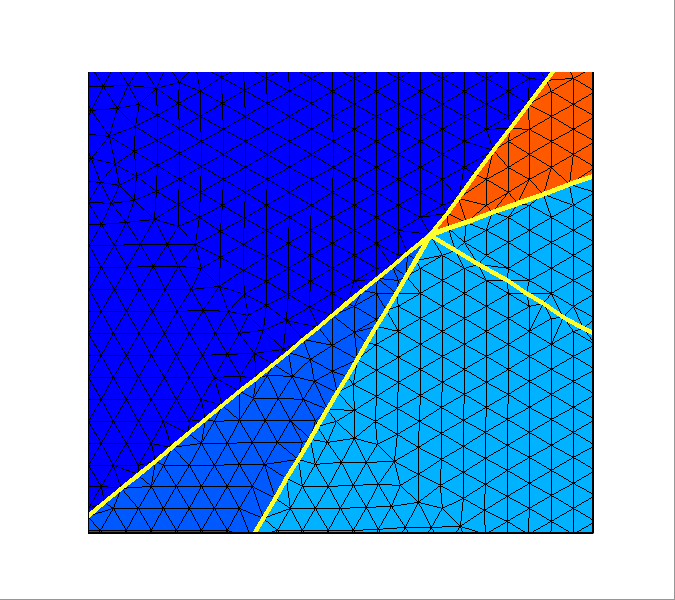}}
\end{center}
\caption{Interaction between two shocks of the same family: entropy iso-contour lines.}  \label{ss2-f5}
\end{figure}

In the hybrid simulation (the middle row of Fig.~\ref{ss2-f5}) only the incident and one of the reflected shocks, resp.\ I1 and R1, have been fitted, whereas all other discontinuities have been captured. This makes the hybrid shock-fitting approach algorithmically simpler than the fully-fitted one, not only because I1 and R1 are treated as a single fitted-shock, but also because the QP is captured and, therefore, does not require any modeling. In the fully-fitted simulation (the bottom row of Fig.~\ref{ss2-f5}), on the contrary, all discontinuities have been fitted, including their interaction in the QP, which has been modelled as described in Sect.~\ref{subsec:SS1}. 
Note, also, that in the fully-fitted simulation the two reflected shocks and the slip-stream {do not reach the domain boundaries, but} have been fitted only up to a preset location, downstream of which it is left to the shock-capturing code to capture {each of these three discontinuities. This choice has been deliberately made to demonstrate the code capability to treat the same discontinuity in a hybrid manner, i.e.\ to ``fit'' only part of it, whereas the remainder is captured.
This requires
an {\it{ad-hoc}} modeling of the pair of shock-points where the fitted-front terminates within the flow-field. Therefore, the present test-case not only  completes the testing of the QP modeling, but also of those subroutines dealing with the aforementioned end-points}.

As mentioned earlier, we rely upon the solution computed in shock-capturing mode (the top row of Fig.~\ref{ss2-f5}) to provide the initial location of all the fitted entities: shocks, slip-line and interaction points, along with a reasonable initial flow-field. By doing so, the initial position of the fitted discontinuities turns out to be very close to the position reached at steady-state. For the present test-case, however, in order to challenge the proposed shock-fitting technique, the initial position of the various discontinuities and their interaction point have been deliberately prescribed in locations that are far from those predicted by the shock-capturing simulation. Fig.~\ref{fig:SS2-sketch2} shows how the position of the various discontinuities and the computational mesh of the fully-fitted solution evolve during the calculation. This test-case clearly demonstrates that the shock-fitting technique is able to move correctly all discontinuities, along with the {QP}, towards their steady state location, even though this is quite far from the initial guess.

Figure~\ref{ss2-f5} shows the entropy field obtained using the shock-capturing (top row), hybrid (middle row) and fully-fitted (bottom row) approaches. All these solutions have been obtained using the \texttt{eulfs} solver. The three frames in the left column of Fig.~\ref{ss2-f5} show the entire computational domain, whereas those in the right column show a detailed view of the entropy distribution superimposed on the computational mesh in the  neighbourhood of the QP.
Further details and results concerning this test-case can be found in~\cite{paciorri2011shock}.
\subsection{Shock-wall interaction: regular reflection.
(\texttt{RegularReflection-1}, \texttt{RegularReflection-2}) } \label{subsec:RR}
\begin{figure}[ht]
\centering
\includegraphics[width=0.75\linewidth]{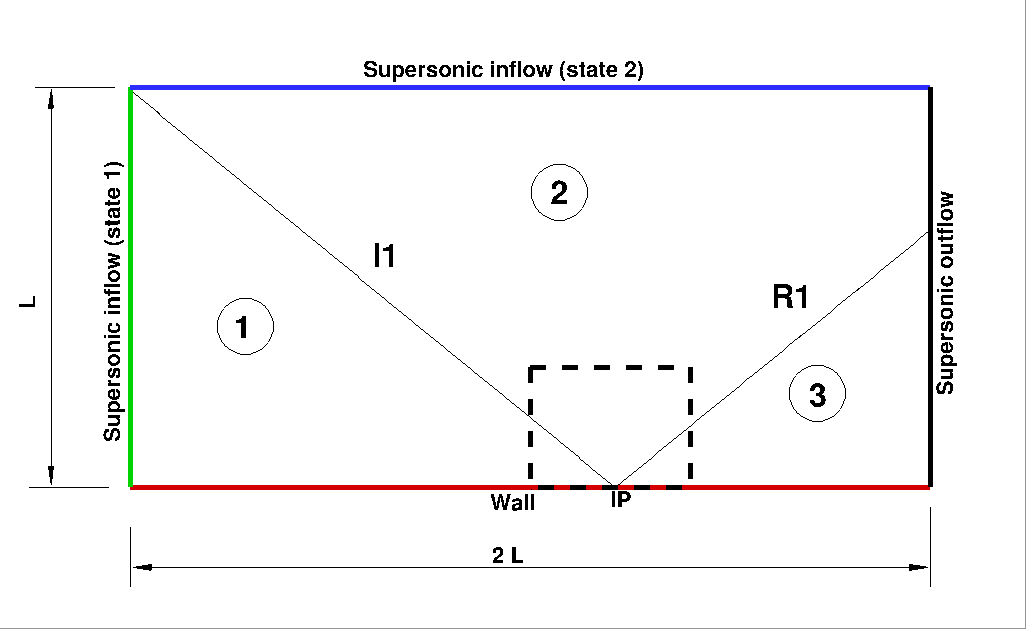}
\caption{Regular reflection: computational domain and boundary conditions.}
\label{fig:RR1-sketch} 
\end{figure}

{When the two incident shocks of opposite families of Sect.~\ref{subsec:SS1} have equal strength, the flow-field is symmetric w.r.t.\ the SS and the flow-field describes the regular reflection of a weak oblique shock from a flat, solid boundary, as sketched in} 
Figure~\ref{fig:RR1-sketch}.
{Regular reflection is possible as long as the flow deflection through the incident shock, I1, is lower than the maximum deflection for the} {Mach number in region 2}.
When I1 impinges (in point IP) on the straight wall, it gives rise to a reflected oblique shock, R1, of the opposite family. The supersonic stream undergoes the same deflection, although opposite in sign, through both shocks. For the given flow parameters ($M_1 = 2$ and $\delta_2 = 10^{\circ}$), Table~\ref{RR-t1} shows the analytically computed Mach number, normalised pressure, density and flow direction (measured w.r.t.\ the wall) in all three regions shown in Fig.~\ref{fig:RR1-sketch}.
\begin{table}[!htb]
\caption{Regular reflection: steady state analytical solution. \label{RR-t1}}
\centerline{
\begin{tabular}{l|rrr}
               & 1    & 2      & 3   \\ \hline
M              &  2.0 &  1.641 &1.287 \\
$\delta$       &   $0^{\circ}$ & $ -10^{\circ}$ &  $  0^{\circ}$ \\
$p/p_1$        &   1.0 & 1.706 &  2.797 \\
$\rho/\rho_1$   &  1.0 &  1.458 &  2.069 \\
\end{tabular}
}
\end{table}

The present test-case has also been computed using both the hybrid and fully-fitted mode; the corresponding computational setups can be found in the \texttt{RegularReflection-1} and \texttt{RegularReflection-2} folders, respectively. {The purpose of the present test-case, when run in fully-fitted mode, is to check the numerical modeling of the regular wall reflection}.

\begin{figure}[ht]
\centering
%\begin{subfigmatrix}{2}
%\subfigure[Shock-capturing]%
%\raisebox{3cm}{shock-capturing} 
 {\includegraphics[scale=0.3]{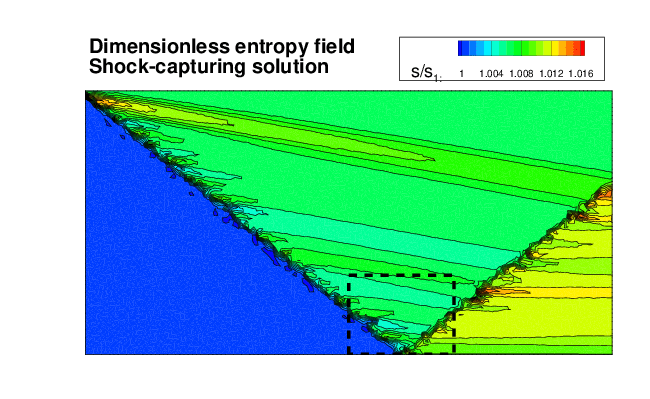}} %\label{ss3-f5a}}
%\subfigure[Shock-capturing]%
 {\includegraphics[scale=0.3]{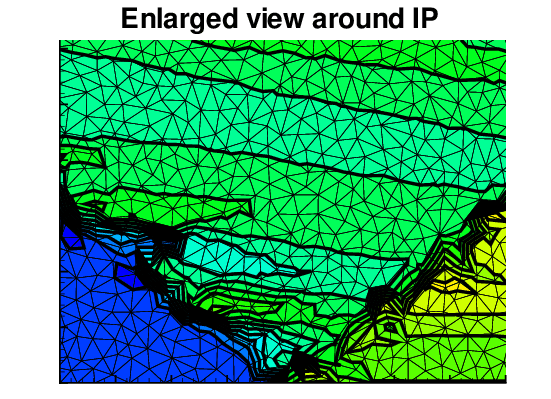}}\\ %\label{ss3-f5aa}}
%\subfigure[Hybrid]%\\
 {\includegraphics[scale=0.3]{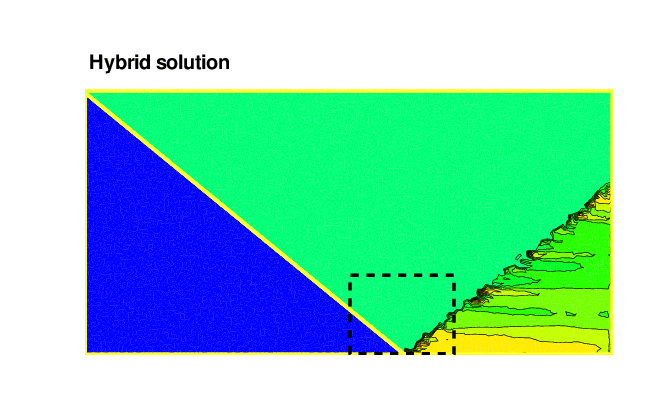}} %\label{ss3-f5b}}
%\subfigure[Hybrid]%
 {\includegraphics[scale=0.3]{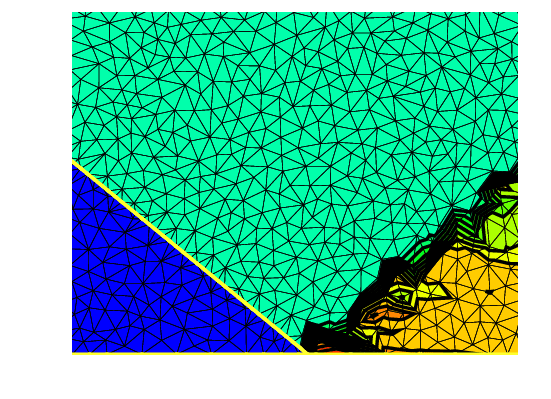}}\\ %\label{ss3-f5bb}}
%\subfigure[Fully-fitted]%\\
 {\includegraphics[scale=0.3]{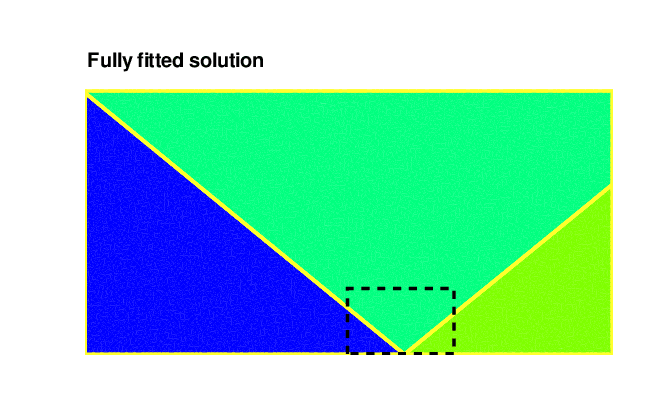}} %\label{ss3-f5c}}
%\subfigure[Fully-fitted]%
 {\includegraphics[scale=0.3]{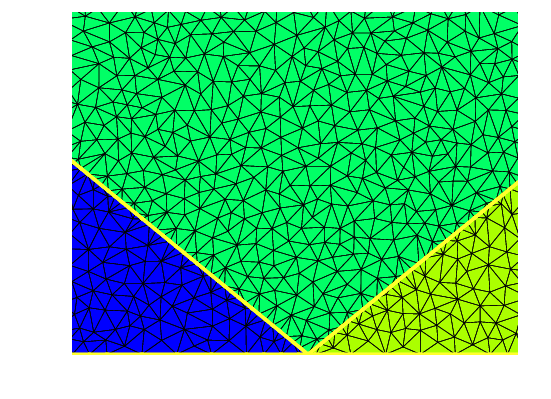}} \\ %\label{ss3-f5cc}}\\
%\end{subfigmatrix}
  \caption{Regular reflection: entropy field computed using the three different shock-modeling options and the \texttt{eulfs} solver.}
  \label{RR-f1}
\end{figure}

The computational domain is a $2\,L \times L$ rectangle, which is shown in Fig.~\ref{fig:RR1-sketch} along with the boundary conditions. The background mesh, {which has been generated using the \texttt{Triangle}~\cite{Triangle,shewchuk1996engineering} mesh generator by constraining the triangle areas not to exceed $0.0002\,L^2$}, is made of 10281 grid-points and 16016 triangles. Unlike the meshes used for the previous test-cases, which were almost entirely made of equilateral triangles, the one used here features a rather irregular distribution of grid-points {and scalene triangles; compare the right column of frames of Figs.~\ref{ss2-f5} and~\ref{RR-f1}}.

Figure~\ref{RR-f1} shows the computed entropy field for all three sets of {calculations: shock-capturing computed on the background mesh (top row), hybrid (middle row) and fully-fitted (bottom row)}. {All three different shock-modeling options make use of the \texttt{eulfs} solver}. Frames in the left column of  Fig.~\ref{RR-f1} show the entire computational domain (the dashed square points to the location of the IP), whereas those in the right column show a zoom of the region surrounding the IP.
{Further details and a more extensive commentary of the results} can be found in~\cite{paciorri2011shock}.
\subsection{Steady Mach reflection (\texttt{MachReflection-1, MachReflection-2})\label{subsec:MM} }
\begin{figure}[H]
\centering
\includegraphics[width=0.750\linewidth]{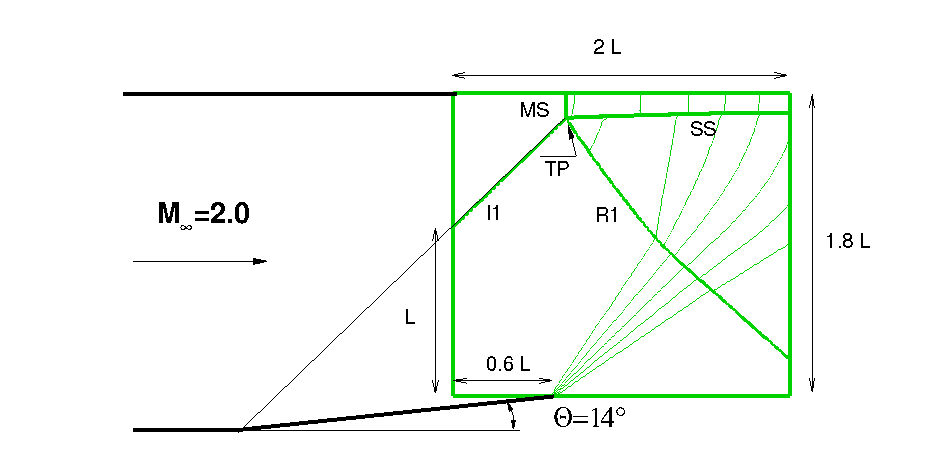}
\caption{Steady Mach reflection: flow configuration.}
\label{fig:MRdomain} 
\end{figure}

{Whenever the flow deflection imposed by a solid surface on an impinging weak oblique shock is larger than the maximum allowable deflection for the {downstream} Mach number, a so-called Mach reflection takes place instead of the regular reflection already examined in Sect.~\ref{subsec:RR}}.
This is schematically shown in Fig.~\ref{fig:MRdomain}: a uniform, supersonic ($M_{\infty}=2$) stream of air undergoes a $\Theta=14^{\circ}$ deflection through the incident shock, I1. Regular reflection of I1 is however impossible for the chosen pair of $M_{\infty},\Theta$ parameters, so that a steady triple-point (TP) arises which joins I1, the reflected shock (R1), the Mach stem (MS) and the slip-stream (SS).

This test-case {has been split into} two different simulations: the \texttt{MachReflection-1} directory contains the hybrid simulation, whereas a fully-fitted simulation {can be run} in the \texttt{MachReflection-2} directory. In the former case the incident shock and the {slip-stream are captured}, whereas the reflected shock and the Mach stem are fitted as a single shock. In the latter case all discontinuities, {as well as the triple-point,} are fitted. From the viewpoint of code-checking the two simulations are very different. The {hybrid simulation} only demonstrates the capability of the algorithm to run in hybrid mode, but tests the same functionalities already checked in the circular cylinder simulation, with the only addition of {the interaction between} a normal shock (the MS) and a {flat} wall. On the contrary, the {fully fitted simulation tests the capability of the \texttt{UnDiFi-2D} code to fit the triple point}.

\begin{figure}[h]
\centering
\subfloat[\texttt{eulfs} shock-capturing]{%}
\includegraphics[trim=1cm 1.9cm 1cm 3cm,clip=true,scale=0.22]{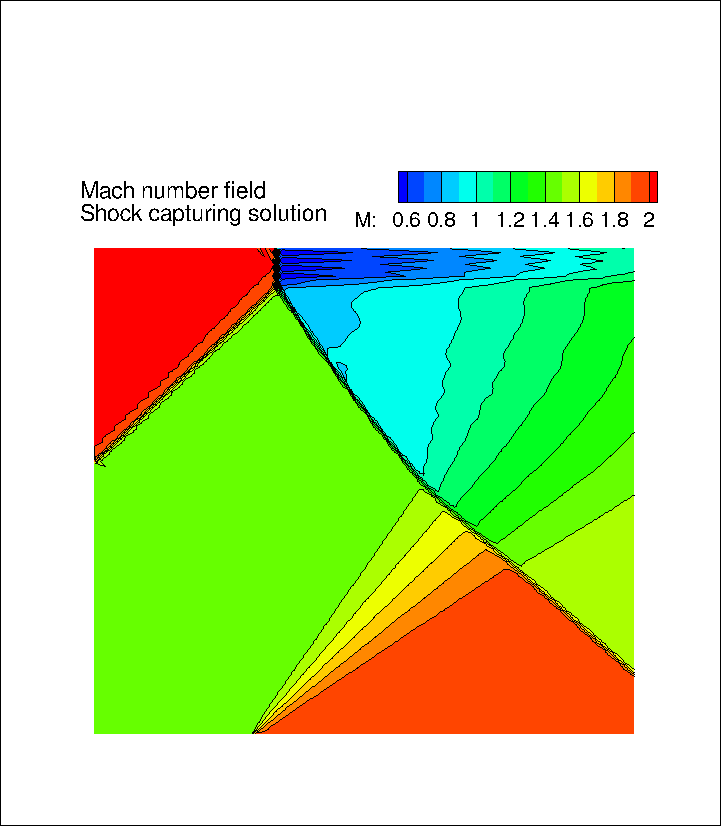}\label{fig:MRsols1a} }
\subfloat[\texttt{eulfs} hybrid]{%}
\includegraphics[trim=0.5cm 0.5cm 0.5cm 0cm,clip=true,scale=0.49]{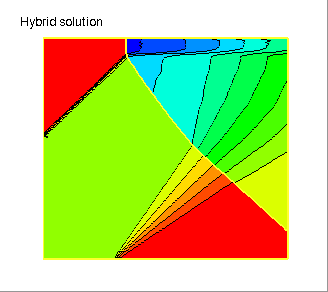}\label{fig:MRsols1b} }
\subfloat[\texttt{eulfs} fully fitted]{%
\includegraphics[trim=0.5cm 0.5cm 0.5cm 0cm,clip=true,scale=0.49]{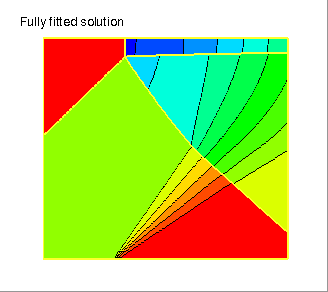}\label{fig:MRsols1c} }\\
\subfloat[\texttt{NEO} shock-capturing]{%}
\includegraphics[trim=1cm 1.9cm 1cm 3cm,clip=true,scale=0.22]{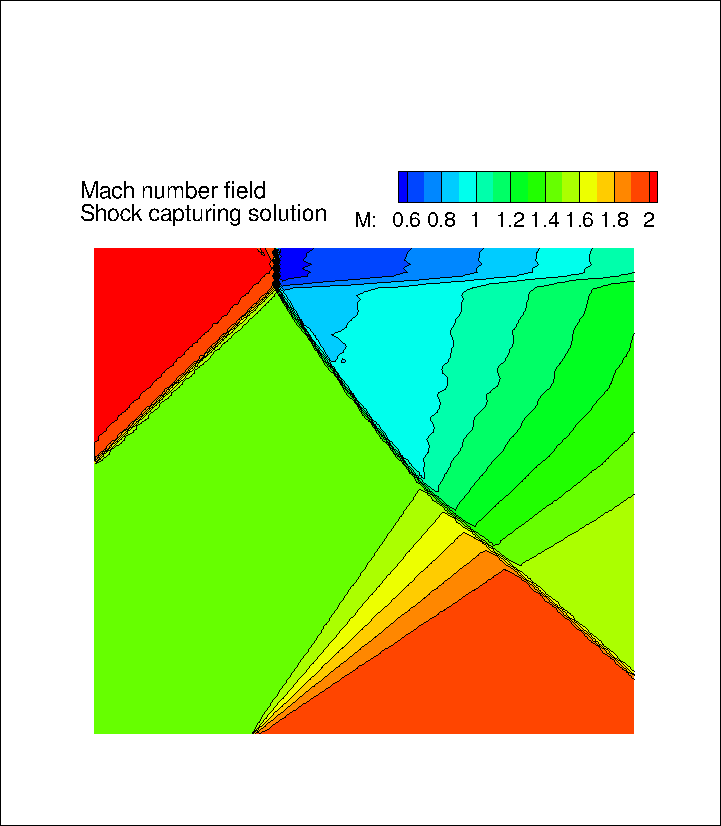}\label{fig:MRsols2a}}
\subfloat[\texttt{NEO} hybrid]{%}
\includegraphics[trim=1cm 1cm 1cm 0cm,clip=true,scale=0.32]{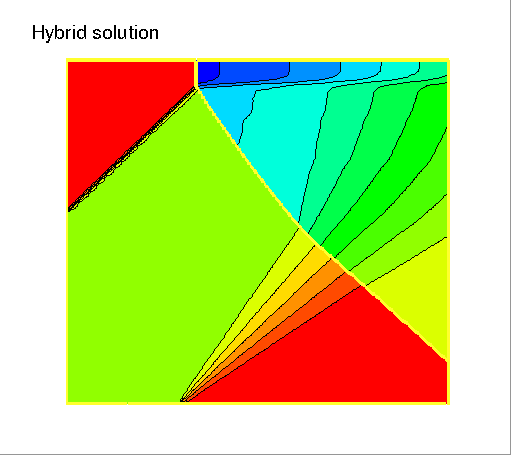}\label{fig:MRsols2b}}
\subfloat[\texttt{NEO} fully fitted]{%
\includegraphics[trim=1cm 1cm 1cm 0cm,clip=true,scale=0.32]{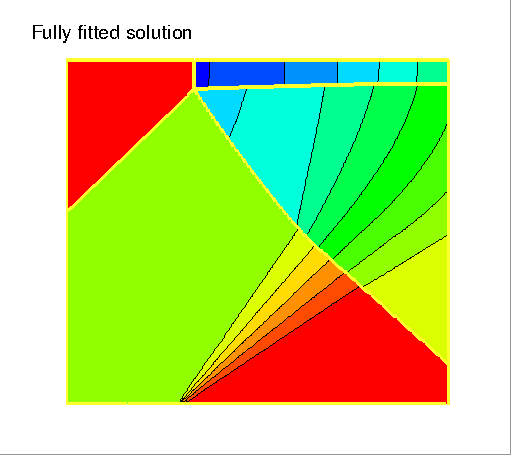}\label{fig:MRsols2c}}
\caption{Mach reflection: Mach number iso-contour lines {computed using three different shock-modeling options and the two different shock-capturing codes}.}
\label{fig:MRsols12} 
\end{figure}

The computational domain used for both simulations has been marked in green in Fig.~\ref{fig:MRdomain}. Inside this area a background grid of almost equilateral triangles has been generated using the \texttt{delaundo}~\cite{mullerdelaundo} mesh generator by specifying a uniform distribution of grid-points along the domain boundaries {with spacing $h = 0.0167\,L$, see Fig.~\ref{fig:MRdomain}}. Table~\ref{tab:MR} reports the number of triangles and grid-points of the background mesh along with those of the computational meshes at {steady-state for both} the hybrid and the fully-fitted simulations; the corresponding number of shock-points is also shown.
\begin{center}
\begin{table}[H]
\caption{Grid-points and triangles of the background and computational meshes.}
\label{tab:MR}
\begin{centering}
\begin{tabular}{cccccccccc}
\hline 
   \multicolumn{2}{c}{Background} &  & \multicolumn{7}{c}{Computational meshes at steady-state}   \tabularnewline
%\cline{2-6} 
   \multicolumn{2}{c}{Mesh} & &  \multicolumn{3}{c}{Hybrid simulation} & &\multicolumn{3}{c}{Fully fitted simulation}  \tabularnewline
\cline{1-2}  \cline{4-6} \cline{8-10}
 Cells & Nodes &  &  Cells & Nodes & Shock-points & & Cells & Nodes & Shock-points \tabularnewline
\hline 
29214 & 14833 &  & 29292 & 17633 & 142 & & 29365 & 19633 & 294 \tabularnewline
\hline
\end{tabular}
\end{centering}
\end{table}
\end{center}
{Shock-capturing, hybrid and fully-fitted calculations obtained using the two different Residual Distribution codes, \texttt{NEO} and \texttt{eulfs}, are compared in Fig.~\ref{fig:MRsols12}}. 

Even though the two solvers {use very similar numerical recipes, as explained in Sect.~\ref{sec:RDmethods}, it can be seen that the two shock-capturing solutions, Figs.~\ref{fig:MRsols1a} and~\ref{fig:MRsols2a}, exhibit non-negligible differences, in particular downstream of the Mach stem. These differences are significantly reduced in the hybrid simulations, Figs.~\ref{fig:MRsols1b} and~\ref{fig:MRsols2b}, and have almost disappeared in the fully fitted ones, Figs.~\ref{fig:MRsols1c} and~\ref{fig:MRsols2c}.}

Further analyses about these numerical solutions can be found in~\cite{paciorri2009shock-fitting,paciorri2011shock}, whereas the algorithmic details concerning the treatment of the triple point are reported in~\cite{ivanov2010computation}.
\subsection{Planar source flow (\texttt{Q1D}\label{subsec:Q1D}) }
\begin{figure}[ht]
\centering
%\subfigure[Shock-capturing]%
\subfloat[]{\includegraphics[scale=0.25]{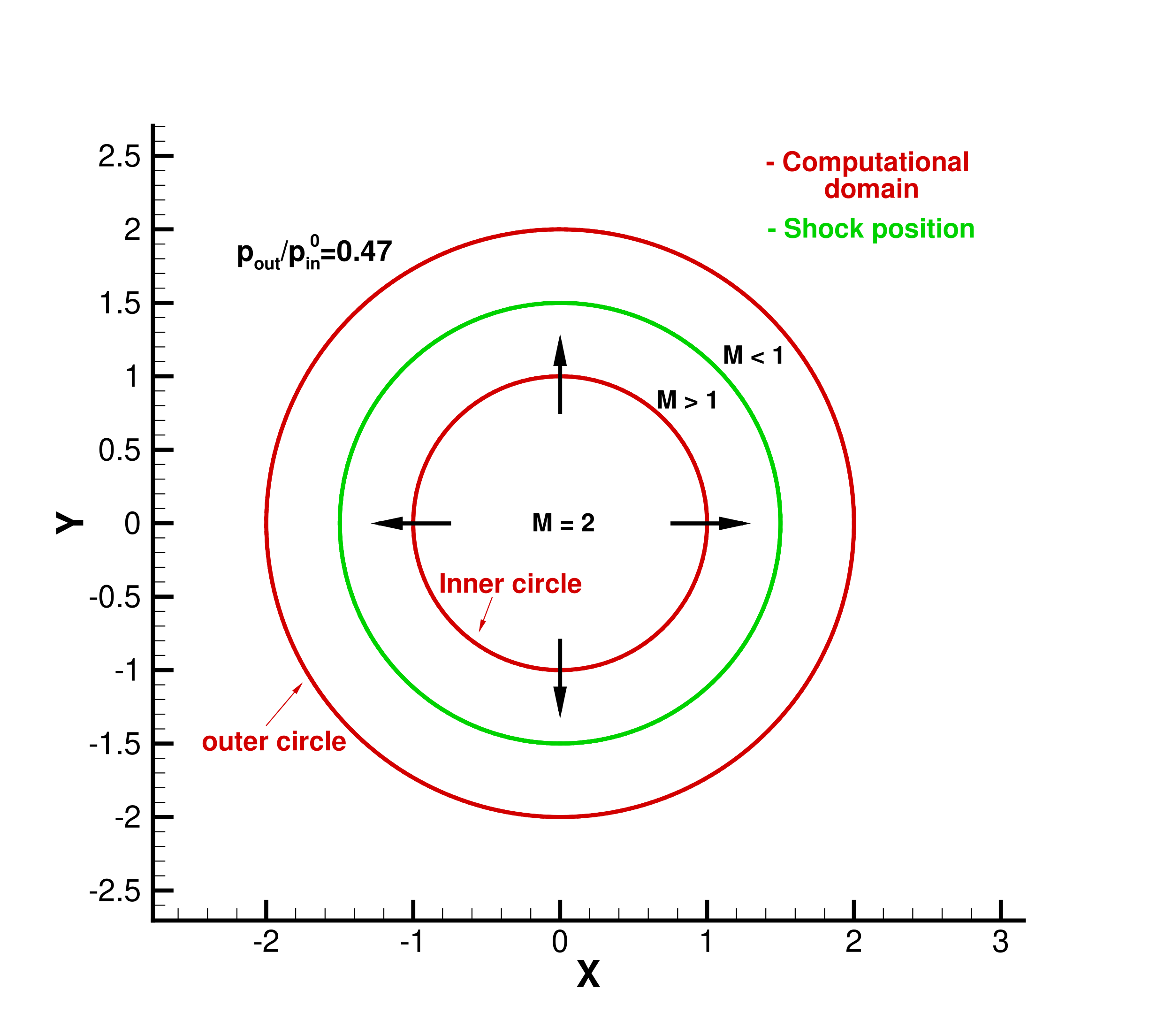}\label{ss3-f5a}}
%\subfigure[Shock-capturing]%
\subfloat[]{\includegraphics[scale=0.23]{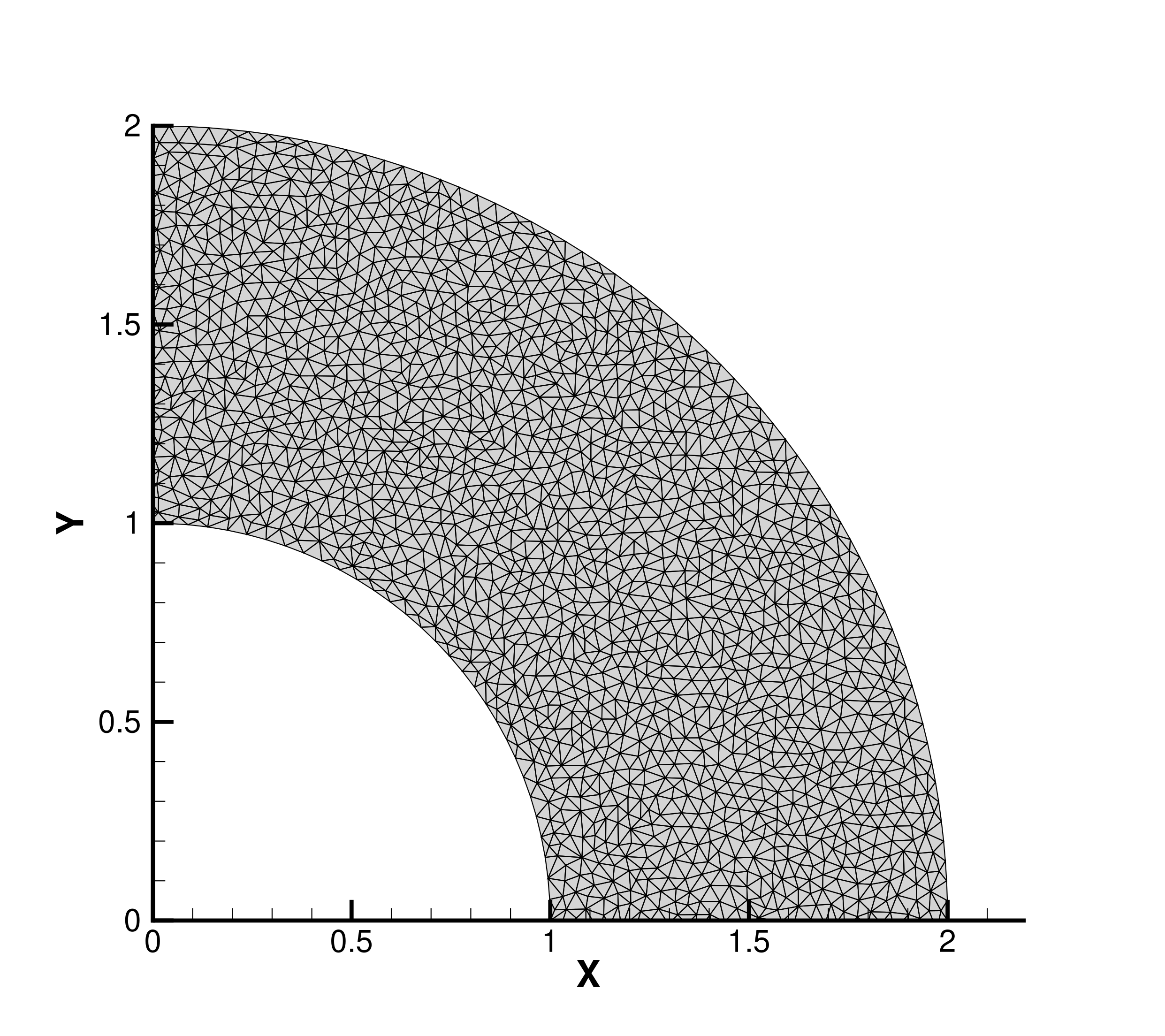}\label{ss3-f5b}}
%\subfigure[Hybrid]%\\
\subfloat[]{\includegraphics[scale=0.23]{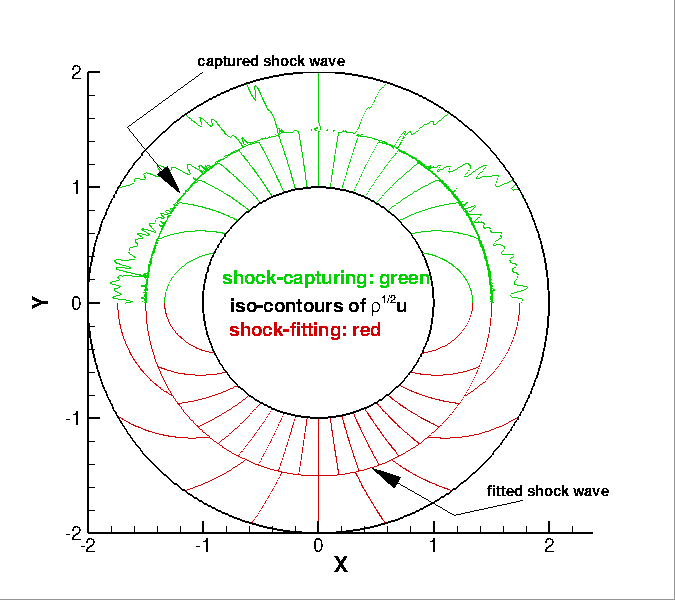}\label{ss3-f5c}}
\caption{ Planar, transonic source flow: computational domain, mesh and solutions.}
  \label{Q1D-f1}
\end{figure}

This test-case consists in a supersonic planar source flow that originates from the {centre of the inner red circle of Fig.~\ref{ss3-f5a} and, passing through a normal shock (the green circle in Fig.~\ref{ss3-f5a}), turns into a subsonic source flow}.
%Due to the availability of an analytical solution on not uniform regions, this test case can be used as an useful verification test by performing  convergence analysis based on the exact errors. 
Assuming that the analytical velocity field has a purely radial velocity component, it may be easily verified that the governing PDEs, written in a polar coordinate system, become identical to those governing a compressible quasi-one-dimensional variable-area flow:
\begin{equation}
\label{eq:arearule}
        \frac{1}{M}\bigg[\frac{2}{\gamma+1}\bigg(1\,+\,\frac{\gamma-1}{2}M^2\bigg)\bigg]^{\frac{\gamma+1}{2(\gamma-1)}}
\,=\,\frac{A}{A^*}
\end{equation}
provided that the nozzle area $A = A\left(r\right)$ varies linearly with the radial distance, $r$, from the {source. Equation~(\ref{eq:arearule})} allows to calculate the solution in the smooth flow regions both upstream and downstream of the normal shock, whereas {pre- and post-shock conditions are linked through} the R-H jump relations. 

The computational domain consists in the annulus sketched in Fig.~\ref{ss3-f5a}: the ratio between the radii of the outer and inner circles ($L = r_{in}$) has been set equal to $r_{out} /r_{in} = 2$ {and the exact solution features} a supersonic inlet flow ($M_{in} = 2)$  along the inner circle and a ratio between the outlet static and inlet total pressures $p_{out} / p^0_{in} = 0.47$ such that the shock forms at $r_{sh} /r_{in} = 1.5$.
The Delaunay mesh, which is partially shown in Fig.~\ref{ss3-f5b}, is made
of 6916 grid-points and 13456 triangles and has been generated using \texttt{Triangle}~\cite{Triangle,shewchuk1996engineering} in such a way that no {systematic} alignment {occurs} between the edges of the triangulation and the shock-edges, thus making the discrete problem truly two-dimensional. 
{Due to radial symmetry, the shock is a closed curve, as shown in Figs.~\ref{ss3-f5a} and~\ref{ss3-f5c}. From an algorithmic viewpoint,
this is obtained by overlapping the two end-points of the shock-mesh.
Therefore, beside being a powerful tool for code verification, thanks
to the availability of an exact solution, the present test-case checks that the two end-points of the shock-mesh are correctly joined to form a single closed shock}.

A qualitative comparison between the shock-capturing and shock-fitting solutions computed using the \texttt{eulfs} solver is shown in Fig.~\ref{ss3-f5c}, which shows the iso-contours of the third component, $\sqrt{\rho}u$, of the parameter vector.
It is evident that the two solutions are identical within the
supersonic, shock-upstream region, whereas noticeable {differences}
are visible within the entire subsonic, shock-downstream region.
{This is because} the captured shock wave has given rise to
remarkable oscillations downstream of the shock wave.
Full details {concerning this test-case} can be found in~\cite{bonfiglioli2014convergence}.

\subsection{Shock {formation due to the coalescence of compression waves}  
(\texttt{CoaSHCK}) 
\label{subsec:CS}}
\begin{figure}[ht]
\centering
\subfloat[Flow sketch]{\includegraphics[width=0.45\linewidth]{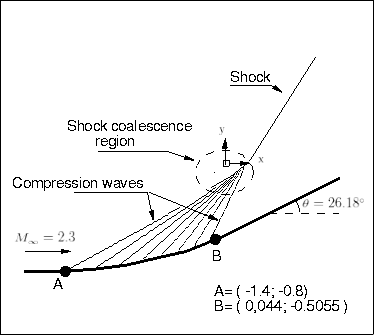}\label{fig:CoaSHCK-1a}}\quad\quad
\subfloat[Computational domain]{\includegraphics[width=0.45\linewidth]{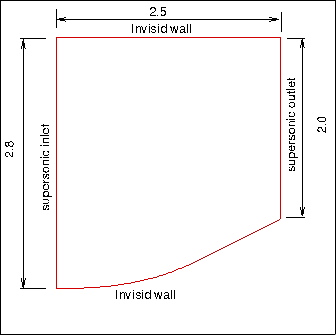}\label{fig:CoaSHCK-1b}}
\caption{{Shock formation due to the coalescence of compression waves.}}
\label{fig:CoaSHCK-1} 
\end{figure}

{Figure~\ref{fig:CoaSHCK-1a} shows a supersonic stream ($M_{\infty} = 2.3$) being deflected by a  convex wall. The wall is straight up to point A and beyond point B, whereas the shape of the wall smoothly changes between A and B according to the following cubic polynomial law:
\begin{equation}\label{eq:walldef}
    y=0.039923\,x^3+0.25144\,x^2+0.46928\,x-0.52683 \quad\quad -1.4 \le x \le 0.044
\end{equation}
As sketched in Fig.~\ref{fig:CoaSHCK-1a}, the Mach waves that originate in A and B bound a simple-wave compression region made of straight characteristic lines that merge into a shock-wave at some distance from the wall.}
Figure~\ref{fig:CoaSHCK-1b} shows the computational domain and the boundary conditions. The background mesh, which is {made} of 25531 triangles and 12965 {grid-points, has been generated using the} \texttt{delaundo} mesh generator by specifying a uniform distribution of the boundary nodes with spacing $h = 0.025$. 

\begin{figure}[ht]
\centering
\subfloat[Shock-capturing.]{\includegraphics[width=0.45\linewidth]{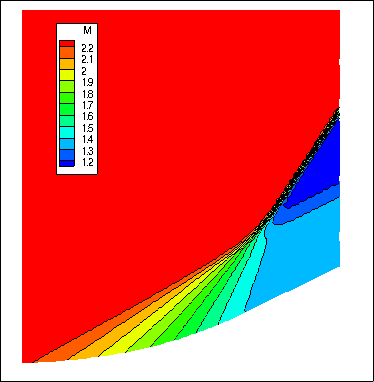}\label{fig:CoaSHCK-2a}}\quad\quad
\subfloat[Shock-fitting.]{\includegraphics[width=0.45\linewidth]{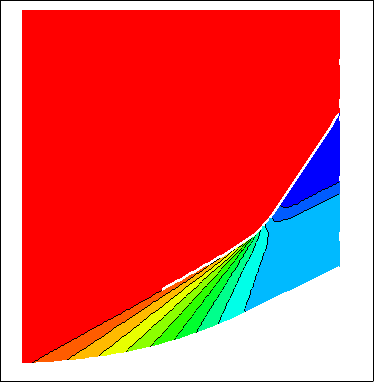}\label{fig:CoaSHCK-2b}}
\caption{{Shock formation due to the coalescence of compression waves: Mach iso-contour lines}.} 
\label{fig:CoaSHCK-2} 
\end{figure}

Figure~\ref{fig:CoaSHCK-2} compares the Mach iso-contour lines computed with the shock-capturing and shock-fitting approaches using the \texttt{eulfs} CFD solver.

{Inspection of Fig.\ref{fig:CoaSHCK-2b}, which refers to the shock-fitting calculation, clearly reveals that the fitted shock-mesh (shown using a white solid line) starts well ahead of the point where the characteristics merge into the shock. This is because, in Moretti's words~\cite{10.1007/BFb0019764}: ``Premature fitting of the shock in the region where compression waves tend to coalesce is not harmful at all, provided that the shock behaves as one of the characteristic surfaces coalescing into a finite discontinuity''. Modeling shock-formation as described by Moretti requires an {\em ad-hoc}
calculation of the normal to the shock within the end-point of the shock-mesh, so that the present test-case checks that this functionality works correctly}.

\subsection{Transonic flow past a NACA0012 profile (\texttt{NACA0012\_M080\_A0})} \label{subsec:NACA}
\begin{figure}[ht!]
\centering
\subfloat[Background triangulation.]{\includegraphics[width=0.45\linewidth]{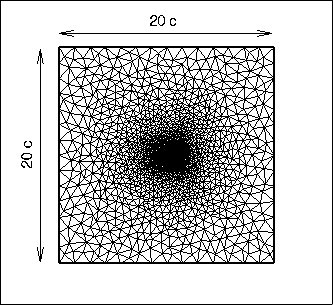}\label{fig:NACA-f1a}}\quad\quad
\subfloat[{Computational} mesh at steady-state.]{\includegraphics[width=0.45\linewidth]{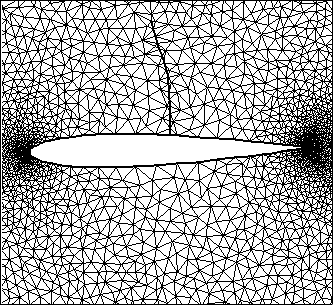}\label{fig:NACA-f1b}}\\
\subfloat[Zoom around the profile.]{\includegraphics[width=0.45\linewidth]{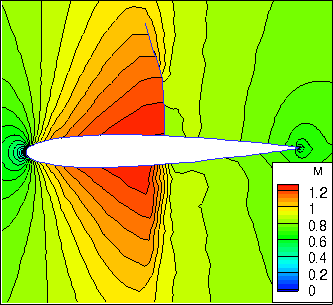}\label{fig:NACA-f2a}}\quad\quad
\subfloat[Close-up of the tip of the shock.]{\includegraphics[width=0.45\linewidth]{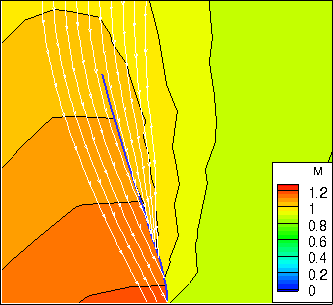}\label{fig:NACA-f2b}}
\caption{Transonic flow past the NACA0012 profile: meshes and Mach number iso-contour lines.}
\label{fig:NACA-f1} 
\end{figure}

The present test-case {deals with} the two-dimensional, inviscid, transonic flow past the NACA 0012 airfoil at $\alpha_{\infty} = 0^{\circ}$ degrees angle of attack and free-stream Mach number equal to $M_{\infty} = 0.80$.
The profile is placed in the middle of a $20c \times 20c$ square domain where {the characteristic length} $c$ is the profile's chord, see Fig.~\ref{fig:NACA-f1}. 
A symmetric flow-field has been chosen since it allows to compare the  shock-capturing and shock-fitting approaches using a single numerical simulation: as shown in Fig.~\ref{fig:NACA-f1b}, the shock on the upper side of the profile has been fitted, whereas the one on the lower side has been captured.
{Figure~\ref{fig:NACA-f1a} shows the background triangulation which has been generated using \texttt{Triangle}~\cite{Triangle,shewchuk1996engineering} and features 9912 triangles and 5024 mesh-points, 78 of which are placed along the airfoil's profile. Grid-points have been clustered close to the airfoil (see Fig.~\ref{fig:NACA-f1a}), in particular close to the leading-edge and trailing-edge (see Fig.~\ref{fig:NACA-f1b})}. 
Figure~\ref{fig:NACA-f1b} shows the {computational} mesh at steady state: the mesh above the profile has been modified by the shock-fitting algorithm to accommodate the {shock-mesh} and it only differs from the background triangulation in the immediate neighbourhood of the discontinuity; the mesh below the profile coincides with the background triangulation.

{As shown in Fig.~\ref{fig:NACA-f1b}, the shock-mesh is bounded by the foot and the tip of the shock. A close-up of the Mach number iso-contour lines close to the tip is shown in Fig.~\ref{fig:NACA-f2b},} which reveals that the shock is formed through the coalescence of those characteristic curves of the steady Euler equations that form an angle $\mu$ with respect to the streamline,  $\mu$ being the Mach angle. These same features could hardly be seen in a shock-capturing solution, unless a much finer, or feature-adapted, mesh were used. 

{The numerical treatment of the tip of the shock is identical to that described in Sect.~\ref{subsec:CS} so that the
present test-case checks that the shock-fitting algorithm handles correctly the foot of the shock, i.e.\ a normal shock impinging on} a curved wall.

Finally, Fig.~\ref{fig:NACA-f2a} allows to compare the fitted and captured shocks using grids of nearly identical spatial resolution: it is clear that shock-fitting provides a much more realistic shock thickness than shock-capturing.
Further details concerning this test case can be found in~\cite{bonfiglioli2018unstructured}.
\subsection{Shock-vortex interaction\label{subsec:SV} (\texttt{ShockVortex})}
This {last} test-case, {which is used to verify that \textit{UnDiFi-2D} works correctly also when dealing with unsteady flows, features the interaction between a moving vortex and a standing shock}.

\begin{figure}[H]
\begin{center}
\includegraphics[width=0.85\linewidth]{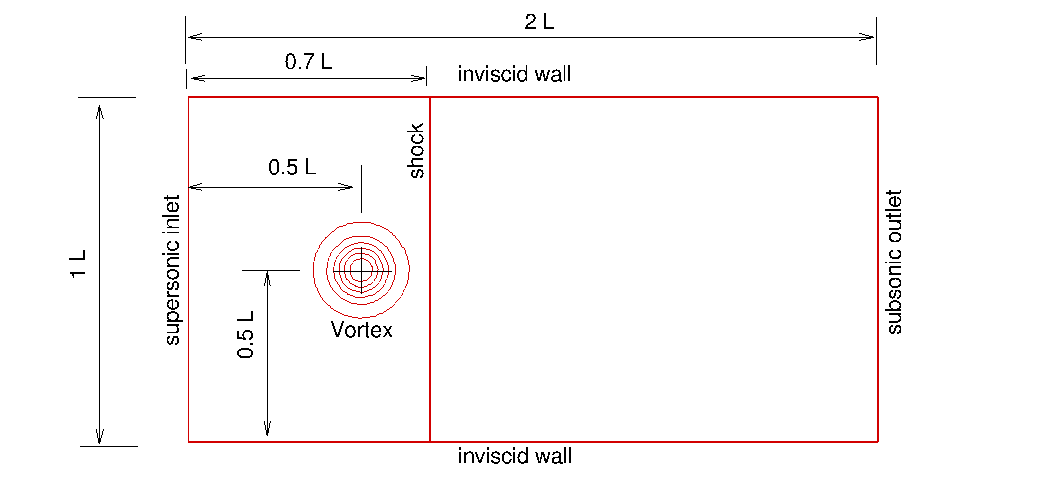}
%\subfloat[Initial and boundary conditions.]{\includegraphics[width=0.65\linewidth]{figs/SVdomain.png}}
%\subfloat[Plot of Eq.~(\ref{eq:tangentialV}).]{\includegraphics[width=0.25\linewidth]{figs/vortex.png}}
\caption{Shock--vortex interaction: computational domain and initial condition.}
\label{fig:SV1} 
\end{center}
\end{figure}

Figure~\ref{fig:SV1} shows the computational domain along with the boundary and initial conditions. 

{The flow-field is initialized by adding the perturbation velocity field induced by the vortex to the uniform flow past a steady normal shock. As shown in Fig.~\ref{fig:SV1}, at the initial time $t = 0$, the vortex is located 0.2 unit lengths $L$ ahead of the standing shock}.
{Using a cylindrical reference frame attached to the vortex core, the perturbation velocity field reads}:
\begin{subequations}
\begin{eqnarray}
\label{eq:tangentialV}
\widetilde{u}_{\theta}&=&-\epsilon|\boldsymbol{u}_{\infty}|\tau\exp^{\alpha\left(1-\tau^{2}\right)} \\  
\label{eq:radialV}
\widetilde{u}_{r}&=&0
\end{eqnarray}
\end{subequations}
{where the dimensionless radial distance from the vortex core, 
$\text{\ensuremath{\tau}=\ensuremath{\frac{r}{r_{c}}}}$, has been used in
Eq.~\eqref{eq:tangentialV}, with $r_c = 0.05\,L$.} {The two dimensionless parameters $\alpha$ and $\epsilon$, which respectively control the width and magnitude of the velocity perturbation, %that appear in Eq.~\eqref{eq:tangentialV}
are mutually related via the shock and
vortex Mach numbers}:
\begin{equation}\label{eq:machV}
\epsilon = \frac{M_{v}}{M_{s}} \frac{\sqrt{2\alpha} }{\exp^{\left(\alpha-\frac{1}{2}\right)}}
\end{equation}
where:
\begin{equation}
    M_s = \frac{|\boldsymbol{u}_{\infty}|}{a_{\infty}} \qquad
    M_v = \frac{\max|u_{\theta}|}{a_{\infty}}
\end{equation}
{For the chosen pair of shock and vortex Mach numbers: $M_s = 2$, $M_{v} = 0.2$, which gives rise to a {\em weak} shock-vortex interaction, according to the nomenclature of~\cite{key-7}, the vortex strength  $\epsilon \approx 8.6\,10^{-2}$ follows from Eq.~(\ref{eq:machV}), having set $\alpha = 0.204$}.

\begin{figure}[H]
\centering
\includegraphics[scale=0.275]{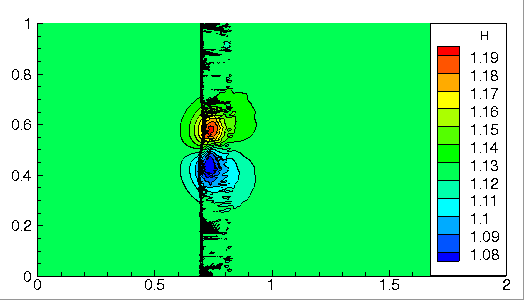}
\includegraphics[scale=0.275]{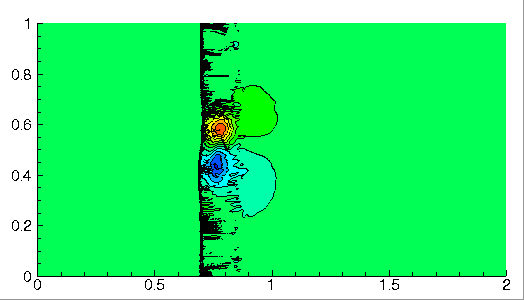}
\includegraphics[scale=0.275]{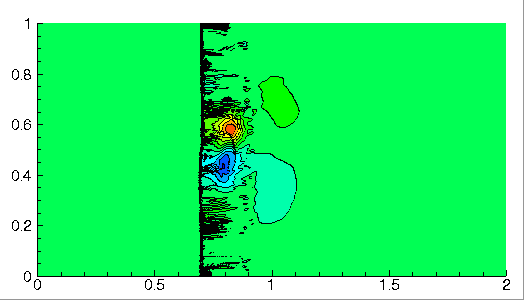} \\
\includegraphics[scale=0.275]{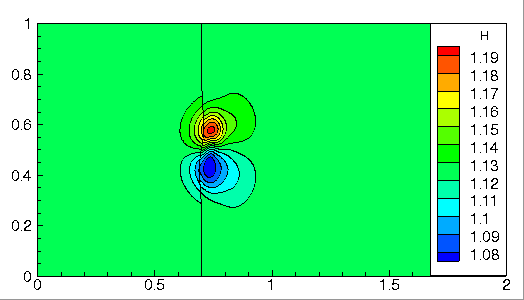}
\includegraphics[scale=0.275]{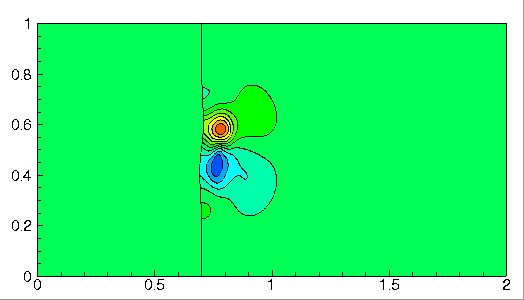}
\includegraphics[scale=0.275]{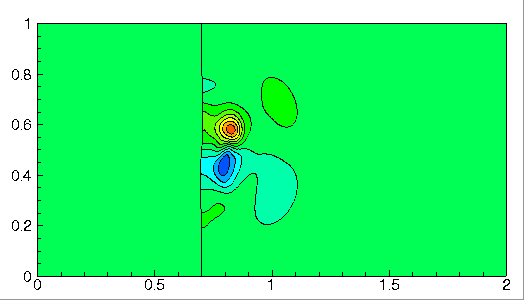}
\caption{Shock-vortex interaction: total enthalpy contours at times $t = 0.3$ (left), $t = 0.4$ (centre) and $t = 0.5$ (right) using {the LDA scheme; shock-capturing in the upper row of frames and shock-fitting in the lower row}.
%15 equally spaced contour levels obtained with the shock-capturing (top) and with shock-fitting (bottom). (a) $t = 0.3$ LDA (capturing). (b) $t = 0.4$ LDA (capturing). (c) $t = 0.5$ LDA (capturing). (d) $t = 0.2$ LDAN (capturing). (e) $t = 0.3$ LDAN (capturing). (f) $t = 0.4$ LDAN (capturing). (g) $t = 0.3$ LDA (fitting). (h) $t = 0.4$ LDA (fitting). (i) $t = 0.5$ LDA (fitting).
}
\label{fig:SV2} 
\end{figure}

The {Delaunay triangulation} of the
computational domain was generated using \texttt{Triangle}; {it features 217569 grid-points and 433664 triangles and a mesh spacing along the boundaries equal to $h / L = 0.00375$}.
The shock-capturing and shock-fitting simulations were performed using the \texttt{NEO} solver. 
A qualitative comparison {between the two shock-modeling options} is given in Fig.~\ref{fig:SV2},
{where total enthalpy iso-contour lines at three subsequent time instants are shown: shock-capturing on the top row and shock-fitting on the bottom row}.
In particular, besides the oscillations related to the approximation of the shock, we can see clearly that the contours downstream of the discontinuity are much less smooth in the captured solution. 

The fitted computations, on the other hand, show very nice and smooth contours.
Further details about this test-case and further simulations {dealing with shock-vortex} interactions can be found in~\cite{bonfiglioli2016unsteady,campoli2017and}.

\section{Concluding remarks and {future} perspectives}
\label{sec:development}

In this article {we have presented} the state-of-the-art of a 10-year-long development of a shock-fitting technique for unstructured meshes.
%, highlighting its strengths, but also its current weaknesses. The analysis of a few selected applications allowed us to underline the advantages that fitted shocks offer over captured ones, when unstructured grids are used. It has also been shown that the use of unstructured grids allows to relieve some of the shortcomings that plagued the classical shock-fitting technique, originally developed within the structured-grid environment. Adequate room has also been given to the current limitations of the technique.
%underlining that these are not intrinsic to the methodology, and presenting some preliminary attempts aimed at their overcoming. 

{We have described the key algorithmic features of the technique, which has been} implemented in an open-source code, available in a dedicated repository. {A fairly extensive selection of ready-to-run test-cases demonstrates the features and current} capabilities of the code, {which we have shown to be able to deal with compressible flows featuring either isolated or mutually interacting discontinuities. The superior quality of fitted shock-waves over captured ones has also been emphasized.

{A number of unsolved issues remain to be addressed, as detailed hereafter.}

{In all flow-cases presented in this paper the effects of viscosity have been neglected. When dealing with viscous flow, shocks cease to be discontinuities in a mathematical sense, and have a finite thickness. At high Reynolds number, however, the width of a shock is comparable to the {\em microscopic} length-scales, hence orders of magnitude smaller than any {\em macroscopic} length scale, such as Kolmogorov's, for instance. Under this circumstances, it is still appropriate to ignore the inner shock-structure and consider shock-waves as having zero thickness}. %Even so, the presence of boundary-layers makes viscous flows very different from inviscid ones as far as the interaction between shock-waves and solid boundaries is concerned, because the near-wall boundary-layer is subsonic and, therefore, shock-waves cannot reach a wall when the effects of viscosity are accounted for.} 
The capability of dealing with shock/boundary-layer interactions is work in progress~\cite{Assonitis2020124} and it is likely going to be introduced in a (near) future release of the project}.

{When dealing with complex shock-shock and/or shock-boundary interactions in {\em steady} flows, a preliminary shock-capturing calculation, followed by an automatic shock-detection and shock-pattern identification strategy~\cite{paciorri2020accurate}, is capable of supplying a reasonably good initial guess to the shock-fitting algorithm.}

{When dealing with {\em un-steady} flows, however, things get much harder and the currently unsolved issues that remain to be addressed are well summarized in a 1986 paper by Glimm and co-workers~\cite{chern1986front}:
\begin{quote}
\begin{enumerate}
    \item Treating changes of the topology of regions bounded by fronts from simply connected to multiply connected regions.
    \item Treating the disappearance of weakening fronts and the appearance of new fronts at boundaries or at collisions of other fronts
\end{enumerate}
\end{quote}
}
%In its current As a matter of fact, the current code is not yet capable of simulating all types of 2D flows and two are the main limitations of the code. The present shock-fitting code, indeed, is unable to deal with viscous flows with shock/boundary-layer interactions and unsteady flows featuring changes in the topology of the shock-pattern. %number of shocks and flow lines, and in the number and type of shock-shock and shock-wall interactions, occur.
%{\color{blue}}
%The first limitation could be removed in a short amount of time.
%In fact, interactions between shock wave and boundary layer have recently been calculated using the shock-fitting code in~\cite{Assonitis2020124} and an improved version of the code to be uploaded into the repository is currently under verification and testing and we are confident that it will be released shortly. Overcoming the second limitation will require more effort and time.
{As far as the first issue is concerned, changes in the shock topology may occur in un-steady flows for several reasons. For example, new discontinuities may appear when a shock-wave moving along a wall encounters an abrupt  change in the slope of the wall or when two discontinuities begin to interact. 
As far as the second issue is concerned, beside the formation of a shock due to the steepening of compression waves, a mechanism which we have examined in Sect.~\ref{subsec:CS}, an existing shock may progressively weaken and finally disappear}. {In order to be able}
to manage all these topological changes it will be necessary to develop {new algorithmic tools} capable of detecting the occurrence {of a change in the shock-topology} and modify {accordingly} the fitted discontinuities and {their mutual} interactions.
Some tests have been carried out on specific flow configurations, as shown in~\cite{bonfiglioli2017development,paciorri2017basic}, but the development of a {general purpose tool} will not be trivial and will probably require the {use} of advanced, {multi-disciplinary} techniques such as those used in~\cite{paciorri2020accurate}. 
The effort which will be necessary to complete the development of the {unstructured, shock-fitting technique} and bring it to {full} maturity, is not modest and probably requires the merging of different skills and expertise. In order to be successful, it will be necessary to broaden the audience of {developers} and technical expertise involved. With such a goal in mind, we launch the present project.
%Our hope is that over the next 10 years the unstructured shock-fitting can reach full maturity and be considered as a valid alternative to shock-capturing techniques, thereby achieving the goal that has been so actively pursued by Gino Moretti during his scientific life.
%To the authors knowledge there is no open-source, actively maintained and well-documented shock-fitting code other than \textit{UnDiFi} available online. The present paper was born with the aim of making the shock-fitting technique more popular among the CFD community. In order to reach this goal, by means of a thorough presentation, a description of the most important algorithmic features was given, and a \texttt{github} repository has been made publicly available.

\subsection{Key features}
Usability, modularity, maintenance and enhancement are the key features of the \textit{UnDiFi-2D} project. In particular, a comprehensive documentation is provided by means of exhaustive source files comments,
%that are parsed through \texttt{doxygen} free software producing high quality \texttt{html} and \texttt{latex} documentation pages.} 
an in-depth user-manual and a selection of ready-to-run test-cases. % are included in the repository to show the features of the code. 
{Moreover}, in order to demonstrate code modularity, the shock-fitting {algorithm has been} coupled with two different CFD solvers (\texttt{NEO} and \texttt{EulFS}) that can be alternatively selected by the user {at run-time}. 
\subsection{Known issues}
An overview of the currently open issues and future lines of development has been given {in the opening of Sect.~\ref{sec:development}}. It is also worth emphasizing that \textit{UnDiFi-2D} is developed and maintained by a team of {CFD developers, rather than computer scientists}, meaning that \textit{UnDiFi-2D} is designed to be as simple as possible. {For instance}, no effort has been made to improve the efficiency of the algorithm by introducing any form of parallelism. At present, the main computational bottleneck is due to the {use of disk} I/O for communicating among the different building blocks of the algorithm, as described in Sect.~\ref{subsec:UnDiFi}. In this sense, a better performing infrastructure may be devised in future releases, but up to now it has been considered as a secondary research branch with respect to the algorithmic development. {Finally}, modern features of the Fortran language may be exploited in the future to enhance the intrinsic modularity of the algorithm.

\subsection{Perspectives}
The reported examples of applications and the corresponding bibliographic references, have shown that \textit{UnDiFi-2D} is a mature project able to already provide superior results in terms of accuracy for many test-cases with respect to {state-of-the-art} shock-capturing codes. 
%In~\cite{bonfiglioli2013three-dimensional} a real application (e.g. complex gas dynamics application) has been presented. 
However, \textit{UnDiFi-2D} is a relatively young project with some important limitations. Nevertheless, the software platform that we have presented in the article represents a solid starting point for developments to overcome these limitations {in a collaborative framework}.

%\section*{Acknowledgments}
%This study is supported by Saint Petersburg State University (project No. 6.37.206.2016). L. Campoli acknowledges that he is employed by Saint Petersburg State University in the frame of post-doctoral fellowship 02/7-79-226.
%The author would like to thank Renato Paciorri and Aldo Bonfiglioli for the time they spent sharing their knowledge about such fascinating technique and for providing the most updated version of the discontinuity-fitting algorithm code. Special thanks are due to Mario Ricchiuto for sharing his residual distribution solver and for the help in coupling the codes.

%\bibliographystyle{spphys}    
\bibliographystyle{elsarticle-num}
%\section*{Bibliography}
\bibliography{refs.bib}

\begin{thebibliography}{10}
\expandafter\ifx\csname url\endcsname\relax
  \def\url#1{\texttt{#1}}\fi
\expandafter\ifx\csname urlprefix\endcsname\relax\def\urlprefix{URL }\fi
\expandafter\ifx\csname href\endcsname\relax
  \def\href#1#2{#2} \def\path#1{#1}\fi

\bibitem{doi:10.1063/1.168744}
H.~G. Weller, G.~Tabor, H.~Jasak, C.~Fureby, A tensorial approach to
  computational continuum mechanics using object-oriented techniques, Computers
  in Physics 12~(6) (1998) 620--631.
\newblock \href {https://doi.org/10.1063/1.168744}
  {\path{doi:10.1063/1.168744}}.

\bibitem{palacios_stanford_2013}
F.~Palacios, J.~Alonso, K.~Duraisamy, M.~Colonno, J.~Hicken, A.~Aranake,
  A.~Campos, S.~Copeland, T.~Economon, A.~Lonkar, T.~Lukaczyk, T.~Taylor,
  Stanford {University} {Unstructured} ({SU}{\textasciicircum}2): {An}
  open-source integrated computational environment for multi-physics simulation
  and design, in: 51st {AIAA} {Aerospace} {Sciences} {Meeting} including the
  {New} {Horizons} {Forum} and {Aerospace} {Exposition}, American Institute of
  Aeronautics and Astronautics, 2013.
\newblock \href {https://doi.org/10.2514/6.2013-287}
  {\path{doi:10.2514/6.2013-287}}.

\bibitem{doi:10.1063/1.1699639}
J.~VonNeumann, R.~D. Richtmyer, A method for the numerical calculation of
  hydrodynamic shocks, Journal of Applied Physics 21~(3) (1950) 232--237.
\newblock \href {https://doi.org/10.1063/1.1699639}
  {\path{doi:10.1063/1.1699639}}.

\bibitem{Moretti1998}
G.~Moretti, Intellectual consumerism, Meccanica 33~(5) (1998) 524--531.
\newblock \href {https://doi.org/10.1023/A:1004376728032}
  {\path{doi:10.1023/A:1004376728032}}.

\bibitem{emmons-TN-1944}
H.~W. Emmons, \href{https://apps.dtic.mil/dtic/tr/fulltext/u2/b814416.pdf}{The
  numerical solution of compressible fluid flow problems}, NACA-TN 932, NASA
  (1944).
\newline\urlprefix\url{https://apps.dtic.mil/dtic/tr/fulltext/u2/b814416.pdf}

\bibitem{emmons1948flow}
H.~W. Emmons, \href{https://ntrs.nasa.gov/citations/19930082372}{Flow of a
  compressible fluid past a symmetrical airfoil in a wind tunnel and in free
  air}, NACA-TN 1746 (1948).
\newline\urlprefix\url{https://ntrs.nasa.gov/citations/19930082372}

\bibitem{moretti:66}
G.~Moretti, M.~Abbett, A time-dependent computational method for blunt body
  flows., AIAA Journal 4 (1966) 2136--41.
\newblock \href {https://doi.org/10.2514/2.6898} {\path{doi:10.2514/2.6898}}.

\bibitem{glimm1981front}
J.~Glimm, E.~Isaacson, D.~Marchesin, O.~McBryan, Front tracking for hyperbolic
  systems, Advances in Applied Mathematics 2~(1) (1981) 91--119.
\newblock \href {https://doi.org/https://doi.org/10.1016/0196-8858(81)90040-3}
  {\path{doi:https://doi.org/10.1016/0196-8858(81)90040-3}}.

\bibitem{GLIMM1985259}
J.~Glimm, C.~Klingenberg, O.~McBryan, B.~Plohr, D.~Sharp, S.~Yaniv, Front
  tracking and two-dimensional riemann problems, Advances in Applied
  Mathematics 6~(3) (1985) 259 -- 290.
\newblock \href {https://doi.org/https://doi.org/10.1016/0196-8858(85)90014-4}
  {\path{doi:https://doi.org/10.1016/0196-8858(85)90014-4}}.

\bibitem{chern1986front}
I.-L. Chern, J.~Glimm, O.~Mcbryan, B.~Plohr, S.~Yaniv, Front tracking for gas
  dynamics, Journal of Computational Physics 62~(1) (1986) 83 -- 110.
\newblock \href {https://doi.org/https://doi.org/10.1016/0021-9991(86)90101-4}
  {\path{doi:https://doi.org/10.1016/0021-9991(86)90101-4}}.

\bibitem{moretti2002thirty-six}
G.~Moretti, Thirty-six years of shock fitting, Computers {\&} Fluids 31~(4-7)
  (2002) 719--723.
\newblock \href {https://doi.org/10.1016/s0045-7930(01)00072-x}
  {\path{doi:10.1016/s0045-7930(01)00072-x}}.

\bibitem{pandolfi2001numerical}
M.~Pandolfi, D.~D'Ambrosio, Numerical instabilities in upwind methods: analysis
  and cures for the ``carbuncle'' phenomenon, Journal of Computational Physics
  166~(2) (2001) 271--301.
\newblock \href {https://doi.org/10.1006/jcph.2000.6652}
  {\path{doi:10.1006/jcph.2000.6652}}.

\bibitem{doi:10.2514/2.835}
M.~H. Carpenter, J.~H. Casper, Accuracy of shock capturing in two spatial
  dimensions, AIAA Journal 37~(9) (1999) 1072--1079.
\newblock \href {https://doi.org/10.2514/2.835} {\path{doi:10.2514/2.835}}.

\bibitem{Johnsen20101213}
E.~Johnsen, J.~Larsson, A.~V. Bhagatwala, W.~H. Cabot, P.~Moin, B.~J. Olson,
  P.~S. Rawat, S.~K. Shankar, B.~Sj\"ogreen, H.~Yee, X.~Zhong, S.~K. Lele,
  Assessment of high-resolution methods for numerical simulations of
  compressible turbulence with shock waves, Journal of Computational Physics
  229~(4) (2010) 1213 -- 1237.
\newblock \href {https://doi.org/10.1016/j.jcp.2009.10.028}
  {\path{doi:10.1016/j.jcp.2009.10.028}}.

\bibitem{pirozzoli2011numerical}
S.~Pirozzoli, Numerical methods for high-speed flows, Annual Review of Fluid
  Mechanics 43~(1) (2011) 163--194.
\newblock \href {https://doi.org/10.1146/annurev-fluid-122109-160718}
  {\path{doi:10.1146/annurev-fluid-122109-160718}}.

\bibitem{zhong2012direct}
X.~Zhong, X.~Wang, Direct numerical simulation on the receptivity, instability,
  and transition of hypersonic boundary layers, Annual Review of Fluid
  Mechanics 44~(1) (2012) 527--561.
\newblock \href {https://doi.org/10.1146/annurev-fluid-120710-101208}
  {\path{doi:10.1146/annurev-fluid-120710-101208}}.

\bibitem{ROMICK2017210}
C.~M. Romick, T.~D. Aslam, High-order shock-fitted detonation propagation in
  high explosives, Journal of Computational Physics 332 (2017) 210 -- 235.
\newblock \href {https://doi.org/https://doi.org/10.1016/j.jcp.2016.11.049}
  {\path{doi:https://doi.org/10.1016/j.jcp.2016.11.049}}.

\bibitem{nasuti2017steady}
F.~Nasuti, M.~Onofri, Steady and Unsteady Shock Interactions by Shock Fitting
  Approach, Springer International Publishing, Cham, 2017, pp. 33--55.
\newblock \href {https://doi.org/10.1007/978-3-319-68427-7_2}
  {\path{doi:10.1007/978-3-319-68427-7_2}}.

\bibitem{Rawat20106744}
P.~S. Rawat, X.~Zhong, On high-order shock-fitting and front-tracking schemes
  for numerical simulation of shock-disturbance interactions, Journal of
  Computational Physics 229~(19) (2010) 6744 -- 6780.
\newblock \href {https://doi.org/10.1016/j.jcp.2010.05.021}
  {\path{doi:10.1016/j.jcp.2010.05.021}}.

\bibitem{paciorri2009shock-fitting}
R.~Paciorri, A.~Bonfiglioli, A shock-fitting technique for 2d unstructured
  grids, Computers \& Fluids 38~(3) (2009) 715 -- 726.
\newblock \href {https://doi.org/10.1016/j.compfluid.2008.07.007}
  {\path{doi:10.1016/j.compfluid.2008.07.007}}.

\bibitem{ivanov2010computation}
M.~S. Ivanov, A.~Bonfiglioli, R.~Paciorri, F.~Sabetta,
  \href{https://doi.org/10.1007/s00193-010-0266-y}{Computation of weak steady
  shock reflections by means of an unstructured shock-fitting solver}, Shock
  Waves 20~(4) (2010) 271--284.
\newline\urlprefix\url{https://doi.org/10.1007/s00193-010-0266-y}

\bibitem{paciorri2011shock}
R.~Paciorri, A.~Bonfiglioli, Shock interaction computations on unstructured,
  two-dimensional grids using a shock-fitting technique, Journal of
  Computational Physics 230~(8) (2011) 3155 -- 3177.
\newblock \href {https://doi.org/http://dx.doi.org/10.1016/j.jcp.2011.01.018}
  {\path{doi:http://dx.doi.org/10.1016/j.jcp.2011.01.018}}.

\bibitem{bonfiglioli2014convergence}
A.~Bonfiglioli, R.~Paciorri, Convergence analysis of shock-capturing and
  shock-fitting solutions on unstructured grids, AIAA J. 52~(7) (2014)
  1404--1416.
\newblock \href {https://doi.org/10.2514/1.J052567}
  {\path{doi:10.2514/1.J052567}}.

\bibitem{pepe2014shock-fitting}
R.~Pepe, A.~Bonfiglioli, A.~D'Angola, G.~Colonna, R.~Paciorri, Shock-fitting
  versus shock-capturing modeling of strong shocks in nonequilibrium plasmas,
  Plasma Science, IEEE Transactions on 42~(10) (2014) 2526--2527.
\newblock \href {https://doi.org/10.1109/TPS.2014.2324493}
  {\path{doi:10.1109/TPS.2014.2324493}}.

\bibitem{pepe2014towards}
R.~Pepe, A.~Bonfiglioli, R.~Paciorri, A.~Lani, J.~Garicano-Mena, C.~F.
  Ollivier-Gooch,
  \href{http://wccm-eccm-ecfd2014.org/admin/files/filePaper/p3220.pdf}{Towards
  a Modular Approach for Unstructured Shock-Fitting}, International Center for
  Numerical Methods in Engineering, Barcelona, Spain, 2014.
\newline\urlprefix\url{http://wccm-eccm-ecfd2014.org/admin/files/filePaper/p3220.pdf}

\bibitem{bonfiglioli2016unsteady}
A.~Bonfiglioli, R.~Paciorri, L.~Campoli, Unsteady shock-fitting for
  unstructured grids, International Journal for Numerical Methods in Fluids
  81~(4) (2016) 245--261.
\newblock \href {https://doi.org/10.1002/fld.4183}
  {\path{doi:10.1002/fld.4183}}.

\bibitem{campoli2017and}
L.~Campoli, P.~Quemar, A.~Bonfiglioli, M.~Ricchiuto, Shock-fitting and
  predictor-corrector explicit ale residual distribution, in: M.~Onofri,
  R.~Paciorri (Eds.), Shock Fitting: Classical Techniques, Recent Developments,
  and Memoirs of Gino Moretti, Springer International Publishing, Cham, 2017,
  pp. 113--129.
\newblock \href {https://doi.org/10.1007/978-3-319-68427-7_5}
  {\path{doi:10.1007/978-3-319-68427-7_5}}.

\bibitem{Assonitis2020124}
A.~Assonitis, R.~Paciorri, A.~Bonfiglioli, Numerical simulation of shock
  boundary layer interaction using shock fitting technique, Lecture Notes in
  Mechanical Engineering (2020) 124--134\href
  {https://doi.org/10.1007/978-3-030-41057-5_10}
  {\path{doi:10.1007/978-3-030-41057-5_10}}.

\bibitem{lani2017sf}
A.~Lani, V.~De~Amicis, Sf: An open source object-oriented platform for
  unstructured shock-fitting methods, in: M.~Onofri, R.~Paciorri (Eds.), Shock
  Fitting: Classical Techniques, Recent Developments, and Memoirs of Gino
  Moretti, Springer International Publishing, Cham, 2017, pp. 85--112.
\newblock \href {https://doi.org/10.1007/978-3-319-68427-7_4}
  {\path{doi:10.1007/978-3-319-68427-7_4}}.

\bibitem{coolfluid-web-page}
A.~Lani, al., \href{https://github.com/andrealani/COOLFluiD/wiki}{{COOLF}lui{D}
  {W}iki page} (2020).
\newline\urlprefix\url{https://github.com/andrealani/COOLFluiD/wiki}

\bibitem{Scovazzi3}
T.~Song, A.~Main, G.~Scovazzi, M.~Ricchiuto, The shifted boundary method for
  hyperbolic systems: Embedded domain computations of linear waves and shallow
  water flows, Journal of Computational Physics 369 (2018) 45--79.
\newblock \href {https://doi.org/10.1016/j.jcp.2018.04.052}
  {\path{doi:10.1016/j.jcp.2018.04.052}}.

\bibitem{ciallella2020extrapolated}
M.~Ciallella, M.~Ricchiuto, R.~Paciorri, A.~Bonfiglioli, Extrapolated shock
  tracking: Bridging shock-fitting and embedded boundary methods, Journal of
  Computational Physics 412 (2020) 109440.
\newblock \href {https://doi.org/https://doi.org/10.1016/j.jcp.2020.109440}
  {\path{doi:https://doi.org/10.1016/j.jcp.2020.109440}}.

\bibitem{bonfiglioli2013three-dimensional}
A.~Bonfiglioli, M.~Grottadaurea, R.~Paciorri, F.~Sabetta, An unstructured,
  three-dimensional, shock-fitting solver for hypersonic flows, Computers \&
  Fluids 73~(0) (2013) 162--174.
\newblock \href {https://doi.org/10.1016/j.compfluid.2012.12.022}
  {\path{doi:10.1016/j.compfluid.2012.12.022}}.

\bibitem{zaide2014inserting}
D.~W. Zaide, C.~F. Ollivier-Gooch, Inserting a curve into an existing two
  dimensional unstructured mesh, in: Proceedings of the 22nd International
  Meshing Roundtable, Springer, 2014, pp. 93--107.

\bibitem{zaide2016inserting}
D.~W. Zaide, C.~F. Ollivier-Gooch, Inserting a surface into an existing
  unstructured mesh, Int. J. Numer. Methods Eng 106~(6) (2016) 484--500.
\newblock \href {https://doi.org/10.1002/nme.5132}
  {\path{doi:10.1002/nme.5132}}.

\bibitem{zaide2017inserting}
D.~Zaide, C.~Ollivier-Gooch, Inserting a Shock Surface into an Existing
  Unstructured Mesh, Springer International Publishing, Cham, 2017.
\newblock \href {https://doi.org/10.1007/978-3-319-68427-7_7}
  {\path{doi:10.1007/978-3-319-68427-7_7}}.

\bibitem{Liu2017}
J.~Liu, D.~Zou, A Shock-Fitting Technique for ALE Finite Volume Methods on
  Unstructured Dynamic Meshes, Springer International Publishing, Cham, 2017,
  pp. 131--149.
\newblock \href {https://doi.org/10.1007/978-3-319-68427-7_6}
  {\path{doi:10.1007/978-3-319-68427-7_6}}.

\bibitem{zou2017shock}
D.~Zou, C.~Xu, H.~Dong, J.~Liu, A shock-fitting technique for cell-centered
  finite volume methods on unstructured dynamic meshes, Journal of
  Computational Physics 345 (2017) 866 -- 882.
\newblock \href {https://doi.org/https://doi.org/10.1016/j.jcp.2017.05.047}
  {\path{doi:https://doi.org/10.1016/j.jcp.2017.05.047}}.

\bibitem{Chang2019}
S.~Chang, X.~Bai, D.~Zou, Z.~Chen, J.~Liu, An adaptive discontinuity fitting
  technique on unstructured dynamic grids, Shock Waves 29~(8) (2019)
  1103--1115.
\newblock \href {https://doi.org/10.1007/s00193-019-00913-3}
  {\path{doi:10.1007/s00193-019-00913-3}}.

\bibitem{ZOU2021104847}
D.~Zou, A.~Bonfiglioli, R.~Paciorri, J.~Liu, An embedded shock-fitting
  technique on unstructured dynamic grids, Computers \& Fluids 218 (2021)
  104847.
\newblock \href
  {https://doi.org/https://doi.org/10.1016/j.compfluid.2021.104847}
  {\path{doi:https://doi.org/10.1016/j.compfluid.2021.104847}}.

\bibitem{DAQUILA2021110096}
L.~M. D'Aquila, B.~T. Helenbrook, A.~Mazaheri, A novel stabilization method for
  high-order shock fitting with finite element methods, Journal of
  Computational Physics (2021) 110096\href
  {https://doi.org/10.1016/j.jcp.2020.110096}
  {\path{doi:10.1016/j.jcp.2020.110096}}.

\bibitem{Zahr2018b}
M.~Zahr, P.-O. Persson, An optimization-based approach for high-order accurate
  discretization of conservation laws with discontinuous solutions, Journal of
  Computational Physics 365 (2018) 105 -- 134.
\newblock \href {https://doi.org/https://doi.org/10.1016/j.jcp.2018.03.029}
  {\path{doi:https://doi.org/10.1016/j.jcp.2018.03.029}}.

\bibitem{zahr2020implicit}
M.~J. Zahr, A.~Shi, P.-O. Persson, Implicit shock tracking using an
  optimization-based high-order discontinuous galerkin method, Journal of
  Computational Physics (2020) 109385\href
  {https://doi.org/10.1016/j.jcp.2020.109385}
  {\path{doi:10.1016/j.jcp.2020.109385}}.

\bibitem{Corrigan2019moving}
A.~Corrigan, A.~D. Kercher, D.~A. Kessler, A moving discontinuous {G}alerkin
  finite element method for flows with interfaces, International Journal for
  Numerical Methods in Fluids 89~(9) (2019) 362--406.
\newblock \href {https://doi.org/10.1002/fld.4697}
  {\path{doi:10.1002/fld.4697}}.

\bibitem{Corrigan2020}
A.~D. Kercher, A.~Corrigan, D.~A. Kessler, The moving discontinuous galerkin
  finite element method with interface condition enforcement for compressible
  viscous flows, International Journal for Numerical Methods in Fluids
  n/a~(n/a).
\newblock \href {https://doi.org/https://doi.org/10.1002/fld.4939}
  {\path{doi:https://doi.org/10.1002/fld.4939}}.

\bibitem{Frontier++}
\href{http://www.ams.sunysb.edu/~linli/FronTier++_Manual/index.html}{\texttt{FronTier++}:
  a \texttt{C/C++} library for front tracking}.
\newline\urlprefix\url{http://www.ams.sunysb.edu/~linli/FronTier++_Manual/index.html}

\bibitem{glimm1998front}
J.~Glimm, M.~Graham, J.~Grove, X.~Li, T.~Smith, D.~Tan, F.~Tangerman, Q.~Zhang,
  Front tracking in two and three dimensions, Computers \& Mathematics with
  Applications 35~(7) (1998) 1--11, advanced Computing on Intel Architectures.
\newblock \href {https://doi.org/https://doi.org/10.1016/S0898-1221(98)00028-5}
  {\path{doi:https://doi.org/10.1016/S0898-1221(98)00028-5}}.

\bibitem{glimm1998three}
J.~Glimm, J.~W. Grove, X.~L. Li, K.-m. Shyue, Y.~Zeng, Q.~Zhang,
  Three-dimensional front tracking, SIAM Journal on Scientific Computing 19~(3)
  (1998) 703--727.
\newblock \href {https://doi.org/10.1137/S1064827595293600}
  {\path{doi:10.1137/S1064827595293600}}.

\bibitem{glimm1999simple}
J.~Glimm, J.~W. Grove, X.~L. Li, N.~Zhao, Simple front tracking, in: Nonlinear
  partial differential equations ({E}vanston, {IL}, 1998), Vol. 238 of Contemp.
  Math., Amer. Math. Soc., Providence, RI, 1999, pp. 133--149.
\newblock \href {https://doi.org/10.1090/conm/238/03544}
  {\path{doi:10.1090/conm/238/03544}}.

\bibitem{du2006simple}
J.~Du, B.~Fix, J.~Glimm, X.~Jia, X.~Li, Y.~Li, L.~Wu, A simple package for
  front tracking, Journal of Computational Physics 213~(2) (2006) 613--628.
\newblock \href {https://doi.org/https://doi.org/10.1016/j.jcp.2005.08.034}
  {\path{doi:https://doi.org/10.1016/j.jcp.2005.08.034}}.

\bibitem{SHE2016383}
D.~She, R.~Kaufman, H.~Lim, J.~Melvin, A.~Hsu, J.~Glimm, Chapter 15 -
  front-tracking methods, in: R.~Abgrall, C.-W. Shu (Eds.), Handbook of
  Numerical Methods for Hyperbolic Problems, Vol.~17 of Handbook of Numerical
  Analysis, Elsevier, 2016, pp. 383--402.
\newblock \href {https://doi.org/https://doi.org/10.1016/bs.hna.2016.07.004}
  {\path{doi:https://doi.org/10.1016/bs.hna.2016.07.004}}.

\bibitem{flashcode}
\href{http://flash.uchicago.edu/site/flashcode/}{The {FLASH} code}, {F}lash
  Center for Computational Science at the University of Chicago (2021).
\newline\urlprefix\url{http://flash.uchicago.edu/site/flashcode/}

\bibitem{Triangle}
J.~R. Shewchuk, \href{http://www.cs.cmu.edu/~quake/triangle.html}{Triangle: A
  two-dimensional quality mesh generator and delaunay triangulator.} (2005).
\newline\urlprefix\url{http://www.cs.cmu.edu/~quake/triangle.html}

\bibitem{shewchuk1996engineering}
J.~R. Shewchuk, Triangle: {E}ngineering a {2D} {Q}uality {M}esh {G}enerator and
  {D}elaunay {T}riangulator, in: M.~C. Lin, D.~Manocha (Eds.), Applied
  Computational Geometry: Towards Geometric Engineering, Vol. 1148 of Lecture
  Notes in Computer Science, Springer-Verlag, 1996, pp. 203--222, from the
  First ACM Workshop on Applied Computational Geometry.
\newblock \href {https://doi.org/10.1007/BFb0014497}
  {\path{doi:10.1007/BFb0014497}}.

\bibitem{Burkardt++}
\href{https://people.sc.fsu.edu/~jburkardt/index.html}{John {B}urkardt's home
  page}, last accessed: January 19, 2021.
\newline\urlprefix\url{https://people.sc.fsu.edu/~jburkardt/index.html}

\bibitem{bonfiglioli2000fluctuation}
A.~Bonfiglioli, Fluctuation splitting schemes for the compressible and
  incompressible {E}uler and {N}avier-{S}tokes equations, Int. J. Comput. Fluid
  Dyn. 14~(1) (2000) 21--39.
\newblock \href {https://doi.org/10.1080/10618560008940713}
  {\path{doi:10.1080/10618560008940713}}.

\bibitem{bonfiglioli2013mass-matrix}
A.~Bonfiglioli, R.~Paciorri, A mass-matrix formulation of unsteady fluctuation
  splitting schemes consistent with {Roe's} parameter vector, International
  Journal of Computational Fluid Dynamics 27~(4-5) (2013) 210--227.
\newblock \href {https://doi.org/10.1080/10618562.2013.813491}
  {\path{doi:10.1080/10618562.2013.813491}}.

\bibitem{arpaia2014ale}
L.~Arpaia, M.~Ricchiuto, R.~Abgrall, An {ALE} formulation for explicit
  {Runge-Kutta} residual distribution, Journal of Scientific Computing (2014)
  1--46\href {https://doi.org/10.1007/s10915-014-9910-5}
  {\path{doi:10.1007/s10915-014-9910-5}}.

\bibitem{ricchiuto2010explicit}
M.~Ricchiuto, R.~Abgrall, Explicit {Runge-Kutta} residual distribution schemes
  for time dependent problems: Second order case, Journal of Computational
  Physics 229~(16) (2010) 5653 -- 5691.
\newblock \href {https://doi.org/10.1016/j.jcp.2010.04.002}
  {\path{doi:10.1016/j.jcp.2010.04.002}}.

\bibitem{ricchiuto2015explicit}
M.~Ricchiuto, An explicit residual based approach for shallow water flows,
  Journal of Computational Physics 280 (2015) 306 -- 344.
\newblock \href {https://doi.org/https://doi.org/10.1016/j.jcp.2014.09.027}
  {\path{doi:https://doi.org/10.1016/j.jcp.2014.09.027}}.

\bibitem{IDOLIKE}
H.~N. (as K.~Masatsuka), I Do Like CFD, 2013, iSBN 9781304827937, {\tt
  www.cfdbooks.com/}.

\bibitem{anderson}
J.~Anderson, \href{https://books.google.it/books?id=xwY8PgAACAAJ}{Fundamentals
  of Aerodynamics}, McGraw-Hill Education, 2010.
\newline\urlprefix\url{https://books.google.it/books?id=xwY8PgAACAAJ}

\bibitem{onofri2017shock}
M.~Onofri, R.~Paciorri, Shock Fitting: Classical Techniques, Recent
  Developments, and Memoirs of Gino Moretti, Springer, 2017.

\bibitem{TADMOR1986211}
E.~Tadmor, A minimum entropy principle in the gas dynamics equations, Applied
  Numerical Mathematics 2~(3) (1986) 211 -- 219, special Issue in Honor of Milt
  Rose's Sixtieth Birthday.
\newblock \href {https://doi.org/10.1016/0168-9274(86)90029-2}
  {\path{doi:10.1016/0168-9274(86)90029-2}}.

\bibitem{ar2017}
R.~Abgrall, M.~Ricchiuto, High-Order Methods for CFD, John Wiley \& Sons, Ltd,
  2017, pp. 1--54.
\newblock \href {https://doi.org/10.1002/9781119176817.ecm2112}
  {\path{doi:10.1002/9781119176817.ecm2112}}.

\bibitem{dr2017}
H.~Deconinck, M.~Ricchiuto, Residual Distribution Schemes: Foundations and
  Analysis, John Wiley \& Sons, Ltd, 2017, pp. 1--53.
\newblock \href {https://doi.org/10.1002/0470091355.ecm054}
  {\path{doi:10.1002/0470091355.ecm054}}.

\bibitem{abg99b}
R.~Abgrall, K.~Mer, B.~Nkonga, A {L}ax--{W}endroff type theorem for residual
  schemes, in: M.~Haffez, J.~Chattot (Eds.), Innovative methods for numerical
  solutions of partial differential equations, World Scientific, 2002, pp.
  243--266.
\newblock \href {https://doi.org/10.1142/9789812810816}
  {\path{doi:10.1142/9789812810816}}.

\bibitem{abg2001d}
R.~Abgrall, P.~L. Roe, High order fluctuation schemes on triangular meshes,
  Journal of Scientific Computing 19~(1) (2003) 3--36.
\newblock \href {https://doi.org/10.1023/A:1025335421202}
  {\path{doi:10.1023/A:1025335421202}}.

\bibitem{Roe:81}
P.~Roe, Approximate riemann solvers, parameter vectors, and difference schemes,
  Journal of Computational Physics 43~(2) (1981) 357--372.
\newblock \href {https://doi.org/10.1016/0021-9991(81)90128-5}
  {\path{doi:10.1016/0021-9991(81)90128-5}}.

\bibitem{deconinck93}
H.~Deconinck, P.~Roe, R.~Struijs, A multidimensional generalization of {Roe's}
  flux difference splitter for the euler equations, Computers {\&} Fluids
  22~(2-3) (1993) 215--222.
\newblock \href {https://doi.org/10.1016/0045-7930(93)90053-c}
  {\path{doi:10.1016/0045-7930(93)90053-c}}.

\bibitem{crd02}
A.~Cs\'{\i}k, M.~Ricchiuto, H.~Deconinck, A conservative formulation of the
  multidimensional upwind residual distribution schemes for general nonlinear
  conservation laws, Journal of Computational Physics 179~(1) (2002) 286--312.
\newblock \href {https://doi.org/https://doi.org/10.1006/jcph.2002.7057}
  {\path{doi:https://doi.org/10.1006/jcph.2002.7057}}.

\bibitem{rcd05}
M.~Ricchiuto, A.~Cs\'{\i}k, H.~Deconinck, Residual distribution for general
  time-dependent conservation laws, Journal of Computational Physics 209~(1)
  (2005) 249--289.
\newblock \href {https://doi.org/10.1016/j.jcp.2005.03.003}
  {\path{doi:10.1016/j.jcp.2005.03.003}}.

\bibitem{abg99f}
R.~Abgrall, Toward the ultimate conservative scheme~: Following the quest,
  Journal of Computational Physics 167~(2) (2001) 277--315.
\newblock \href {https://doi.org/10.1006/jcph.2000.6672}
  {\path{doi:10.1006/jcph.2000.6672}}.

\bibitem{vanderweide}
E.~van~der Weide, H.~Deconinck, Positive matrix distribution schemes for
  hyperbolic systems, in: Computational Fluid Dynamics, Wiley, New York, 1996,
  pp. 747--753.

\bibitem{abg2001c_steady}
R.~Abgrall, M.~Mezine, Construction of second-order accurate monotone and
  stable residual distribution schemes for steady problems, Journal of
  Computational Physics 195~(2) (2004) 474 -- 507.
\newblock \href {https://doi.org/https://doi.org/10.1016/j.jcp.2003.09.022}
  {\path{doi:https://doi.org/10.1016/j.jcp.2003.09.022}}.

\bibitem{Roe:87}
P.~L. Roe, Linear advection schemes on triangular meshes, Tech. Rep. CoA 8720,
  Cranfield Institute of Technology (1987).

\bibitem{Roe:90}
P.~L. Roe, ``optimum'' upwind advection on a triangular mesh, Tech. Rep. ICASE
  90-75, ICASE (1990).

\bibitem{sidilkover}
P.~Roe, D.~Sidilkover, Optimum positive linear schemes for advection in two and
  three dimensions., {SIAM} J. Numer. Anal. 29~(6) (1992) 1542--1568.
\newblock \href {https://doi.org/10.1137/0729089} {\path{doi:10.1137/0729089}}.

\bibitem{pd97b}
H.~Paillere, H.~Deconinck, Multidimensional upwind residual distribution
  schemes for the 2d euler equations, in: H.~Deconinck, B.~Koren (Eds.), Notes
  on Numerical Fluid Mechanics, Vieweg-Verlag, Braunschweig, Germany, 1997, pp.
  51--112.

\bibitem{bonfiglioli1997}
A.~Bonfiglioli, H.~Deconinck, Multidimensional upwind residual distribution
  schemes for the 3d euler equations, in: H.~Deconinck, B.~Koren (Eds.), Notes
  on Numerical Fluid Mechanics, Vieweg-Verlag, Braunschweig, Germany, 1997, pp.
  141--185.

\bibitem{hughes2010stabilized}
T.~J.~R. Hughes, G.~Scovazzi, T.~E. Tezduyar, Stabilized methods for
  compressible flows, Journal of Scientific Computing 43~(3) (2010) 343--368.
\newblock \href {https://doi.org/10.1007/s10915-008-9233-5}
  {\path{doi:10.1007/s10915-008-9233-5}}.

\bibitem{hubbard}
M.~Hubbard, P.~Roe, Compact high-resolution algorithms for time-dependent
  advection on unstructured grids, International Journal for Numerical Methods
  in Fluids 33~(5) (2000) 711--736.
\newblock \href
  {https://doi.org/10.1002/1097-0363(20000715)33:5<711::AID-FLD27>3.0.CO;2-O}
  {\path{doi:10.1002/1097-0363(20000715)33:5<711::AID-FLD27>3.0.CO;2-O}}.

\bibitem{csikjournal}
A.~Cs\'{\i}k, H.~Deconinck, Space-time residual distribution schemes for
  hyperbolic conservation laws on unstructured linear finite elements,
  International Journal for Numerical Methods in Fluids 40~(3‐4) (2002)
  573--581.
\newblock \href {https://doi.org/10.1002/fld.315} {\path{doi:10.1002/fld.315}}.

\bibitem{doru_CF}
D.~Caraeni, L.~Fuchs, Compact third-order multidimensional upwind scheme for
  {N}avier-{S}tokes simulations, Theoretical and Computational Fluid Dynamics
  15~(6) (2002) 373--401.
\newblock \href {https://doi.org/10.1007/s00162-002-0060-2}
  {\path{doi:10.1007/s00162-002-0060-2}}.

\bibitem{abg2001c}
R.~Abgrall, M.~Mezine, Construction of second-order accurate monotone and
  stable residual distribution schemes for unsteady flow problems., Journal of
  Computational Physics 188~(1) (2003) 16 -- 55.
\newblock \href {https://doi.org/10.1016/S0021-9991(03)00084-6}
  {\path{doi:10.1016/S0021-9991(03)00084-6}}.

\bibitem{HUBBARD2011263}
M.~Hubbard, M.~Ricchiuto, Discontinuous upwind residual distribution: A route
  to unconditional positivity and high order accuracy, Computers \& Fluids
  46~(1) (2011) 263 -- 269.
\newblock \href {https://doi.org/10.1016/j.compfluid.2010.12.023}
  {\path{doi:10.1016/j.compfluid.2010.12.023}}.

\bibitem{ABGRALL-dec}
R.~Abgrall, High order schemes for hyperbolic problems using globally
  continuous approximation and avoiding mass matrices, Journal of Scientific
  Computing 73~(2) (2017) 461--494.
\newblock \href {https://doi.org/10.1007/s10915-017-0498-4}
  {\path{doi:10.1007/s10915-017-0498-4}}.

\bibitem{ABGRALL2019274}
R.~Abgrall, P.~Bacigaluppi, S.~Tokareva, High-order residual distribution
  scheme for the time-dependent euler equations of fluid dynamics, Computers \&
  Mathematics with Applications 78~(2) (2019) 274 -- 297.
\newblock \href {https://doi.org/https://doi.org/10.1016/j.camwa.2018.05.009}
  {\path{doi:https://doi.org/10.1016/j.camwa.2018.05.009}}.

\bibitem{mullerdelaundo}
J.~D. M{\"u}ller,
  \href{http://www.ae.metu.edu.tr/tuncer/ae546/prj/delaundo/}{Grid generation
  tools: Delaunay triangulation with delaundo}.
\newline\urlprefix\url{http://www.ae.metu.edu.tr/tuncer/ae546/prj/delaundo/}

\bibitem{10.1007/BFb0019764}
G.~Moretti, On the matter of shock fitting, in: R.~D. Richtmyer (Ed.),
  Proceedings of the Fourth International Conference on Numerical Methods in
  Fluid Dynamics, Springer, Berlin, Heidelberg, 1975, pp. 287--292.
\newblock \href {https://doi.org/10.1007/BFb0019764}
  {\path{doi:10.1007/BFb0019764}}.

\bibitem{bonfiglioli2018unstructured}
A.~Bonfiglioli, R.~Paciorri, Unstructured shock-fitting calculations of the
  transonic flow in a gas turbine cascade, Aerotecnica Missili {\&} Spazio
  97~(4) (2018) 189--197.
\newblock \href {https://doi.org/10.1007/BF03406053}
  {\path{doi:10.1007/BF03406053}}.

\bibitem{key-7}
F.~Grasso, S.~Pirozzoli, Shock-wave-vortex interactions: Shock and vortex
  deformations, and sound production, Theoretical and Computational Fluid
  Dynamics 13~(6) (2000) 421--456.
\newblock \href {https://doi.org/10.1007/s001620050121}
  {\path{doi:10.1007/s001620050121}}.

\bibitem{paciorri2020accurate}
R.~Paciorri, A.~Bonfiglioli, Accurate detection of shock waves and shock
  interactions in two-dimensional shock-capturing solutions, Journal of
  Computational Physics 406 (2020) 109196.
\newblock \href {https://doi.org/10.1016/j.jcp.2019.109196}
  {\path{doi:10.1016/j.jcp.2019.109196}}.

\bibitem{bonfiglioli2017development}
A.~Bonfiglioli, R.~Paciorri, L.~Campoli, V.~De~Amicis, M.~Onofri, Development
  of an unsteady shock-fitting technique for unstructured grids, in:
  G.~Ben-Dor, O.~Sadot, O.~Igra (Eds.), 30th International Symposium on Shock
  Waves 2, Springer International Publishing, Cham, 2017, pp. 1501--1504.
\newblock \href {https://doi.org/10.1007/978-3-319-44866-4_124}
  {\path{doi:10.1007/978-3-319-44866-4_124}}.

\bibitem{paciorri2017basic}
R.~Paciorri, A.~Bonfiglioli, Basic elements of unstructured shock-fitting:
  Results achieved and future developments, in: M.~Onofri, R.~Paciorri (Eds.),
  Shock Fitting: Classical Techniques, Recent Developments, and Memoirs of Gino
  Moretti, Springer International Publishing, Cham, 2017, pp. 59--84.
\newblock \href {https://doi.org/{10.1007/978-3-319-68427-7_3}}
  {\path{doi:{10.1007/978-3-319-68427-7_3}}}.

\end{thebibliography}


\begin{thebibliography}{10}
%\bibitem{key-1-1}TecIO Library, Tecplot Inc. proprietary library for I/O binary files in Tecplot format, http://www.tecplot.com/downloads/tecio-library/.

\bibitem{key-2-1} Git, a free and open source distributed version
control system, \url{http://git-scm.com}.

\bibitem{key-4-1} \textit{UnDiFi-2D} documentation, \url{https://github.com/UnDiFi/UnDiFi-2D/wiki}.

\bibitem{key-3-1} Github, a web-based hosting service for software
development projects using git versioning system, \url{https://github.com}.

%\bibitem{key-5-1} Doxygen, a documentation system for many programming languages, http://www.stack.nl/dimitri/ doxygen.

\bibitem{key-6-1} Dominici, M. (2014). An overview of Pandoc. TUGboat, 35(1), 44-50.
\end{thebibliography}

\end{document}